\documentclass[11pt]{article}
\pdfoutput=1

\usepackage[utf8]{inputenc}

\usepackage[hidelinks]{hyperref}

\usepackage{xcolor}

\usepackage{caption}
\usepackage{enumerate}
\usepackage{caption}
\usepackage{subcaption}
\newcommand{\he}{\hat{e}}

\newcommand{\operator}[1]{\hat{#1}}

\newcommand{\R}{\mathbb{R}}

\newcommand{\Z}{\mathbb{Z}}

\newcommand{\T}{\mathbb{T}}

\newcommand{\pauli}[1]{
    \ifnum#1=1
        \operator{\sigma}_{x}
    \else
        \ifnum#1=2
           \operator{\sigma}_{y}
        \else
            \ifnum#1=3
                \operator{\sigma}_{z}
            \else
                \errmessage{Incorrect number given to pauli}
            \fi
        \fi
    \fi
}

\usepackage{geometry}
\usepackage{comment}

\usepackage{soul}
\usepackage{subcaption}

\usepackage{hyperref}
\usepackage{amsmath,amssymb,amsthm}
\usepackage{bm}
\usepackage{slashed}
\usepackage{graphicx}
\usepackage[T1]{fontenc}
\usepackage{fix-cm}

\makeatletter
\newcommand*\bigcdot{\mathpalette\bigcdot@{.5}}
\newcommand*\bigcdot@[2]{\mathbin{\vcenter{\hbox{\scalebox{#2}{$\m@th#1\bullet$}}}}}
\makeatother

\newlength{\dummysp}
\usepackage{mwe}

\def\R{{\mathbb R}}
\def\S{{\mathbb S}}
\def\Z{{\mathbb Z}}
\def\T{{\mathbb T}}

\def\tr{\,{\rm tr}\,}

\usepackage{authblk}
\usepackage{hyperref}
\newcommand*{\email}[1]{%
    \normalsize\href{mailto:#1}{#1}\par
    }
 \title{\bf Metamorphosis of  fractional instantons on a twisted $\mathbf{\T^4}$ with a double-trace deformation: a numerical study }

\date{}

\author{\bf  Benjamin Dobozy and Erich Poppitz}

\affil{Department of Physics, University of Toronto\\ 60 St George St, Toronto, ON M5S 1A7, Canada\\ \email{ben.dobozy@mail.utoronto.ca, erich.poppitz@utoronto.ca}}

\begin{document}

\maketitle

\thispagestyle{empty}

\vspace{1.4cm}
\begin{abstract} 
We use numerical minimization of the lattice action of trace-deformed Yang-Mills theory on $\T^4$ with twisted boundary conditions to find the classical minimum action configurations of fractional topological charge. We vary the twists and ratios of torus periods to interpolate between different $\R^{4-k} \times \T^k$  geometries.  This allows us to see how the corresponding minimum action saddle point configurations---monopole-instantons ($k=1$), center vortices ($k=2$), and fractional instantons ($k=3,4$)---morph into each other. We also study how the transition between them depends on the presence of a deformation potential. In particular, we argue that the recent analytic picture of chains of monopole-instantons collimating their flux into center-vortex sheets, while technically relying on the deformation potential, also holds in pure Yang-Mills theory, for tori whose shape causes  the abelianization due to the deformation to align with the one due to the twists. Our results also indicate that with nonzero deformation potential, some transitions between different minimal-action fractional charge configurations may be discontinuous and involve level crossing.  
\end{abstract}

%%%%%%%%%%%%%%%%%%%%%%%%%%%%%%%%%%%%%%%%%%
\setcounter{page}{0}
\setcounter{section}{0} %% Remove this when starting to work on the template.

%\end{paracol}

\tableofcontents

\section{Introduction}

\subsection{Motivation and some background}
\label{sec:motivation}

Understanding the nonperturbative dynamics of four-dimensional non-abelian gauge theories from first principles is a challenging problem lacking a  complete solution. While many approaches exist, none of them can account for all interesting aspects of the physics. This paper is devoted to  one particular direction of research:  the study of the weak-coupling semiclassical dynamics of $SU(N)$ gauge theories on compact spaces of the form $\R^{4-k} \times \T^k_{*}$. Here $\T^k_{*}$ denotes a $k$ dimensional torus with $k=1,2,3,4$. The subscript $*$ indicates the  inclusion of  at least one of 't Hooft twisted boundary conditions \cite{tHooft:1979rtg,tHooft:1981sps} or a double-trace deformation  \cite{Unsal:2008ch}. We use $\Lambda$ to denote the strong coupling scale of the gauge theory. We take  Euclidean signature in all $\R^{4-k} \times \T^{k}_*$ cases, except on a few occasions where the use of  Minkowski signature is explicitly stated.  
 
 If the  size of the torus  $L_{\T^k}$ is small,\footnote{For simplicity, the discussion here only refers to the overall size of the $\T^k$, denoted by $L_{\T^k}$,  tacitly assuming that all periods of the torus are  of the same order. Later, we relax this assumption, considering different ratios of periods of $\T^k$.} such that $L_{\T^k} N \Lambda \ll 1$, one can show that the $\R^{4-k} \times \T^k_{*}$ theory abelianizes: the gauge group is Higgsed, by the boundary conditions or the deformation,   $SU(N) \rightarrow  U(1)^{N-1}$ or $SU(N) \rightarrow  \Z_N$, at a high  scale ${1 \over L_{\T^k} N} \gg \Lambda$, ensuring that the   theory  is weakly coupled. Let us be slightly more precise: for $k=1,2$, the long-distance $\R^3$ or $\R^2$  theory is weakly coupled because of the abelianization. For $k=3,4$, the theory on  $\R\times \T^3$ or $\T^4$ is weakly coupled at small  $L_{\T^k}$ even without  the nontrivial 't Hooft twists or deformation---this is the ``femtouniverse'' of \cite{Bjorken:1979hv,Luscher:1982ma}. However, now  the small-$L_{\T^k}$ and large-$L_{\T^k}$ theories are believed to not be continuously connected in the sense discussed below. The reason for our  cautious statement is that the continuous classical vacuum degeneracy  in compactifications without  't Hooft twists or deformations complicates\footnote{For recent work on this problem in theories with supersymmetry, see \cite{ArabiArdehali:2026kvt}.} the semiclassical study of the ground state properties of the femtouniverse---a task that has not yet been completed, to the best of our knowledge. For this reason, we only study  $\R^{4-k} \times \T^k_{*}$ theories with 't Hooft twists and/or deformation.

 The weak coupling at small $L_{\T^k}$  permits the use of theoretically controlled semiclassical methods to study ground state properties, $\theta$-angle dependence, symmetry realization, spectra, etc.; various aspects are reviewed in \cite{Dunne:2016nmc,Poppitz:2021cxe,Gonzalez-Arroyo:2023kqv}. Remarkably, it is found that the properties of the ground state of the pure gauge theory, obtained analytically at small $L_{\T^k}$, evolve continuously into the ones seen in the strongly-coupled  $L_{\T^k} \rightarrow \infty$, or $\R^4$,  limit. Demonstrating this  continuity, known as ``adiabatic continuity,''  requires, at the current state of the art,  the use of numerical simulations to leave the weakly-coupled small-$L_{\T^k}$ regime; for certain quantities in supersymmetric theories, one can appeal to nonrenormalization theorems (see \cite{Anber:2024mco})  to argue for small to large volume continuity. 
Here, we shall not review further details of  the semiclassical dynamics nor discuss different ideas to connect the semiclassical and the strongly coupled limits; for this, we recommend the reviews cited above. 

The purpose of this paper is to  study  the relation between the nontrivial saddles in the path integral on  $\R^{4-k} \times \T^k_{*}$ for different $k$. These are finite action instantons  contributing to the nonperturbative dynamics. We begin by listing the kind of finite action configurations that give the leading semiclassical contribution to the nonperturbative dynamics. We use the antisymmetric  twist tensor ``$n_{\mu\nu}$,'' defined (mod $N$) in \cite{tHooft:1979rtg}, to denote the imposition of nontrivial  twisted boundary conditions in the $\T^k$ (without, at this point, being explicit about the particular choice made). Likewise, we use ``def.'' to denote the addition of a double-trace deformation  \cite{Unsal:2008ch}. 
The upshot is that, for the different values of $k$, the pattern of gauge symmetry breaking, the kinds of  minimal action saddles, their ``quantum numbers'' (e.g. topological charge $Q$), as well their localization properties, are found to be as follows:
\begin{eqnarray} \label{s1}
\R^3 \times \S^1_{\text{def.}}: &&SU(N) \rightarrow U(1)^{N-1}, \; N \; \text{kinds of monopole-instantons}, \; Q={1 \over N}, \\
&& \text{localized in} \; \R^3 \; \text{and wrapped around the} \; \S^1, \nonumber \\
&& \text{with magnetic charges under} \; U(1)^{N-1} \;\text{labelled by}\; (\alpha_0,\alpha_1, ...,\alpha_{N-1}),\nonumber
\\ \label{t2}
\R^2 \times \T^2_{n_{\mu\nu}}: &&SU(N) \rightarrow \Z_N, \;  \text{center vortices}, \; Q={1 \over N},
\\  
&& \text{localized in} \; \R^2 \; \text{and wrapped around } \; \T^2, \nonumber \\
\label{t3}
\R \times \T^3_{n_{\mu\nu}}: & &SU(N) \rightarrow \Z_N,\; \text{fractional instantons}, \; Q={1 \over N}, 
\\  
&& \text{localized in} \; \R\; \text{and wrapped around } \; \T^3, \nonumber \\ 
\T^4_{n_{\mu\nu}}: &&SU(N) \rightarrow \Z_N, \;  \text{fractional instantons}, \; Q={1\over N}, \label{t4}\\
&& \text{localized or extended in some or all directions, depending on $n_{\mu\nu}$ and ratios of $\T^4$ periods.}  \nonumber
\end{eqnarray}
In each case, we have assumed that the choice of $n_{\mu\nu}$ twists is such that the minimal $Q={1 \over N}$; this  choice shall be made more explicit in the body of the paper.
Let us now  discuss various aspects of (\ref{s1})-(\ref{t4}):
\begin{enumerate}
\item For all values of $k$,  the minimum action semiclassical objects have topological charge $1/N$. 
For $k=1$, this has been known since the classification of finite action configurations on $\R^3 \times \S^1$ of \cite{Gross:1980br}, which implies that $Q=1/N$  at the center-symmetric point \cite{Unsal:2008ch}. For $k=2,3$, the existence of $Q={1/N}$ fractional instantons on $\R \times \T^3_{n_{\mu\nu}}$ and $\R^2 \times \T^2_{n_{\mu\nu}}$ has been realized already in \cite{RTN:1993ilw, Gonzalez-Arroyo:1995ynx,Gonzalez-Arroyo:1998hjb,Montero:2000pb}.\footnote{See  Appendix \ref{appx:classify} for a classification argument of finite action configurations on $\R^2 \times \T^2_{n_{12}}$ (and, by a simple extension, to $\R \times \T^3_{n_{12}}$). While the discussion of this semi-infinite volume limit classification seems absent in the literature, it appears to be known to many and we include it for completeness.} For $k=4$,   't Hooft argued  \cite{tHooft:1979rtg,tHooft:1981sps} that $Q=1/N$ for some choices of twists, later studied in great detail in  \cite{vanBaal:1982ag}.
These instantons are exactly self-dual (or, for $k=1$ with no  supersymmetry, approximately self-dual). Thus, their semiclassical contribution to the path integral scales as $\exp(- {8 \pi^2 \over g^2 N})$, where $g^2$ is the gauge theory coupling at a scale of order $L_{\T^k}$. As indicated, these finite action configurations are localized in the noncompact $\R^{4-k}$ and have size of order $L_{\T^k}$.

\item Notice that for $k=2,3,4$, there is a single, up to moduli, minimal action configuration  with $Q={1/N}$.  For $k=1$, on the other hand, there are $N$ different objects of fractional topological charge, distinguished by their $U(1)^{N-1}$ ``magnetic'' charges. These are labelled by the simple roots, $\alpha_1,...,\alpha_{N-1}$, of $SU(N)$, along with the affine root $\alpha_0$$=$$-(\alpha_1$$+$$\alpha_2$$+$$...$$+$$\alpha_{N-1})$. These monopole-instantons are the $N$ ``constituents'' of the BPST $Q=1$ instanton \cite{Lee:1997vp,Kraan:1998sn}.

\item In each of the cases with $k<4$, a dilute gas of the relevant $Q=1/N$ instantons disorders large Wilson loops in the noncompact $\R^{3}$ (case (\ref{s1})) or  $\R^2$ (case (\ref{t2})), leading to area law and confinement. For $k=1$, see \cite{Unsal:2008ch}. For $k=2$, confinement by center vortices is shown in \cite{Tanizaki:2022ngt}.  For $k=3$, Refs.~\cite{RTN:1993ilw, Gonzalez-Arroyo:1995ynx} showed that tunnelling due to fractional instantons causes the correlator of two winding Wilson loops, separated in $\R$, to obey the area law (this is recently reviewed in Section 6 of \cite{Anber:2025vjo}). 
In all cases, the string tensions in the pure gauge theory, computed in the leading semiclassical approximation (for the case (\ref{s1}), see \cite{Poppitz:2017ivi}) with exponential-only accuracy, are proportional to $\exp(- {8\pi^2 \over g^2 N})$.

\item The $\R^3 \times \S^1_{\text{def.}}$ and $\R^2 \times \T^2_{n_{\mu\nu}}$ cases, (\ref{s1}) and (\ref{t2}), are examples of semiclassically tractable mechanisms of confinement due to monopole-instantons and center vortices, respectively. These mechanisms have been extensively discussed in the lattice community, in the strong coupling regime of a large $\T^4$. There, monopoles and center vortices are identified after appropriate gauge fixing of   lattice configurations belonging to an ensemble generated via Monte Carlo techniques. It is observed that removing the configurations containing center vortices/monopoles from the ensemble destroys the area law. There are many important details;  for these and references, we recommend Greensite's  monograph \cite{Greensite:2011zz}.\footnote{For more recent work  discussing newer developments on the use of effective models of ensembles of unoriented center vortices, see Ref.~\cite{Junior:2025gxg}, also containing an updated list of  references.}

\item We will not dwell much on the $k=4$ case in this Introduction. This is because the study of $Q=1/N$ configurations on  $\T^4_{n_{\mu\nu}}$ is really the subject of most of the rest of the paper. Here, we only mention that fractional instantons on  $\T^4_{n_{\mu\nu}}$ have been used to calculate the gaugino condensate in super-Yang-Mills theory \cite{Anber:2022qsz,Anber:2023sjn,Anber:2024mco}, showing complete agreement (owing to the supersymmetric nonrenormalization theorems) with the $\R^4$ results reviewed in \cite{Dorey:2002ik}.

\item Our final remark is that there are few known analytic solutions with $Q=1/N$ in the geometries (\ref{s1})-(\ref{t4}). In fact, there are only two kinds: the $N$ BPS monopole instantons on $\R^3 \times \S^1$ \cite{Lee:1997vp} and 't Hooft's constant field strength fractional instantons on $\T^4$   \cite{tHooft:1981nnx}. Most of what is known about the saddle points mentioned in (\ref{s1})-(\ref{t4}) has been learned by numerically minimizing lattice actions with 't Hooft twists, largely due to the long-term efforts of the Madrid group  \cite{Gonzalez-Arroyo:2023kqv}.  In its use of numerical tools, this paper is not an exception.
\end{enumerate}
After a further look at the various semiclassical objects appearing in (\ref{s1})-(\ref{t4}) it should perhaps not come as a surprise that  they can be related to each other. This is easiest to contemplate for the $k=2,3,4$ cases, where the only ``quantum number'' is the topological charge $Q=1/N$. Thus, at least naively (however, see Section \ref{sec:morphing2}), one imagines compactifying a center vortex, a 2d sheet wrapped on $\T^2$, as in (\ref{t2}), on an $\S^1$ orthogonal to the sheet,  obtaining  (\ref{t3}); this can then be further compactified to obtain (\ref{t4}).
Conversely, one could begin with a $\T^4$ with appropriately chosen $n_{\mu\nu}$ twists (giving the minimal $Q=1/N$) and then take limits of its periods such that it approximates any one of the cases with $k=1,2,3$. 
This may be especially clear if one wants to obtain the $k=2,3$ cases, where the same twist in the $\T^2$ or $\T^3$  appearing in (\ref{t2}) or (\ref{t3})  can be imposed on the original $\T^4$---so that one obtains the desired $\R^2 \times \T^2_{n_{\mu\nu}}$ or $\R \times \T^3_{n_{\mu\nu}}$ in the limit.\footnote{A twist involving a $\T^4$ direction taken to infinity will not be relevant and  serves only to select the desired boundary condition, one of the many possible,  at  $\R^{2}$ or $\R$ infinity; see Appendix \ref{appx:classify}.}

Obtaining $\R^3 \times \S^1_{\text{def.}}$ from  (\ref{t2})-(\ref{t4}), however, is bound to be a little trickier. First of all, as  mentioned above, now there are $N$ monopole-instantons with different $U(1)^{N-1}$ magnetic charges, as per  (\ref{s1}), instead of a single $Q=1/N$ fractional instanton; we call this  the ``multiplicity problem.''   Second, if there is no deformation on the  $\T^4$ one starts with,  there is no way to strictly obtain $\R^3 \times \S^1_{\text{def.}}$,  a setup where abelianization only occurs because of  the deformation.\footnote{As explained in Section \ref{sec:rvwym}, at finite $\T^4$ periods, abelianization around a localized fractional instanton can occur without deformation potential, due to the twists alone, provided the torus periods satisfy particular relations. In some cases, this abelianization is aligned with the one due to the deformation (however, this does not occur in the strict $\R^3 \times \S^1$ limit; see  \cite{Wandler:2024hsq}).} These issues were tackled by the authors of \cite{Hayashi:2024yjc,Guvendik:2024umd}, who proposed replacing  $\R^3 \times \S^1_{\text{def.}}$ by $\R^2 \times \S^1 \times \S^1_{\text{def.}}$ and realized that  adding a 't Hooft twist in the thus-obtained two-torus  $\S^1 \times \S^1_{\text{def.}}$ solves the ``multiplicity problem.'' This is because an $\S^1$-period translation in the presence of a 't Hooft twist is a center symmetry transformation in the orthogonal $\S^1_{\text{def.}}$ direction. This center symmetry cyclically permutes the 
$N$ different monopole-instantons \cite{Anber:2015wha}, thus causing them to alternate on the covering space of the $\S^1$, in effect creating a single object involving a monopole-instanton and all its $N$ images under the $\S^1$ translation. Refs.~\cite{Hayashi:2024yjc,Guvendik:2024umd} then used  analytic tools familiar from $\R^3 \times \S^1$ studies of deformed Yang-Mills theory  \cite{Unsal:2008ch} and  classical electrodynamics to show that this single object acts like a center vortex: it localizes in the noncompact $\R^2$, is  wrapped around $\S^1 \times \S^1_{\text{def.}}$, and disorders the Wilson loops surrounding it. The resulting picture is   shown, for $N=2$, on our Figure~\ref{fig:picture1}, referring the reader to  Section \ref{sec:rvwcontinuity} for details.

The monopole-instanton/center-vortex continuity is an interesting new development, showing, in a theoretically controlled semiclassical framework, that these two confinement mechanisms are related. We note, however, that the relation between center vortices and monopole-instantons in the strong coupling regime has appeared earlier in the lattice literature---see  \cite{Greensite:2011zz}, in particular Ch.~8 there, where pictures of gauge-fixed lattice configurations appear,  identical to our Fig.~\ref{fig:picture1}.

However, it is satisfying to have a setup where a controllable  analytic argument  is  available: the semiclassical limit on small compact spaces gives a clear sense as to why these configurations dominate the path integral---in contrast with the phenomenological models of the strong coupling regime. Furthermore, weak coupling semiclassics on $\R^{4-k} \times \T^k_{*}$  allows one to treat properties and theories which are either very challenging or simply intractable at the current level of  development of lattice techniques. These include the multibranched structure and associated $\theta$-dependence of the vacuum\footnote{Required by consistency with various generalized anomalies \cite{Gaiotto:2017yup} (these are described in the old-fashioned language close to the one used here in \cite{Cox:2021vsa}).} and  the study of theories with massless fermions in various representations, including chiral gauge theories and general supersymmetric theories, as in  e.g.~\cite{Anber:2017pak,Hayashi:2024gxv,Hayashi:2024qkm,Hayashi:2023wwi,Tanizaki:2022plm}. As a concrete application of the recent developments, the picture of \cite{Hayashi:2024yjc,Guvendik:2024umd}  was used to explain some puzzles regarding confinement in supersymmetric theories \cite{Hayashi:2024psa,Hayashi:2025mgk}. A more speculative remark \cite{Guvendik:2024umd} is that viewing the  monopole-instantons as BPST instanton constituents---recalling Comment 2. after Eqns.~(\ref{s1}-\ref{t4})---suggests that the continuity is a hint that both monopole-instantons and center vortices may lurk inside the gas of four dimensional instantons, in a way waiting to be made more concrete; see \cite{Nguyen:2023rww,Nguyen:2025voy} for related developments.

In conclusion of this overview, we hope to have conveyed the idea that the study of how the different nonperturbative saddles appearing in (\ref{s1})-(\ref{t4}) morph into each other is of interest, as it sheds light on the relation between the  confinement mechanisms operating in the various geometries---and perhaps, ultimately, in the $\R^4$ limit.
We now continue to discuss in more detail the scope and results of this paper.

\subsection{Overview and summary of results}
\label{sec:overview}

In this paper, we further study the relation---or ``metamorphosis''---between the various minimal action instantons in the geometries (\ref{s1})-(\ref{t4}). We focus on $SU(2)$ Yang-Mills theory with a double-trace deformation potential, which we abbreviate as dYM, the deformed Yang-Mills theory  of \"Unsal and Yaffe \cite{Unsal:2008ch}. We use numerical minimization of the lattice action of dYM on $\T^4$, subject to twisted boundary conditions.
To motivate the use of numerics, we note that the analytic tools used in \cite{Hayashi:2024yjc,Guvendik:2024umd} to study the monopole-instanton/center-vortex continuity  apply in the limit where $L_{\S^1} \gg L_{\S^1_{\text{def}}}$ (see Section \ref{sec:rvwcontinuity}). One of our goals is to use numerical methods to study the construction away from this limit. These also  allow us to explore the core of the solution and the shape of the center vortex obtained from the  monopole-instanton chain of Figure~\ref{fig:picture1}. 
In addition, we are also able to study the transition between the other configurations in some detail, e.g. between (\ref{t2}) and (\ref{t3}), as alluded to in Section \ref{sec:motivation}.

The continuity between different fractionally charged semiclassical configurations on $\T^4$ was previously studied using cooling (see \cite{GarciaPerez:1993lic,deForcrand:1995qq,Montero:2000mv}) to find the lattice action minima \cite{Wandler:2024hsq}, albeit in pure gauge theory, without the addition of a deformation potential. An important technical point is that the addition of the nonlocal deformation potential requires the use of a different method  to find the  minima of the action. We reduce the action using a gradient flow method, similar to the Hamiltonian evolution phase of hybrid Monte Carlo, as described in Appendix \ref{appx:flow}. dYM has been the subject of Monte Carlo simulations, e.g.~\cite{Bonati:2018rfg,Athenodorou:2020clr,Bonati:2020lal,Bonati:2025hik,Bonati:2026ljv}, but to the best of our knowledge, ours is the first study devoted to finding classical minimal action instanton configurations in dYM.

{\flushleft \bf Section \ref{sec:setup}: Definitions.} The paper begins by   presenting the lattice action of $SU(2)$ dYM. To describe our results, we now briefly go over the main features of the setup. The double trace deformation is added in the $x_0$ direction of period $L_0$. The scaling of the deformation term, with coefficient equal to $c \over L_0^3$ (see Eqn.~(\ref{action1})) is motivated by the continuum one-loop expression familiar from $\R^3 \times \S^1$ studies \cite{Unsal:2008ch,Unsal:2007jx}. In  our simulations, we consider two choices of the dimensionless coefficient $c$ of the double-trace deformation: $c=1$ and $c=0$. The latter choice is equivalent to pure Yang-Mills, which has been previously studied (e.g.~\cite{Wandler:2024hsq}), and here is used to generate minimum action configurations to compare with $c=1$ dYM vacua. 
The other directions are labelled $x_{1,2,3}$, of respective lengths $L_{1,2,3}$.
 We always impose a nontrivial twist in the $12$ plane, $n_{12}=1$. For  the purpose of finding $|Q| = 1/2$ instantons, we turn on another nontrivial twist, $n_{03}=1$. Finally, in all our simulations we take $L_1=L_2$ and vary $L_1$ and $L_3$, keeping $L_0$ small and fixed. This setup allows us to interpolate between the geometries (\ref{s1})-(\ref{t4}), with the deformation in $x_0$ present or not.
 
 {\flushleft\bf{Section \ref{sec:competing}: Level crossing in dYM.}} We begin by studying the ground state of dYM on a spatial $\T^3_{(L_0, L_1, L_2)\vert_{n_{12}=1}}$, with $x_3$ considered as time and with the only nonzero 't Hooft twist $n_{12}=1$. This single-twist setup explores $Q=0$ minimum action configurations on $\T^4$, i.e. classical ground states. Earlier, two local minima of the energy of dYM $\T^3_{(L_0, L_1, L_2)\vert_{n_{12}}}$ were proposed \cite{Poppitz:2022rxv} as the candidate ground states: the ``flux'' and ``no-flux'' states (see \cite{GarciaPerez:2013idu} for earlier discussion of the flux state) Their relevant properties are described in Section  \ref{sec:competing}. As the names suggest,  there is physical magnetic flux in one of these states (the ``flux'' one, where $SU(2)\rightarrow U(1)$)  and none in the other (the ``no-flux'' one, where $SU(2)\rightarrow \Z_2$). 
 
 The main result of Section \ref{sec:competing} is our Figure~\ref{fig:transition}. It presents numerical evidence that either the ``flux'' or ``no-flux'' state is  the global energy minimum in dYM, for any shape of $\T^3_{(L_0, L_1, L_2)\vert_{n_{12}=1}}$, with a level crossing occurring at a critical value   $\left({L_1/L_0}\right)_c \simeq 1.5$. For $L_1/L_0$ larger than the critical value, the ``flux'' vacuum is the minimum energy state, while for smaller values of the ratio, it is the ``no-flux'' vacuum.
This level crossing has implications on the transition between center vortices on $\R^2 \times \T^2$, as per (\ref{t2}), and fractional instantons  on $\R \times \T^3$, (\ref{t3}), in dYM, the subject of   Section 
\ref{sec:morphing2}.

{\flushleft\bf{Section \ref{sec:rvwym}: Reminder on pure YM theory.}} Here, we recall  some known properties of minimum action configurations in pure YM theory on $\T^4$ with twists, with either a single twist $(Q=0)$ or with two twists $(|Q| = 1/2$). The most important point is that, depending on the ratio between torus periods, abelianization around a  $|Q|=1/2$ instanton localized in some directions can also occur in pure YM theory, as noted in \cite{GarciaPerez:1999hs},   \cite{Wandler:2024hsq}. This implies that there are cases where the abelianization due to the deformation potential in dYM and due to the twist in YM align. We shall see later in the paper that, whenever this happens, there is little qualitative difference between $|Q|=1/2$ instantons  in YM and dYM, with the deformation having the main effect of slightly raising the action above the BPS limit of $4 \pi^2$.

{\flushleft{\bf Section \ref{sec:morphing1}: Center vortex/monopole-instanton continuity.}} Here, we study the transition between fractional instantons on different $\T^4$ approximating   $\R^3_{(x_1, x_2, x_3)} \times \S^1_{(x_0)}$ and $\R^2_{(x_1,x_2)} \times \T^2_{(x_0, x_3)}$. As usual the nonzero twists are $n_{12}$ and $n_{03}$. To study this transition, 
in our numerical minimization,  we keep $L_0$ small, taking  $L_1 \; (= L_2)$ large---but by necessity finite, as opposed to the $\R^2$ of \cite{Hayashi:2024yjc,Guvendik:2024umd}---and vary $L_3$ from large to small, interpolating between lattices approximating  $\R^3 \times \S^1$ and $\R^2 \times \T^2$, our Eqns.~(\ref{s1}) and (\ref{t2}).
Using numerics, 
we are able to relax the condition $L_3 \gg L_0$, needed to justify the  analytical tools of \cite{Hayashi:2024yjc,Guvendik:2024umd}, and thus show that the continuity persists all the way to small $L_3 \gtrsim L_0$. 
The dYM vacuum surrounding the localized $|Q|=1/2$ configurations of Section \ref{sec:morphing1}, for sufficiently large $L_3$, is always the ``flux'' $SU(2) \rightarrow U(1)$ vacuum, as we always have large $L_1 > 1.5 L_0$, hence the level crossing to the ``no-flux'' vacuum plays no role. However, for small $L_3 \sim L_0$, the vacuum surrounding the center vortex in the $12$ plane breaks $SU(2) \rightarrow \Z_2$ instead.\footnote{This is yet another ``no-flux'' vacuum, but this time on a spatial $\T^3_{(L_0,L_1,L_3)\vert_{n_{03}=1}}$, with $L_2$ considered as time (thus relevant for large $L_2$). There are two zero-energy classical vacua, which are the global minima of both YM and dYM (with the deformation   still in the $x_0$ direction) for any ratio of periods $L_{0,1,3}$.}

{\flushleft{W}}e now summarize the results of the numerical study of Section \ref{sec:morphing1}, presented on Figures~\ref{fig25L3}-\ref{fig:YMvsdYM4}:
\begin{enumerate}
\item
For the small  values of $L_3 \gtrsim L_0$, the abelianizations in YM and dYM align (as per the discussion of Section \ref{sec:rvwym}) and there is no  qualitative difference between the corresponding  $|Q|=1/2$ center-vortex configurations. The small-$L_3$ similarity between YM and dYM minimal action configurations is clearly seen on  Figures~\ref{fig25L3}, \ref{fig45L3}, \ref{fig:sixfigure}, and \ref{fig:flux10202025}, the Gaussian fit on Fig.~\ref{fig:collimationfit} of the center-vortex profile (discussed below), as well as on Fig.~\ref{fig:YMvsdYM4}. The difference between YM and dYM appears at  larger values of $L_3$---where the analysis of \cite{Hayashi:2024yjc,Guvendik:2024umd} is operative---here dYM abelianizes, but the YM configurations are delocalized. This is seen upon comparing the two columns of Fig.~\ref{fig:sixfigure}, as well as the two rows of Fig.~\ref{fig:YMvsdYM4}.

\item
To study the collimation of the flux of the monopole-instanton chain into a center vortex, on Figure \ref{fig:fluxcollimating}  we show how the dYM monopole-instanton flux profile  collimates into a localized center vortex, upon increasing  $L_1L_2$. On Figures \ref{fig:a8} and \ref{fig:b8},  for both dYM and YM on the same size lattice, 
we fit the center-vortex profile to a Gaussian and estimate its width, which agrees qualitatively with the exponential falloff estimate, $\exp(- {r \pi \over L_3})$, seen for  $L_3 \gg L_0$  and in the $\R^2_{(x_1,x_2)}$ case (with $r$ the radial coordinate in $\R^2$) in  \cite{Hayashi:2024yjc,Guvendik:2024umd}. We observe that the   geometry where the fit was done is  one where there is little qualitative difference between the YM and dYM configurations, owing to the alignment of abelianizations. 
\item
On Figure~\ref{fig:a4} and \ref{fig:b4}, we numerically demonstrate the disordering of the Wilson loop surrounding the center vortex for the small-$L_3$ lattice whose flux is shown  on  Figure~\ref{fig:a8}. This is similar to the analysis in pure YM from \cite{Wandler:2024hsq}. Finally, Figure~\ref{fig:YMvsdYM4} shows that gauge invariants other than the action density (i.e. various Wilson loops) also evolve smoothly from large $L_3$ to small $L_3$, both in YM and dYM.
\end{enumerate}
The main lesson is that the $L_3 \gg L_0$ continuity between monopole-instantons and center vortices in dYM persists to small $L_3 \gtrsim L_0$, with all qualitative features similar between the two limits. For the smaller   $L_3$ we studied, the same flux collimation effect is also seen in YM theory without deformation, due to the alignment of abelianization due to twist and deformation, as reviewed in Section \ref{sec:rvwym}.\footnote{\label{footnote:twostage1}As explained  in Section \ref{sec:rvwym}, in pure YM with twists, whenever (\ref{asympt}) holds, there is a two-stage abelianization, $SU(2) \rightarrow U(1)$ at a scale $\pi/L_0$, and a subsequent (for $L_3 \gg L_0$) breaking $U(1) \rightarrow \Z_2$ at a lower scale $\pi/L_3$. See \cite{Guvendik:2024umd} for a discussion of the two-stage Higgsing within a monopole-instanton gas effective theory framework.} However, pure YM fails to abelianize at larger values of $L_3$ (see (\ref{asympt}) and the discussion in Section \ref{sec:rvwym} that follows) and  the $|Q|=1/2$ background becomes delocalized.  The alignment of abelianizations in YM and dYM makes it clear, however, that the mechanism of a monopole-instanton chain becoming a center  vortex, while technically relying on the deformation potential \cite{Hayashi:2024yjc,Guvendik:2024umd}, is more general and also applies in pure YM theory, provided the shape of the $\T^4$ is appropriately tuned.

{\flushleft{\bf Section \ref{sec:morphing2}: From monopole-instantons to fractional instantons on $\R \times \T^3$.}} Here we take a different path, studying the transition between fractional instantons on tori approximating $\R^3_{(x_1,x_2,x_3)} \times \S^1_{(x_0)}$ and   $ \R_{(x_3)} \times \T^3_{(x_0,x_1,x_2)}$. As before, the nonzero twists are $n_{12}$ and $n_{34}$. We keep $L_0$ fixed and small, while $L_3$ is fixed and large, and vary $L_1$ $(=L_2)$ from large $L_1 \sim L_3$, to small $L_1 \sim L_0$, interpolating between the two desired geometries, (\ref{s1}) and (\ref{t3}). As opposed to the previous study, here the dYM vacuum surrounding a localized instanton exhibits level crossing from the ``flux,'' at large $L_1 > 1.5 L_0$, to the ``no-flux'' vacuum at small $L_1 < 1.5 L_0$. Since, as per our results of Section \ref{sec:competing}, this is a dYM level-crossing transition, we expect that the nature of $|Q|=1/2$ minimum action solutions will also change discontinuously. Indeed, this is what  our results on 	 Figures~\ref{fig:L345L1varies}, \ref{fig:disorderT3}, \ref{fig:dYMtransition12}, \ref{fig:dYMtransition13} appear to indicate.  

Here, we only give a brief description of the results. For $L_1 < 1.5 L_0$, we find the $\R \times \T^3_{n_{12}}$ fractional instantons studied long ago in \cite{RTN:1993ilw, Gonzalez-Arroyo:1995ynx} (seen in the low-$L_1$ region of Fig.~\ref{fig:L345L1varies}, as discussed there).  These fractional instantons, as per our Eqn.~(\ref{t3}), are well localized in $L_3$. As is also well known, they   disorder the $\tr W_0$ Wilson loop, which approaches values $\pm 2$ away from the solution. Both these features are  seen  on Fig.~\ref{fig:disorderT3}. 

For $L_1 > 1.5 L_0$, on the other hand, we find the flux vacuum monopole-instanton chain configurations described earlier (such a configuration at the transition point is shown on Fig.~\ref{fig:dYMtransition12}).  

The new feature specific\footnote{We recall that along the same path in pure YM theory, for the lattices we study, the transition proceeds, instead, via the maximally delocalized constant-field strength instanton \cite{tHooft:1981nnx}, which has minimal action when $L_1 L_2 = L_0 L_3$, studied in \cite{Wandler:2024hsq}.} to dYM is the transition region near $L_1 \simeq 1.5 L_0$. Here, both the Wilson and deformation action of the minimum action configurations have a peak (as a function of $L_1$), seen on Fig.~\ref{fig:L345L1varies}. For the critical value of $L_1$,  our action minimization algorithm finds two kinds of configurations, of roughly the same total action. These are either monopole-instantons in the flux vacuum, localized in $x_3$ and shown on Fig.~\ref{fig:dYMtransition12}, or the delocalized (in $x_3$) configurations shown on Fig.~\ref{fig:dYMtransition13} and described there in detail. This apparent degeneracy is consistent with a level crossing transition, however, we do not have a detailed understanding of the delocalized configurations; their study requires a more fine-grained resolution of the transition region, which is beyond our scope here. 

\subsection{Outlook}

The study of this paper further confirms that the set of minimal action configurations listed in (\ref{s1})-(\ref{t4}) are related to each other in intricate ways, forming  a  rich set of interconnected saddle points related by changing the twists and ratios of periods of $\T^4$.
In this regard, our study could be furthered by considering in more detail the ``critical'' region of the transition between the monopole-instanton and fractional instanton configurations of (\ref{s1}) and (\ref{t3}), studied in Section \ref{sec:morphing2} and associated with a transition between the flux and no-flux vacua of dYM. This, however, requires a significantly more fine-grained study of the transition and the associated computer resources. Additionally, the gradient flow technique used to generate minimum action configurations could in principle be used to place  upper bounds on the height of the energy barriers between vacua near the critical point. By giving configurations in one vacuum progressively stronger ``kicks'' or fictitious kinetic energy, until the lattice settles into the other vacuum, it could be possible to trace the path through field space the lattice takes. By computing the energy of these intermediate configurations it would be possible to place an upper bound on the height of the energy barrier. 
 
 A more general question---interesting from both mathematics and physics points of view---concerns the nature and moduli of classical self-dual instantons of (fractional) charge $Q=r/N$, $\forall r \in \mathbb{N}$, in $SU(N)$ YM theory on a twisted $\T^4$. For general $r>1$, the moduli space of such field configurations is only understood locally, as in 
 \cite{Anber:2025yub, Poppitz:2026gfa}, using  a combination of numerical and analytic tools. On the numerical side, it would be interesting to study whether one could make further progress, first for $N=2$, for any $r > 1$, by  generalizing the methods of \cite{GarciaPerez:1993lic,deForcrand:1995qq}.
 
\section{The lattice setup of deformed Yang-Mills (dYM) theory}
\label{sec:setup}

We study  Wilson's  lattice gauge theory for an $SU(2)$ gauge group, on a rectangular $\T^4$ lattice of periods $L_\mu$, $\mu=0,1,2,3$.  The theory obtained by the addition of a double-trace deformation \cite{Unsal:2008ch} in the $L_0$ direction with   coordinate $x_0 \in \{1, 2,..., L_0 \}$ converts it into  deformed  Yang-Mills theory (dYM). It is well known that the double trace deformation can be due to massive adjoint fermions \cite{Unsal:2008ch,Unsal:2007jx,Unsal:2007vu} but this interpretation is not relevant for us, as we are focused on studying classical minima.

We also add a  't Hooft flux background defined by six (mod $2$) integers $n_{\mu\nu}=-n_{\nu\mu}$. The latter will be chosen to be only nontrivial in the $03$ and $12$ planes:
\begin{eqnarray}\label{flux1}
n_{03} &=& 1 \; \text{or} \; 0,~~ n_{12} =  1.
\end{eqnarray}
As indicated, we sometimes  turn off the $03$-plane flux by choosing $n_{03}=0$, notably in Section \ref{sec:competing}. Further, in all our studies, we  take $L_1 = L_2$ and vary $L_1$ along with $L_3$, usually keeping $L_0$ the  smallest dimension. Thus, upon comparing different dimensions, we shall often refer to ratios of $L_0, L_3$ to $L_1$  only.
We begin with the lattice action of dYM:
\begin{eqnarray}\label{action1}
S_{total} &=& A(S_{Wilson} + S_{def.}) = A \left( \sum_{x} \sum_{\mu \nu} {\rm tr}\left( \mathbf{1} - B_{\mu \nu}(x)  {\Box}_{\mu \nu}(x) \right) +  {c \over L_0^3} \sum_{\vec{x}} \left| {\rm tr} W_0(\vec{x})\right|^2\right)  ~.
\end{eqnarray}
Here, $\Box_{\mu\nu}(x)= U_{\mu}(x)U_{\nu}(x + \he_\mu) U_{\mu}^\dagger(x + \he_\nu) U^\dagger_\nu(x)$ is the $\mu\nu$-plane plaquette, located at the lattice point $x$,  with $\he_\mu$ a unit vector in the $x_\mu$ direction and $U_\mu(x)$---the $SU(2)$-valued link variables, periodic upon a shift of $x_\mu$ by $L_\mu$. $x_0$   denotes the coordinate in the direction with the double-trace deformation, and sometimes  we use    $\vec{x}$ to label coordinates in the $x_{1,2,3}$ directions. $B_{\mu\nu}$ is the two-form (i.e. plaquette-based) topological background for the $\Z_2^{(1)}$ $1$-form symmetry, or a 't Hooft twist. Explicitly, the fundamental representation Wilson loop $W_0$ winding in the $x_0$ direction, starting from $x_0=1$,  and the two-form  background are:\footnote{Winding Wilson loops $W_\mu$ in the $x_\mu$ directions are defined similar to (\ref{wilsondef}), with $0 \rightarrow \mu$.}
\begin{eqnarray}
 W_0(\vec{x}) &=& \prod\limits_{j=1}^{j=L_0} U_0(j, \vec{x}), \; {\text{for}} \; \vec{x} \in \mathbb{Z}^3,  \label{wilsondef}\\
B_{\mu \nu}(x) &=& \begin{cases}
 (-1)^{n_{\mu \nu}},  & x_\mu = L_\mu, x_\nu = L_\nu \label{twist}\\
1, & {\rm else} \end{cases}.
\end{eqnarray}

The 't Hooft fluxes are    inserted in a corner of the relevant $2$-plane; this choice is inessential as the backgrounds are topological. Whenever both twists in  (\ref{flux1}) are nontrivial, the two-form background consists of two intersecting nondynamical center vortices. When $n_{03}=0$, there is only a single one: a $2$-plane  extending in $x_3,x_4$ and located at $x_1 = L_1, x_2 = L_2$.  
 The double-trace deformation term in (\ref{action1}) is the one proportional to $|\tr W_0|^2$. 
The $L_0^{-3}$ scaling of the  coefficient of the deformation term is motivated by the usual one-loop expression obtained on $\R^3 \times \S^1$ in the continuum. Our studies are for the particular value $c=1$, but we also take the   pure YM limit $c=0$ (whose relevant properties are reviewed in Section \ref{sec:rvwym}). Throughout, as we further discuss in Section \ref{sec:competing},  the $x_0$ direction of extent $L_0$ is associated with the small spatial  circle familiar from continuum studies on $\R^3 \times \S^1$. 

Our purpose is to study the   minima of the classical action, the one in the brackets in (\ref{action1}), upon varying the twists and the ratio of sides of the torus. Thus, the overall coefficient $A$ will be adjusted at will during the simulation and, after computing the action of a minimum action configuration, it is  divided  out, see Appendix \ref{appx:flow}. When referring to our results, we always list the lattice size in the order $(L_0,L_1,L_2,L_3)$, taking $L_1=L_2$. We study minima of the action (\ref{action1}) with  two choices of twists (\ref{flux1}). The first choice only involves a single nontrivial twist, $n_{12}$,
\begin{eqnarray}
&&(\overbrace{L_0, \underbrace{L_1, L_2}_{n_{12}=1}, L_3}^{n_{03} = 0}) \implies \;{\text{minimum action configurations with $Q=0$, or classical vacua on $\T^3_{(L_0,L_1,L_2)\vert_{n_{12}=1}}$}},\nonumber \\ \label{singletwist}
\end{eqnarray}
and is studied in Section~\ref{sec:competing}. The second case is when both twists are nontrivial \cite{tHooft:1979rtg,tHooft:1981sps,vanBaal:1982ag},
\begin{eqnarray}
&&(\overbrace{L_0, \underbrace{L_1, L_2}_{n_{12}=1}, L_3}^{n_{03} = 1}) \implies \; {\text{minimum action configurations with  $|Q|={1 \over 2}$, or fractional instantons}}.  \label{choicesoftwists}
\end{eqnarray}
For $|Q|=1/2$, for $c=0$, the minimal action solutions obey  the BPS bound, ${1 \over A} S_{Wilson} = {1 \over A} S_{BPS} = 4 \pi^2$.

\section{dYM vacua on $\mathbf{\T^3_{{(L_0, L_1, L_2)}\vert_{\; n_{12}=1}}}$: from ``flux'' to ``no-flux'' upon changing the $\mathbf{\T^3}$ shape}
\label{sec:competing}

In this section, we study the ground states of dYM on $\T^3_{{(L_0, L_1, L_2)}\vert_{\; n_{12}=1}}$, upon changing the ratio $L_1/L_0$ (recall that we always take $L_1=L_2$). 
Our lattice minimization of the action (\ref{action1})  shows that the   one of the two continuum states  discussed in \cite{Poppitz:2022rxv}, known to be local minima, is always a global minimum of the energy, with the transition between them being a level crossing. These two states also played important role in the study of \cite{Guvendik:2024umd}. 

We begin with the classical continuum limit of the dYM lattice action (\ref{action1}): \begin{eqnarray} \label{action2}
{S_{total} \over A}\bigg\vert_{cont.;\; n_{03}, n_{12}}&=& \sum\limits_{\mu,\nu} \int\limits_{\T^4} d^4 x {1 \over 2} \text{tr} F_{\mu\nu} F^{\mu\nu} + {c \over  L_{0}^3} \int\limits_{\T^3} d^3 \vec{x}  \; \left| {\rm tr} W_0(\vec{x})\right|^2,
\end{eqnarray}
where $L_{0}$ now denotes the physical size of the $x_0$ dimension.
As indicated  in (\ref{action2}), the continuum fields are subject to twisted boundary conditions determined by the twists (\ref{flux1}), as in  \cite{tHooft:1979rtg,tHooft:1981sps}. The action (\ref{action2}) is the well-known action of  continuum dYM  theory \cite{Unsal:2008ch}.

We now consider $L_3$ as the time direction and minimize the energy on the spatial $\T^3$ of size $L_0, L_1, L_2$, with a nontrivial twist $n_{12}=1$. We take $n_{03}=0$, consistent with the interpretation that we are looking for energy minima on a spatial $\T^3$ with a single twist.
The continuum energy functional, following  from the Minkowski version of (\ref{action2}), see \cite{Poppitz:2022rxv},    in the $A_3=0$ gauge,\footnote{We hope that the choice of $x_3$ to label   the time direction will not be too confusing.} upon setting the momentum variables (electric fields, or derivatives w.r.t. $x_3$  time) to zero and taking generators obeying tr\;$t^a t^b = \delta^{ab}/2$, is:
\begin{eqnarray}\label{energymin}
{E\over A} =   \int\limits_{0}^{L_1} dx_1  \int\limits_{0}^{L_2}dx_2  \int\limits_{0}^{L_0}dx_0 \left({1 \over 2} F_{12}^a F_{12}^a + {1 \over 2} F_{0i}^a F_{0i}^a + {c \over L_{0}^4} \; \left| {\rm tr} W_0(\vec{x})\right|^2\right),
\end{eqnarray}
with a sum over $a=1,2,3$ and $i=1,2$ implied.
The gauge fields obey twisted boundary conditions in the $12$ plane and are periodic in the third spatial direction of length $L_0$. 

There are two competing classical states 
in the $\T^3$ theory with a deformation and a 't Hooft twist $n_{12}=1$, discussed in great detail in \cite{Poppitz:2022rxv}.  
We refer the reader to that reference for details of the choice of gauge for the transition functions   and for   the gauge-field backgrounds associated with these states. 
We call these the ``flux'' and ``no-flux'' state. 

The essential properties of the flux state are: \begin{eqnarray}
\text{``flux''}:  F_{12} &=&  {2 \pi \over L_1 L_2} {\sigma^3 \over 2}, ~F_{01}=F_{02}=0, ~\tr W_0 = 0,~ \text{tr}(W_0 F_{12}) =\pm i {2 \pi \over L_1 L_2}, \label{flux} \\
&&\tr W_1 = 2 \cos {\pi x_2 \over L_2}, \tr W_2 = 2 \cos {\pi x_1 \over L_1},\nonumber 
\end{eqnarray}
where we ignored the arbitrary origin of $x_1, x_2$ in the Wilson lines above and chose a particular gauge \cite{Poppitz:2022rxv} to describe $F_{12}$. Notice that both $W_1$ and $W_2$ undergo a center symmetry transformation upon a period shift of $x_2$ or $x_1$, respectively. Thus, $W_1$ is  antiperiodic under upon a $x_2$ translation and v.v.. The antiperiodicity of $W_1$ w.r.t. a period translation in $x_2$, etc., is a general property due to the nontrivial 't Hooft twist $n_{12}$.

For the ``no-flux'' vacuum, we have instead:
\begin{eqnarray}
\text{``no-flux''}: F_{12} &=& F_{01}=F_{02}=0, ~ \tr W_0 = \pm 2~,  \label{noflux} \\ 
&& \tr W_1 = \tr W_2 = \tr W_1 W_2 = 0~.\nonumber 
\end{eqnarray}

Both the flux and no-flux classical ground states are two-fold degenerate, the degenerate states related by the ``broken''   $1$-form $\Z_2$ center symmetry in the $x_0$ direction, which acts as $W_0 \rightarrow - W_0$.
From (\ref{energymin}), we immediately obtain for the energies of the  states (\ref{flux}, \ref{noflux}):
\begin{eqnarray}\label{energymin1}
{E_{\text{flux}}\over A} &=& 2 \pi^2 {L_0 \over L_1 L_2}, \nonumber 
\\
{E_\text{no-flux}\over A} &=& 4 c {L_1 L_2 \over L_0^3} \implies {E_{\text{flux}} \over E_{\text{no-flux}}} = {\pi^2 \over 2 c} \left({L_0  \over \sqrt{L_1 L_2}}\right)^4.
\end{eqnarray}
Thus, the classical flux state has lower energy at small $L_0/\sqrt{L_1L_2}$, while the no-flux has lower energy  for large $L_0/\sqrt{L_1L_2}$. The two energies are of the same order when $\sqrt{L_1 L_2}/L_0 = (\pi^2/2)^{1\over 4}$, for $c=1$, thus we expect a transition between these two ground states at a critical value of the ratio\begin{eqnarray}\label{transition}
{L_1 \over L_0}\bigg\vert_{crit.} = \left({\pi^2 \over 2}\right)^{1\over 4} \simeq 1.49, \end{eqnarray}
where we took  $L_1 = L_2$. We can summarize the ``phase structure'' implied by (\ref{transition})   as:
\begin{eqnarray}\label{phases}
    ~ \text{``${L_1 \over L_0} > 1.5 \leftrightarrow$   flux vacuum''} ~~ \text{and} ~~  \text{``${L_1 \over L_0} < 1.5 \leftrightarrow$  no-flux vacuum''}.
\end{eqnarray}
Numerically, we found that   (\ref{transition}, \ref{phases}) correctly locate  the transition between the flux and no-flux vacua on the lattices we study: a minimization of our lattice action (\ref{action1}) with $c=1$, a trivial $n_{03}=0$, and $n_{12}=1$, exhibits this transition precisely near this value (see Figure~\ref{fig:transition} below).  

To connect the above analysis to the lattice minimization of the Euclidean action (\ref{action1}), we note that the two vacua (\ref{flux}, \ref{noflux}) are associated with the following actions, found by simply multiplying the energies (\ref{energymin1}) by the extent of the time direction $L_3$:
\begin{eqnarray}\label{energymin2}
{S_{\text{flux}}\over A} &=& 2 \pi^2 {L_3 L_0 \over L_1 L_2}\bigg\vert_{L_1 = L_2} = 4 {L_3 \over L_0} \;  {\pi^2 \over 2} \left( {L_0 \over L_1 } \right)^2, \\
{S_{\text{no-flux}}\over A} &=& 4   {L_3 L_1 L_2 \over L_0^3}\bigg\vert_{L_1 = L_2}  = 4 {L_3 \over L_0} \; \left( { L_1  \over  L_0}\right)^2~.\nonumber
\end{eqnarray}

For use below, we also give the lattice definition of the gauge invariant $U(1)$ flux, given in the continuum in the last term of the first line in (\ref{flux}):\footnote{The terms $\sim \text{tr}  \;W_0$ vanish in the flux vacuum (\ref{flux}). Now, a  general $SU(2)$ group element is $W_0 = \cos \alpha + i\sin \alpha\; \hat{n} \cdot \vec{\sigma}$. The inclusion of $\text{tr}\; W_0$ in (\ref{field_strength}) is to eliminate the $\alpha$ dependence for tr $W_0 \ne 0$, or $\alpha \ne \pi/2$. To explain, we note that in our further study we will use the definition (\ref{field_strength}) to study the $U(1)$ field around a  monopole-instanton. Near its core, $\cos \alpha \ne 0$,  only approaching zero asymptotically. Naturally, near $\alpha =0$, the definition (\ref{field_strength}) breaks down, as $|\text{tr} \;W_0| = 2$ and the theory is nonabelian.}
\begin{eqnarray} 
F^{U(1)}_{\mu \nu} = \frac{\text{tr} \left( \Box_{\mu \nu}W_0 \right)  - \text{tr} W_0 }{\sqrt{1-\left(\frac{1}{2}\text{tr} W_0 \right)^2 }} ~.\label{field_strength}
\end{eqnarray}
For small $L_0$, as implied by (\ref{phases}), the integral of  $F^{U(1)}_{12}$ over the $12$ plane is indeed equal to $\pm i 2\pi$ in the flux vacua (\ref{flux}).

{\flushleft{B}}efore   continuing to  show that the ``phase transition'' (\ref{phases}) is corroborated by our numerical minimization, let us make some comments regarding the flux and no-flux states:
\begin{enumerate}
\item The flux and no-flux states are local minima of the energy functional in dYM on $\T^3$ with $n_{12}=1$. For the no-flux state in pure YM theory, this has been known since \cite{Witten:1982df} (see \cite{GonzalezArroyo:1987ycm} for a calculation of the massive spectrum). The argument trivially generalizes to dYM. For the flux state, see \cite{Unsal:2020yeh} and, for details of the massive spectrum   \cite{Poppitz:2022rxv}.

However, we stress that
 there is no proof, but only  heuristic arguments \cite{Unsal:2020yeh,Poppitz:2022rxv} (for example, continuity  at  $L_1 L_2 \rightarrow \infty$, where the flux state approaches the dYM ground state on $\R^3 \times \S^1$) that one of these two states represent the   global minimum of the energy in dYM for any shape of $\T^3$. Below, see Figure~\ref{fig:transition}, by minimizing the lattice action of dYM for various shapes of $\T^3$, we numerically find that, indeed, either the flux or the no-flux state represents the true ground state in the appropriate regime (\ref{phases}). This is one of the new results of this paper.

\item Next, we note that in the flux vacuum the theory classically abelianizes. To see this, we note that upon a choice of gauge, we can describe the flux vacuum (\ref{flux}) by taking $\langle W_0 \rangle = \pm i \sigma_3$,  along with the value of $F_{12}^3$ given in (\ref{flux}). The winding Wilson loop $W_0$ acts as a unitary Higgs field, whose vev breaks $SU(2) \rightarrow U(1)$ at a scale $\pi/L_0$ (the vev $i \sigma_3$ only commutes with the $U(1)$ Cartan subgroup and the non-Cartan gauge bosons have mass of order $1/L_0$).  Then, the last term of (\ref{flux}) is the usual gauge invariant    definition  of the flux in an unbroken $U(1)$ via the Higgs field (with lattice version (\ref{field_strength})). The massless and massive spectra are explicitly given in \cite{Poppitz:2022rxv}.

The quantum theory in the flux vacuum remains weakly coupled at  $L_0 \Lambda \ll \pi$, with $\Lambda$ the strong coupling scale of the theory (in this limit, the scale of the breaking is $1/L_0 \gg \Lambda$, ensuring weak coupling). Thus, the classical flux vacuum is close to the true quantum vacuum for such values of $L_0$.  

As we saw above, the flux state is the preferred classical ground state when $L_0/\sqrt{L_1 L_2} \ll 1$ (see eqns.~(\ref{transition}, \ref{phases})) and   certainly in the large  $L_1 L_2$ limit   (e.g.~$\R^3 \times \S^1$). From the above, it remains close to the quantum ground state when, in addition, $L_0$ obeys $L_0 \Lambda \ll 1$.

\item Similarly, we could describe the no-flux vacuum (\ref{noflux}) by taking the Higgs field vev $\langle  W_0 \rangle = \pm {\mathbf 1}$, a unit matrix commuting with all generators, apparently implying that there is no gauge symmetry ``breaking.'' However, there are two more unitary Higgs fields, whose expectation values are $\langle \tr W_1 \rangle = \langle \tr W_2 \rangle = 0$. These correspond to non-commuting unitary adjoint Higgs fields $W_1$ and $W_2$ whose vevs  can be taken the shift and clock matrices, e.g. $i\sigma_1$ and $i \sigma_3$ for $SU(2)$, thus breaking $SU(2) \rightarrow \Z_N$.\footnote{As further elaborated in Section \ref{sec:rvwym}, if one takes $L_1 \ne L_2$ the  breaking proceeds in two stages.}

In the absence of a deformation (i.e. in pure YM theory), the no-flux vacuum is the classical ground state of the $\T^3_{(L_0,L_1,L_2)}$ theory with $n_{12}=1$, for any $L_0, L_1, L_2$  \cite{Witten:1982df}. This is clear from the fact that all field strengths in (\ref{noflux}) vanish---hence, in the absence of a deformation, the classical energy is zero, the minimum possible one. Thus, it is also clear  that this vacuum is expected to be the lowest energy classical state when the deformation contribution is subdominant, i.e. at  $L_0/\sqrt{L_1L_2} \gg 1$, or  the more precise eqn.~(\ref{phases}).

For general $L_1, L_2$  the quantum theory in the no-flux vacuum is strongly coupled. However, at small $\sqrt{L_1 L_2}$, such that $\sqrt{L_1 L_2}  \;\Lambda \ll 1$, the gauge group, as described above, can be considered broken to $\Z_N$ by the 't Hooft boundary conditions. Taking $L_1 = L_2$, for such values of $L_1$, all fields have mass $\sim 1/L_1 \gg \Lambda$ (see  \cite{GonzalezArroyo:1987ycm}). The classical  theory at distances larger than the inverse mass gap  is a $\Z_2$ TQFT, a theory whose Hilbert space consists of the two degenerate no-flux states (\ref{noflux}).  Quantum mechanically, the degeneracy of the two states is lifted by tunnelling associated with fractional instantons. The area law of the appropriate Wilson loop, e.g. the long distance correlator of two $\tr W_0$ operators on $\T^3 \times \R$, is a consequence of the lifting of this degeneracy. This can be described in different ways: in  semiclassical  terms, \cite{RTN:1993ilw, Gonzalez-Arroyo:1995ynx}, recently also reviewed in section 6 of \cite{Anber:2025vjo},   or, in  modern language, as the deformation of the $\Z_2$ TQFT  by a term leading to the restoration of the $\Z_2^{(1)}$ $1$-form symmetry  \cite{Nguyen:2024ikq}.
\end{enumerate}
 \begin{figure}[h]
 \centerline{
\includegraphics[width=\textwidth]{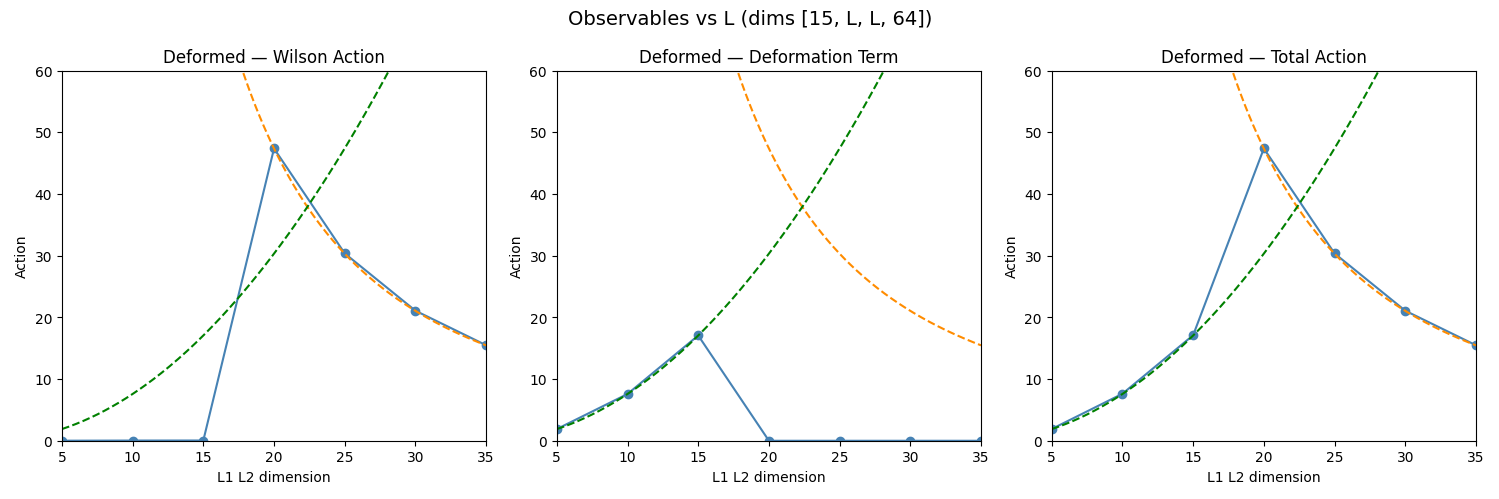}}
\caption{\small The Wilson action, the deformation action, and the total action, computed for the numerically determined  minimal action configuration for dYM on a $(15, L, L, 64)$ lattice, for different $L=5,...,35$. On each plot, we show the two theoretical  continuum values of the actions for the flux and no-flux states of eqn.~(\ref{energymin2}) as function of $L =L_1=L_2$: the curve rising towards the origin is the action of the flux vacuum, $S_{\text{flux}}/A$, which is energetically  favoured at large $L$, while the curve rising to the right shows the action of the no-flux vacuum, $S_{\text{no-flux}}/A$, favoured at small $L$.  The total action plot shows that there is a transition from the no-flux vacuum, at $L < L_c$, to the flux vacuum, at $L > L_c$ at $L_c = 1.5 L_0 \sim 20$. The agreement of the action determined numerically with the theoretical curves is very good, see also  Table  \ref{tab:15_x_x_64_one_twist} for the numerical values.}
 \label{fig:transition}
\end{figure} 

Now, on Figure~\ref{fig:transition}, we present the numerical evidence for the transition (\ref{phases}) between the two minima of the dYM action with $n_{12}=1$ and $n_{03}=0$. As  seen from the numerical data,  there is a transition at a critical value $L_c \sim 20$ ($22.5$ is the theoretical continuum value of eqn.~(\ref{phases})). This provides numerical evidence that, indeed, one of the two local minima, the flux  (\ref{flux}) and no-flux (\ref{noflux}) one, is the dYM ground state on $\T^3_{(x_0,x_1,x_2)\vert_{n_{12}=1}}$ in the appropriate parts of the ``phase diagram.'' We notice that since both vacua are local minima, the transition is ``first order:''  there is level crossing and the flux and the no-flux state are degenerate at the  transition point. Our limited numerical study here  does  not locate precisely this point (at any rate, the continuum critical value (\ref{transition}) is irrational), nor can it measure the height of the energy barrier separating the two.
The point shown at $L_1=20$ has vanishing deformed and nonvanishing Wilson action, i.e.~belongs to the flux vacuum. 
The numerical values obtained for the Wilson action, deformation action, and total action are compared to the theoretical predictions (\ref{energymin2}) in Table \ref{tab:15_x_x_64_one_twist}, showing good agreement.

The level crossing and associated transition from the flux to the no-flux state evident on the r.h. plot of  Figure~\ref{fig:transition} will be important in our Section \ref{sec:morphing2}, when we study the transition from monopole instantons on $\R^3 \times \S^1$ to fractional instantons on $\R \times \T^3$.

 \begin{table}[ht]
\centering
\begin{tabular}{c|c|c|c||c|c}
$L$ & Wilson Action & Deformation Term & Total Action & $ \;\; S_{flux} \;\; $ & $ \;\; S_{no\text{-}flux} \;\; $ \\
\hline
5 & 0.01155 & 1.896 & \underline{1.908} & 758 & \underline{1.896} \\
10 & 0.03513 & 7.585 & \underline{7.62} & 189.5 & \underline{7.585} \\
15 & 0.04314 & 17.07 & \underline{17.11} & 84.22 & \underline{17.07} \\
20 & 47.38 & 5.125 $\times 10^{-14}$& \underline{47.38} & \underline{47.37} & 30.34 \\
25 & 30.37 & 3.706 $\times 10^{-8}$ & \underline{30.37} & \underline{30.32} & 47.41 \\
30 & 21.08 & 3.62 $\times 10^{-8}$ & \underline{21.08} & \underline{21.06} & 68.27 \\
35 & 15.49 & 2.676 $\times 10^{-7}$ & \underline{15.49} & \underline{15.47} & 92.92 \\
\end{tabular}
\caption{Measured Wilson action, deformation term action, and total action values for the $[15, L, L, 64]$ lattice, alongside the theoretical predictions $S_{flux} = 2\pi^2 \frac{L_0 L_3}{L^2}$ and $S_{no\text{-}flux} = 4 L^2 \frac{L_3}{L_0^3}$. For better visualization, we have underlined the measured value of the total action and the theoretical prediction it best agrees with. The numerical data shown here are  used to make the plots on Figure \ref{fig:transition}.}
\label{tab:15_x_x_64_one_twist}
\end{table}

\section{Review of minimum action configurations of  pure YM with different  $\mathbf{(n_{03}, n_{12})}$  on ${\mathbf{\T^4}}$ of varying shape} 
\label{sec:rvwym}

Here, we turn off the deformation and recall some pertinent features of pure YM theory on $\T^4$ with twists (\ref{flux1}). The reason for this review is that the features of pure YM theory on a twisted $\T^4$ described here  are important for the interpretation of some our results of the following Sections, notably the comparison between dYM and YM observables.

First, when only a single twist is nonzero, there are $2^2$ classical zero action configurations in pure YM theory on $\T^4$ \cite{Witten:1982df,Gonzalez-Arroyo:1997ugn}. Using gauge invariant terms, these are easy to describe. We begin with $n_{12}=1$ being the only nonzero twist, where the zero action configurations are:
\begin{eqnarray}\label{state12}
(n_{03}=0, n_{12}=1): S_{min.}=0, \tr W_1 = \tr W_2 = 0, \tr W_0 = \pm 2, \tr W_3= \pm 2, \tr W_1 W_2 = 0, 
\end{eqnarray}
with all signs  uncorrelated.
This is simply the no-flux vacuum of dYM described in Section \ref{sec:competing} (twice repeated, i.e. by taking the time direction to be either $x_3$ or $x_0$). In  pure YM, the action vanishes due to the absence of deformation potential.  The zero-action configurations (\ref{state12}) are related by the action of the classically broken $\Z_2$ center symmetries in the $0$ and $3$ directions, the ones in the plane with vanishing twist.
Likewise, when only $n_{03}=1$ is nonzero,   the roles of the $12$ and $03$ directions are reversed, giving rise to four configurations related by the broken $\Z_2$ center symmetry in the $1$ and $2$ directions: 
 \begin{eqnarray}\label{state03}
(n_{03}=1, n_{12}=0): S_{min.}=0, \tr W_1 = \pm 2, \tr W_2 = \pm 2, \tr W_0 = \tr W_3 = 0, \tr W_0 W_3 = 0.
\end{eqnarray}

Next, when both twists are nonzero, the action saturates the BPS bound for a $|Q|=1/2$ instanton, $S_{BPS}= 4\pi^2$ \cite{tHooft:1979rtg,tHooft:1981sps,vanBaal:1982ag}. The corresponding  instanton, apart from the particular case\footnote{In the tuned $L_1 L_2 = L_0 L_3$ case, the action density of the solution in pure YM is uniform through the entire $\T^4$ \cite{tHooft:1981nnx}.} of an (almost) tuned torus shape, $L_1L_2 = L_0 L_3$, is generically found to be localized in at least some of the $\T^4$ directions. The details of the localization, however, depend on the specific choices of shape of the torus. We consider  two explicit examples in the paragraphs after eqn.~(\ref{asympt}) below, taken from \cite{Wandler:2024hsq}. 
A generic feature seen is that, whenever there are  directions in which the instanton  is  localized, the minimum action configuration approaches, as one moves away from the core of the instanton,  one of the two zero action density  configurations,  (\ref{state12}) or (\ref{state03}). Clearly, this is necessary if the finite action solution is to persist as a localized finite-action ``blob'' in the limit  when the size of these directions (the ones in which the solution is localized) is taken to infinity.

 The rule-of-thumb observation of \cite{Wandler:2024hsq}  is that the asymptotics of the minimum action $|Q|=1/2$ configuration away from the core of the instanton depends on the shape of the torus  as follows:
 \begin{eqnarray}\label{asympt}
(n_{03}=1, n_{12}=1): S_{min.}= 4 \pi^2, \;\begin{array}{c}  \text{asymptotics away} \cr \text{ from the core}   \end{array} \rightarrow \; \left\{\begin{array}{cc} \tr W_{1,2}= 0, |\tr W_{0,3}|=2,&\text{for} \; L_1 L_2 \ll L_0 L_3, \cr  |\tr W_{1,2}|= 2, \tr W_{0,3}=0,&\text{for} \; L_1 L_2 \gg L_0 L_3. \end{array} \right. \nonumber \\
\end{eqnarray}
We now give a heuristic argument in favour of the top line above. Let us argue that $L_1L_2 \ll L_0 L_3$ leads to the asymptotics stated.\footnote{The argument proceeds with the obvious changes for the opposite case $L_1 L_2 \gg L_0 L_3$.}  Begin by assuming the opposite of (\ref{asympt}): suppose that away from the core, the solution localized in  $L_0, L_3$ (and maybe also in one of $L_{1}, L_2$, if  one of the two is large) approaches the configuration with $|\tr W_{1,2}| = 2$. As we already mentioned in our discussion after (\ref{flux}), the nonzero twist in the $12$ plane implies that $\tr W_1$ must change sign as $x_2$ traverses a period $L_2$ and $\tr W_2$ must change sign as $x_1$ traverses a period $L_1$. But since at least one of $L_1$ or $L_2$ is small, say $L_1$, this change of sign implies that $\tr W_{2}$ exhibits a large variation (from $+2$ to $-2$) over a small distance, which should  increase  the action cost. This action cost is avoided if the asymptotics is simply $\tr W_2=0$, i.e. if the asymptotics is the one shown on the top line in (\ref{asympt}). On the other hand, since $L_{3}$ is large, the fact that $\tr W_{0}$   must change sign (again, from $+2$ to $-2$) upon traversing a large distance $L_3$ does not result in large action cost; likewise  for $\tr W_3$. 

{\flushleft{Let us}} now discuss two examples of the use of (\ref{asympt}), which will be relevant further:\begin{enumerate}
\item  Consider the theory with two twists and take a $\T^4$ with $L_0 \ll L_{1, 2, 3}$, but obeying $L_0 L_3 \ll L_1 L_2$.
According to the bottom line in (\ref{asympt}), away from the core of the solution with $|Q|=1/2$,  localized in the large $\T^3_{(x_1,x_2,x_3)}$, we have that  $\tr W_{0,3} =  0$, while $|\tr W_{1,2}| = 2$. Notice that $\tr W_0 = 0$ means, semiclassically, that the theory abelianizes, at a scale $1/L_0$. This is because we can take, asymptotically, $W_0 = i  \sigma_3 = \exp( i {\sigma_3\over 2} L_0 A^3_0)$, thus the ``Higgs field'' vev is $A_0^3 ={ \pi \over L_0}$. Notice, however, that there is a second unitary Higgs field which also has  an asymptotic vev, since also $\tr W_3 = 0$  far away from the solution. The corresponding Higgs vev can be taken $A_3^2 = {\pi \over L_3}$, thus not commuting with the first one---as the vacuum (\ref{state03}) breaks $SU(2)$ to $\Z_2$---but its effect is small since $L_3 \gg L_0$. This hierarchy thus realizes a two-stage Higgsing in the vacuum surrounding the instanton,  \begin{eqnarray}\label{twostage}
SU(2) \underbrace{\rightarrow}_{\pi/L_0} U(1) \underbrace{\rightarrow}_{\pi/L_3} \Z_2, \; \text{with} \; \pi/L_0 \gg \pi/L_3.
\end{eqnarray}

 The $|Q|=1/2$ solution in the   $L_0 L_3 \ll L_1 L_2$, $L_0 \ll L_3$,  geometry was numerically studied in \cite{Wandler:2024hsq} (and much earlier in \cite{GarciaPerez:1999hs}) and it was shown that it, indeed, has the properties   of a monopole instanton on $\R^3_{(x_1, x_2, x_3)} \times \S^1_{(x_0)}$. It is localized in $\R^3$ and extended in the small $\S^1$. Its long-distance abelian field is given by the projection of the $SU(2)$ field onto the direction set by the asymptotics of $W_0$, as in (\ref{field_strength}).   The $L_0$ extent fixes the size of the core of the solution, and the abelian magnetic flux and charge  are as expected. This is clearly seen on  the plots of the action density, asymptotics of Wilson loops, and magnetic field/charge for this solution: see, respectively, 
Figs.~16, 17, 18 in  \cite{Wandler:2024hsq},\footnote{We use the labeling convention of our eqn.~(\ref{choicesoftwists}).  Notice that in ref.~\cite{Wandler:2024hsq}, the $x_0$ and $x_3$ labels are flipped.} for a $\T^4$ with $(L_0, 48, 48, 48)$ for $L_0=6, 9, 12$.  
\item   Now consider  the same two-twist theory, taking $L_0 L_3 \gg L_1 L_2$, but with equal  $L_1=L_2$. The resulting $|Q|=1/2$ finite action configuration is localized in the large $\T^2_{(x_0, x_3)}$,  with a core size determined by the small $L_1$. The asymptotics of the configuration is as in the top line in (\ref{asympt}), with $\tr W_{1,2}=0$ and $|\tr W_{0,3}|=2$, i.e. the vacuum  (\ref{state12}). Notice that because $L_1=L_2$, here the breaking pattern is 
\begin{eqnarray}\label{onestage}
SU(2) \underbrace{\rightarrow}_{\pi/L_1} \Z_2.
\end{eqnarray} 

The instanton is a center vortex, localized in the large $L_{0,3}$. It disorders the Wilson loop, see  \cite{Gonzalez-Arroyo:1998hjb,Montero:2000pb}. In 
the recent \cite{Wandler:2024hsq}, using a $\T^4$ of size $(48, 12, 12, 48)$, the localization in the $03$ plane is   shown  on Fig.~13,   the disordering effect on the Wilson loop surrounding the center vortex is demonstrated on Fig.~14, while Fig.~15 shows that the winding Wilson loops asymptotics at large $x_{0,3}$ precisely follows our Eqn.~(\ref{asympt}), i.e. the solution approaches at large distances the zero action density background of our Eqn.~(\ref{state12}).
 \end{enumerate}

The  above properties of pure   YM theory with both twists nonzero imply that in either of the  limits shown in (\ref{asympt}), abelianization occurs already in YM theory without deformation. For us, the most important lesson is that, for $L_1 L_2 \gg L_0 L_3$, abelianization in YM occurs, with $\tr W_0=0$ (at $L_3 \gg L_0$), exactly as in the flux phase of dYM (\ref{flux}). This implies (and we shall see numerical confirmation of this expectation)  that in this limit, the $|Q|=1/2$ solutions in dYM and YM are  similar, as the abelianization forced by the twist and by the deformation align. The main effect of the deformation will be seen to lift the action above the BPS limit, $S> 4 \pi^2$.

\section{The ``flux'' vacuum of dYM: monopole-instantons  and their transmutation into center-vortices}
\label{sec:morphing1}

We begin our study of the fractionally charged instantons in dYM by considering the flux vacuum  on $\T^3_{(x_0, x_1,x_2)}$ with $n_{12}=1$. As per eqn.~(\ref{phases}), it is the state of lowest energy at sufficiently small $L_0$,  $L_0 < { L_1 \over 1.5}$.  To  numerically study instantons with $|Q| = 1/2$, we also turn on $n_{03}=1$, 
as in eqn.~(\ref{choicesoftwists}). 
 
\subsection{Brief review of the analytic study of the monopole/center vortex continuity} 
\label{sec:rvwcontinuity}

Before we begin with our numerical study of this transition,  we shall review the compelling analytical picture describing the transition of monopole-instantons in dYM on $\R^3_{(x_{1}, x_{2}, x_{3})} \times \S^1_{(x_0)}$ to center vortices on $\R^2_{(x_1, x_2)} \times \T^2_{(x_0, x_3)}$, with a 't Hooft twist in $\T^2_{(x_0, x_3)}$.\footnote{We  label all coordinates according to the conventions of the present paper, eqn.~(\ref{choicesoftwists}).} This picture   was recently developed in the continuum in \cite{Hayashi:2024yjc,Guvendik:2024umd}. 

\begin{figure}[h]
 \centering
 \captionsetup{width=\textwidth}
\includegraphics[width=7cm]{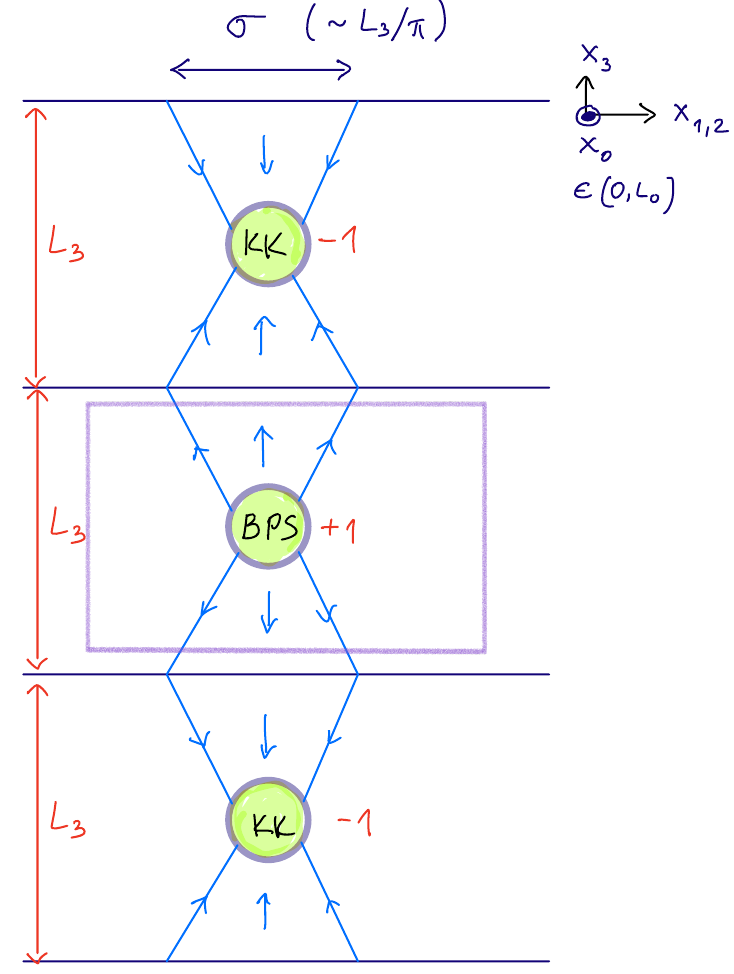}
\caption{A sketch of the lining up of BPS (charge $\alpha_1$) and its image KK (charge $\alpha_0 = - \alpha_1$, $N=2$) monopole-instantons along the compact $x_3$ direction and the collimation of their flux into a center vortex sheet. {\small There is cylindrical symmetry in the $12$ plane and the small $x_0$ direction perpendicular to the page is not shown (there is little $x_0$ dependence in the actual field configuration). Blue arrows indicate the long-range abelian ``magnetic'' field. The flux of the charge $+1$ BPS monopole-instanton is absorbed by its charge $-1$ KK image upon a period translation, etc. The infinite chain of images collimates the flux into a finite radius $\sigma$ ($\sim L_3/\pi$) in the $12$ plane. The configuration behaves like a center vortex sheet, wrapped around the $x_3$ and $x_0$ directions, and localized in $x_1, x_2$. For the actual configurations, compare the fundamental domain of the $13$ plane (indicated by a square surrounding the BPS monopole-instanton)  with the field distribution on Fig.~\ref{fig:flux10202025}. That the radial profile of the collimated flux in the $12$-plane is Gaussian is shown on  Fig.~\ref{fig:fluxcollimating}. The center vortex maximal radius $\sigma$ is determined in  Fig.~\ref{fig:a8} and the disordering of  Wilson loops is shown on Fig.~\ref{fig:a4}.}}
 \label{fig:picture1}
\end{figure}

A side remark is due first, however. The picture of monopole flux collimating to create center vortices has been discussed previously in the lattice literature, see \cite{Greensite:2011zz}, in particular Ch.~8 there, where pictures like our Fig.~\ref{fig:picture1}, obtained after appropriate gauge fixing, appear. The novelty of the observations of \cite{Hayashi:2024yjc,Guvendik:2024umd} is that  in the deformed theory, one can argue for the relation between confinement mechanisms using  an  analytical treatment within a valid semiclassical approximation. Apart from being satisfactory on its own, this semiclassical construction was later useful \cite{Hayashi:2024psa,Hayashi:2025mgk} in explaining some puzzles regarding confinement in supersymmetric theories.

 To describe the construction \cite{Hayashi:2024yjc,Guvendik:2024umd}, we first recall that there are two kinds of monopole instantons in $SU(2)$ dYM theory on $\R^3_{(x_{1}, x_{2}, x_{3})} \times \S^1_{(x_0)}$, both of the same topological charge $Q=1/2$, 
but with opposite magnetic charges $\pm 1$ \cite{Lee:1997vp,Kraan:1998sn,Kraan:1998pm}; an introduction/review is in \cite{Poppitz:2021cxe}. These are sometimes called  BPS and KK monopole-instantons respectively  (and, for convenience, we adopt these names here), and can be thought of as the two constituents of a $Q=1$ instanton. 
It is also well known that in dYM, the size of the core of monopole-instantons, an object localized in $\R^3_{(x_{1}, x_{2}, x_{3})}$ and extended in $\S^1_{(x_0)}$,  is set by $L_0$, the size of the circle.
At distances larger than $L_0$ in $\R^3$, the long distance field of a monopole instanton is abelian  and can be  described, using 3d abelian duality, in terms of a dual photon. 

Ref.~\cite{Hayashi:2024yjc,Guvendik:2024umd} used this long-distance description. The construction begins by making  the $x_3$ direction of $\R^3_{(x_1, x_2, x_3)} \times \S^1_{(x_0)}$   compact, of size $L_3$, and further  imposing a 't Hooft twist $n_{03}$ in the resulting  $\T^2_{(x_0, x_3)}$. Using the abelian long-distance description of the monopole instantons and the dual photon's transformation  under center symmetry \cite{Anber:2015wha}, they argued that the $n_{03}=1$ twist causes the BPS and KK monopole instantons to line up along the  $x_3$ direction. The   twist causes a BPS monopole-instanton of charge $+1$, after traversing a period $L_3$, to convert  into a KK monopole-instanton of charge $-1$. In other words,  the image of the BPS monopole upon a translation on a period $L_3$ in the $x_3$ direction is the KK monopole, whose image after another $L_3$ translation is a BPS monopole instanton, etc. Thus, on the covering space of  $\S^1_{(x_3)}$, there is an infinite chain of  alternating BPS and KK monopole-instantons of charges .../$+1$/$-1$/$+1$/..., etc.\footnote{The lining up of BPS and KK monopole-instantons due to the twist was already   understood in the lattice studies of \cite{GarciaPerez:1999hs}.} 
The effect of this chain of alternating charges, computed in  \cite{Hayashi:2024yjc,Guvendik:2024umd}   using the long-distance abelian description of the monopole-instantons, is that  the magnetic  flux collimates into   the $x_1$,$x_2$ plane (i.e.~in $\R^2_{(x_1,x_2)}$) and has maximum size\footnote{As a function of $x_3$, with the maximum size reached for $x_3$ taken half-way between the monopole instanton and its image.} of order $L_3/\pi$, i.e.~determined by $L_3$. The use of the long-distance approximation to the monopole-instanton field requires $L_3 \gg L_0$ (the core size), i.e. the picture assumes the hierarchy of scales: 
\begin{eqnarray}
L_0 \ll L_3 \ll \sqrt{L_1 L_2} ~~(\rightarrow \infty), \label{hierarchy1}
\end{eqnarray}
where we indicated that they considered infinite $L_1, L_2$. In other words, the analytic discussion  holds on an asymmetric $\T^2_{(x_0, x_3)}$.
Further, ref.~\cite{Hayashi:2024yjc,Guvendik:2024umd} showed that on $\R^2_{(x_1,x_2)}$, this    localized ``blob'' of  magnetic flux of size $\sim L_3/\pi$  was, in fact, a center vortex. Concretely, a $12$-plane fundamental representation Wilson loop  which encloses the flux gets multiplied by $-1$, compared to a Wilson loop that does not enclose it. 

In summary, what has been achieved is to show that  center-vortex semiclassical confinement on $\R^2_{(x_1,x_2)} \times \T^2_{(x_0,x_3)}$  with unit 't Hooft flux  and $L_0, L_3$ obeying (\ref{hierarchy1}),  and the monopole-instanton confinement in dYM on $\R^3_{(x_1,x_2,x_3)} \times \S^1_{(x_0)}$ are  continuously connected, via the deformation  described in the previous two paragraphs. The collimation of the flux of monopole instantons into center vortices is illustrated on Figure~\ref{fig:picture1}.

\subsection{Numerical study of the monopole/center vortex continuity} 

Via numerical minimization, we are able to relax the condition $L_3 \gg L_0$ and show that the continuity between monopole-instantons and center vortices persists to smaller $L_3$ all the way down to a symmetric $\T^2_{(x_0,x_3)}$, where $L_3  \sim L_0$. Numerics will also allow us to explore the core of the instanton and to compare dYM and pure YM solutions.

\begin{figure}[h]
 \centering
  \captionsetup{width=\textwidth}
\includegraphics[width=15cm]{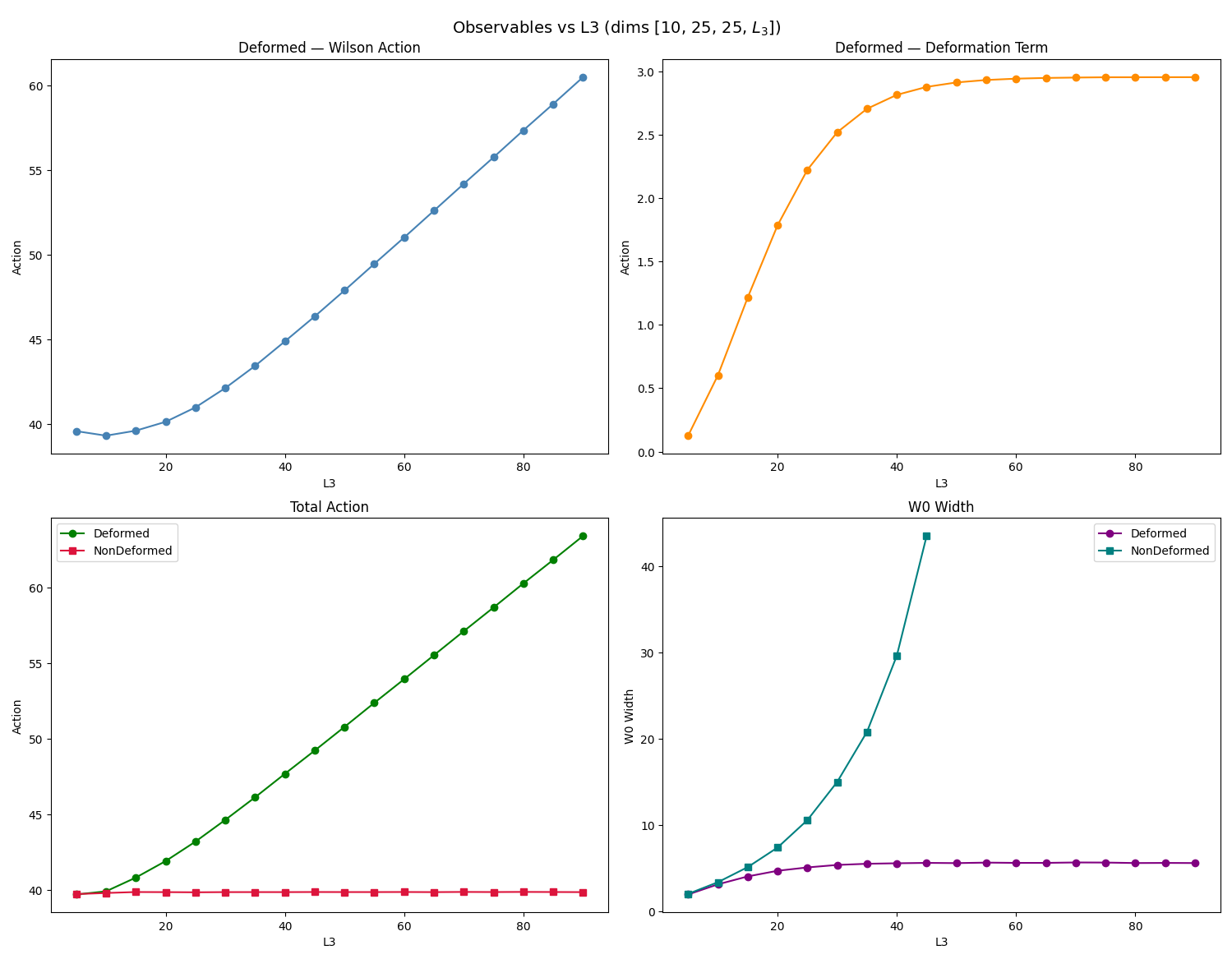}
\caption{\small Properties of the  $|Q| = {1 \over 2}$ fractional instantons for  a lattice of size $(10, 25, 25, L_3)$, for $5 < L_3 < 90$. \underline{\it Top row:}  separate plots of the Wilson action and the deformation action for dYM. \underline{\it Bottom row:}   the total action and the width of the instanton, determined by $\tr W_0$, for dYM and YM. For a discussion, see the text. For pure YM, at large $L_3$, the $12$-plane width of the Higgs field (as per Fig.~\ref{fig:HiggsWidth}) becomes larger than $L_1$, which really signifies failure to abelianize, as seen on  the r.h.s.~plots of Fig.~\ref{fig:sixfigure}. }
 \label{fig25L3}
\end{figure} 

   \begin{figure}[h]
 \centering
  \captionsetup{width=\textwidth}
\includegraphics[width=15cm]{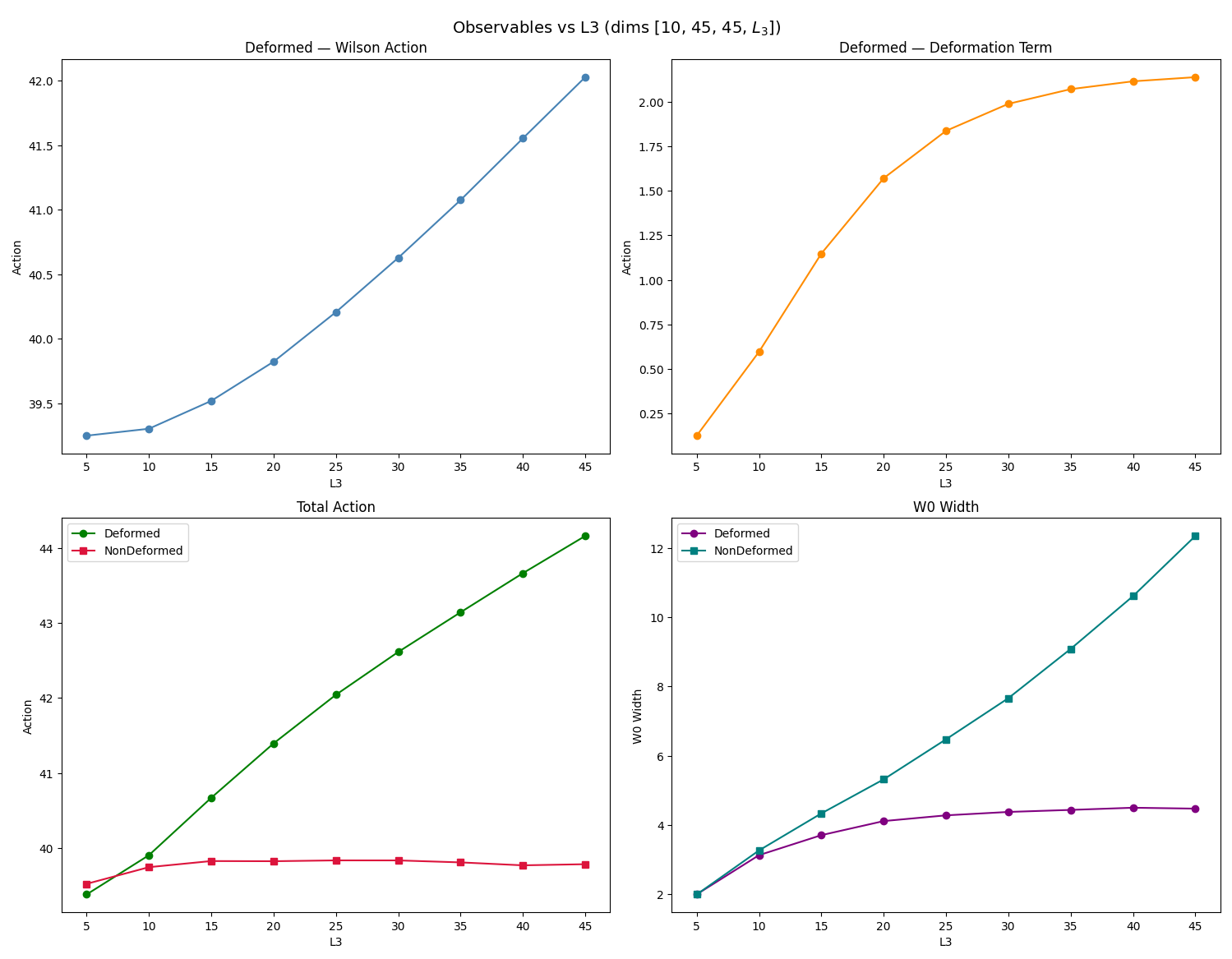}
\caption{\small Similar to Fig.~\ref{fig25L3} but for $L_1=45$: properties of the  $|Q| = {1 \over 2}$ fractional instantons for  a lattice of size $(10, 45, 45, L_3)$, for $5 < L_3 < 45$. \underline{\it Top line:}  separate plots of the Wilson action and the deformation action for dYM. \underline{\it Bottom line:}   the total action and the width of the instanton, determined by $\tr W_0$, for dYM and YM.  } \label{fig45L3}
\end{figure} 
We begin our discussion with
  Figures \ref{fig25L3} and \ref{fig45L3}, where we plot various quantities for dYM theory with $n_{03}=n_{12}=1$, for $L_0=10$, and two different values of $L_1$: $L_1=25$ (on Fig.~\ref{fig25L3}) and $L_1=45$ (on Fig.~\ref{fig45L3}), as a function of $L_3$. On the top line of both figures, we separately plot the dYM Wilson action and the action due to the deformation term. On the bottom line in each figure, we compare the total action for dYM to  the one for YM ($c=0$) as well as the corresponding widths of the localization of the ``Higgs field'' $\tr W_0$ in the $12$ plane around the instanton (i.e. the core of the instanton). All quantities are shown as function of $L_3$ varying in the ranges shown.

\begin{figure}[p]
 \centering
  \captionsetup{width=\textwidth}
 \begin{subfigure}[b]{0.45\textwidth}
 \centering
\includegraphics[width=\textwidth]{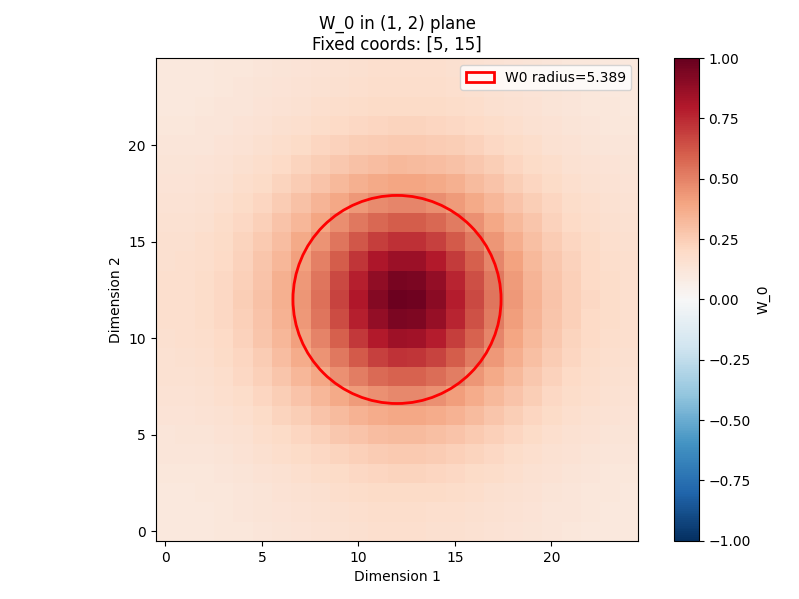}
\caption{\small $\tr W_0$ localization radius for a dYM monopole-instanton on a $(10, 25, 25, 30)$ lattice.  } \label{fig:a7}
\end{subfigure}
\begin{subfigure}[b]{0.45\textwidth}
\centering
\includegraphics[width=\textwidth]{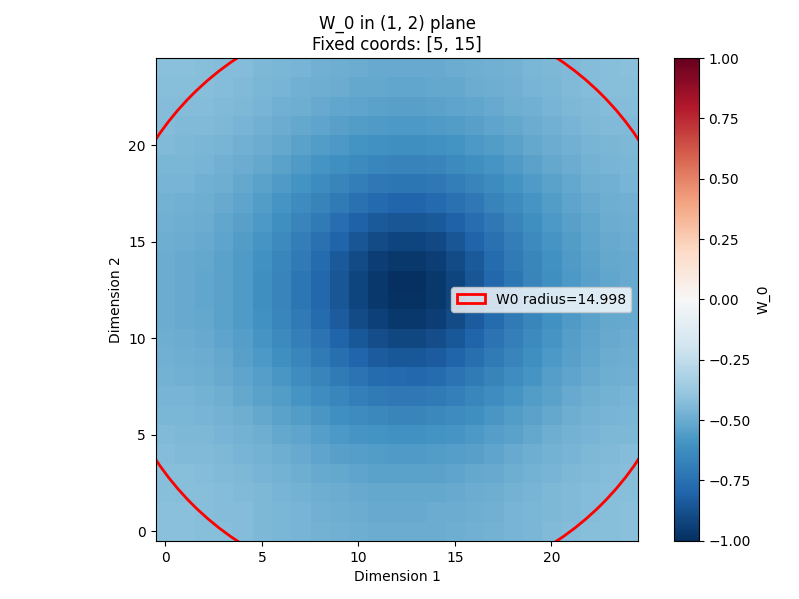}
\caption{\small   $\tr W_0$ localization radius radius   for a YM $|Q|=1/2$ instanton on a $(10, 25, 25, 30)$ lattice. } \label{fig:b7}
\end{subfigure}\\
\begin{subfigure}[b]{0.9\textwidth}
\includegraphics[width=\textwidth]{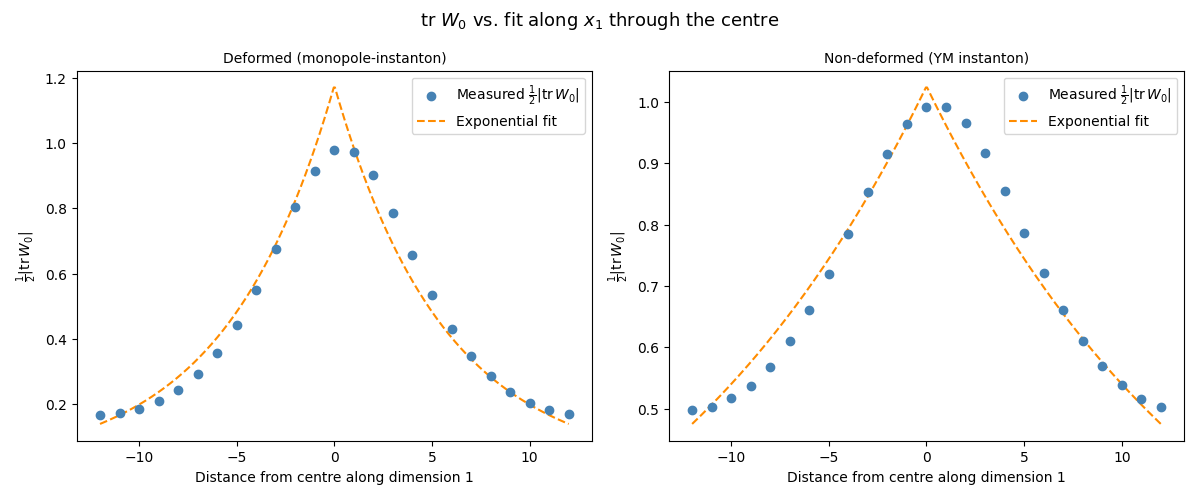}
\caption{\small ${1 \over 2} |\tr W_0|$, for the lattice configurations shown above,  as a function of the distance from the  maximum  along $x_1$ (of total length $L =25$), with the other coordinates kept at the point of maximum action. On the left we show the dYM configuration vs. the fit function. On the right,    the YM profile decreases much more slowly, leading to an unrealistically---see Figure \ref{fig:badfit} for further discussion---larger radius determined by the fit.}\label{fig:hw2}
\end{subfigure}
\caption{\small Defining the localization radius of the ``Higgs field'': $\tr W_0(x_1,x_2)$ is fitted to the radially symmetric function $f(|x|) = a e^{- b |x|}$. The ``localization radius'' is then defined as the value of $r$ where $\int_{|x| < r}d^2 x f(|x|) = {1 \over 4} \int\limits_{0}^{L_1} dx_1 \int\limits_{0}^{L_2} d x_2  f(|x|)$. The large radius for the YM background shown on the top right figure  is because the fitting function does not represent well the actual  minimum action configuration. Admittedly  the fit to a simple exponential is simplistic, but sufficient for our purposes (see also Figure \ref{fig:badfit}).}
\label{fig:HiggsWidth}
\end{figure}

\begin{figure}[h]
 \centering
  \captionsetup{width=\textwidth}
\includegraphics[width=.5\textwidth]{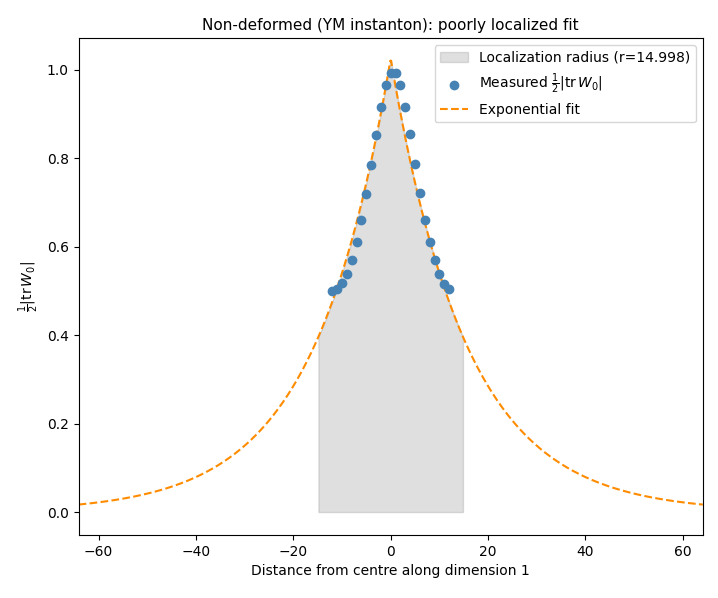}
\caption{\small Expanded version of the  plot on the   r.h.s.~of Figure \ref{fig:hw2}: ${1 \over 2} |\tr W_0|$ as a function of the distance (in $x_1$) to the maximum action point, for the YM configuration on a $(10,25,25,10)$ lattice. The shaded area shows the localization radius  determined by procedure of the caption of Figure \ref{fig:HiggsWidth}, using the area under the simple exponential function. The obtained   value $r \simeq 15$, unrealistic for $L_1 = 25$, is due to the poor fit of the function to the actual configuration: it is evident, both on   Figure \ref{fig:hw2} and here, that the measured configuration ``flattens out'' before the simple exponential function. }
 \label{fig:badfit}
\end{figure}

\begin{figure}[p]
 \centering
  \captionsetup{width=\textwidth}
 \begin{subfigure}[b]{0.45\textwidth}
 \centering
\includegraphics[width=\textwidth]{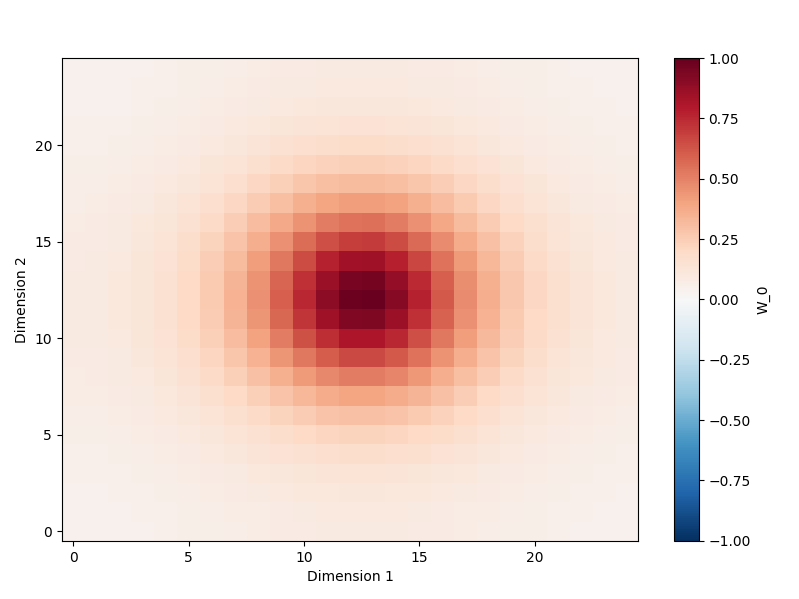}
\caption{\small (10, 25, 25, 15): dYM, $\tr W_0$ in $12$ plane} \label{fig:a}
\end{subfigure}
\begin{subfigure}[b]{0.45\textwidth}
\centering
\includegraphics[width=\textwidth]{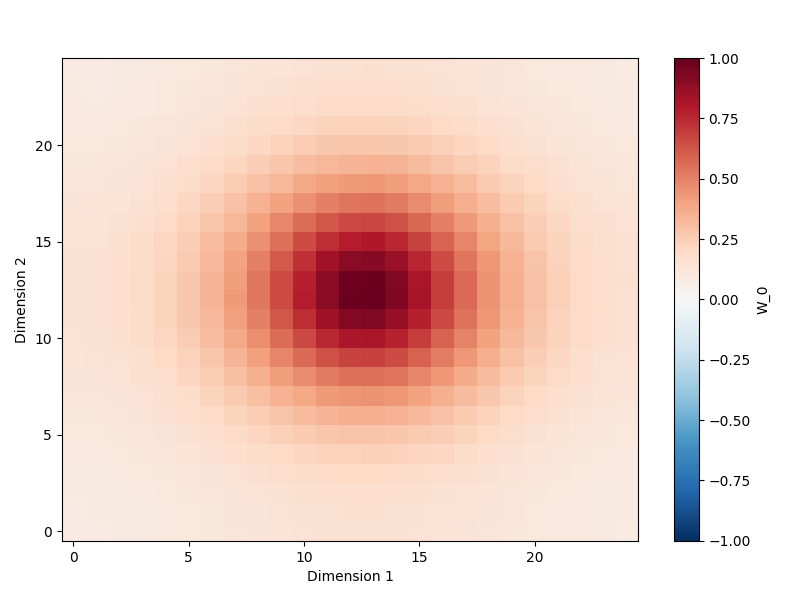}
\caption{\small   (10, 25, 25, 15): YM, $\tr W_0$ in $12$ plane} \label{fig:b}
\end{subfigure}\\
 \begin{subfigure}[b]{0.45\textwidth}
 \centering
\includegraphics[width=\textwidth]{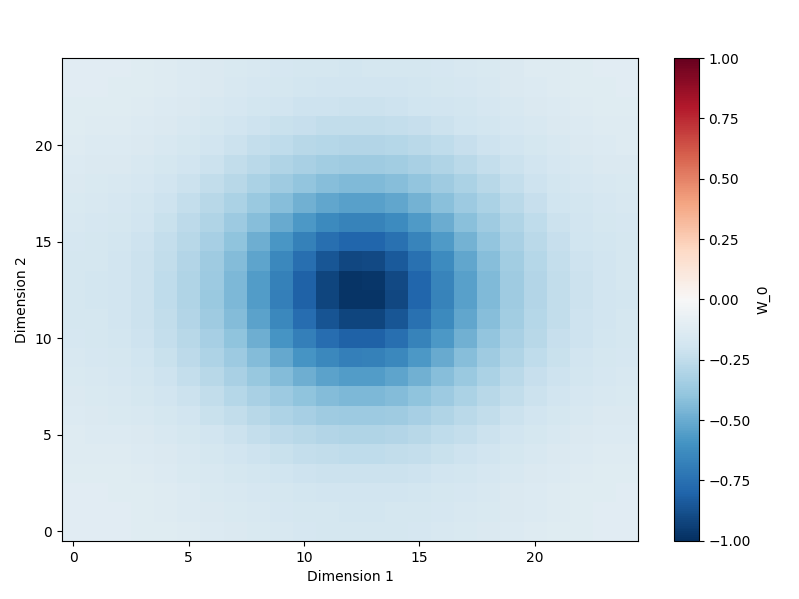}
\caption{\small (10, 25, 25, 40): dYM, $\tr W_0$ in $12$ plane} \label{fig:c}
\end{subfigure}
\begin{subfigure}[b]{0.45\textwidth}
\centering
\includegraphics[width=\textwidth]{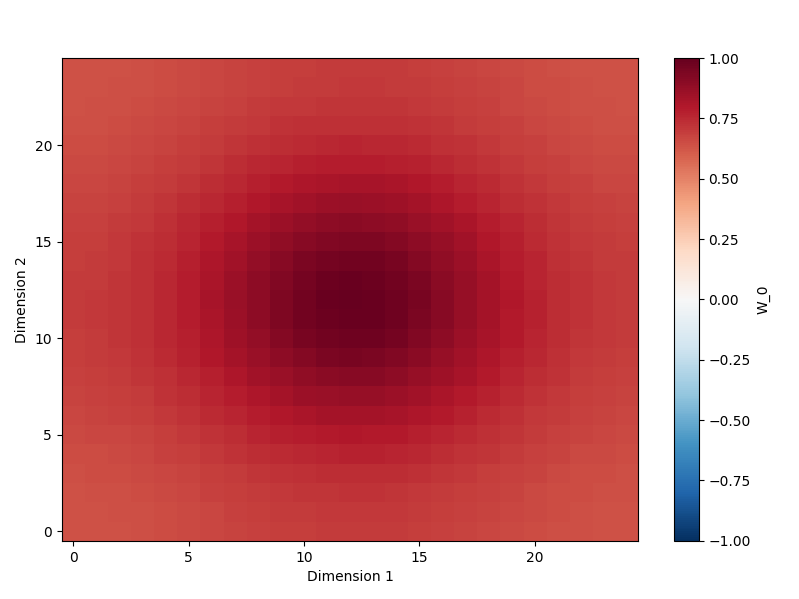}
\caption{\small (10, 25, 25, 40):  YM, $\tr W_0$ in $12$ plane} \label{fig:d}
\end{subfigure}\\
 \begin{subfigure}[b]{0.45\textwidth}
 \centering
\includegraphics[width=\textwidth]{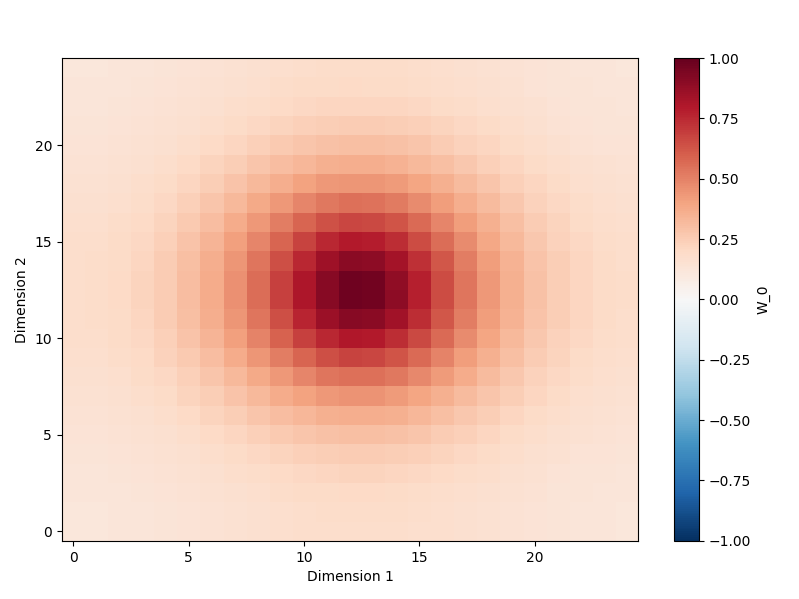}
\caption{\small (10, 25, 25, 70): dYM, $\tr W_0$ in $12$ plane} \label{fig:e}
\end{subfigure}
\begin{subfigure}[b]{0.45\textwidth}
\centering
\includegraphics[width=\textwidth]{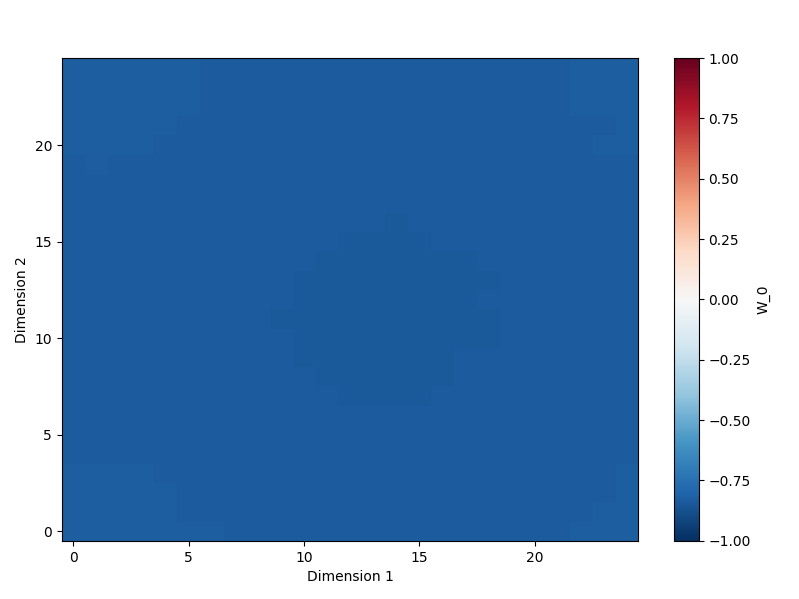}
\caption{\small  (10, 25, 25, 70): YM, $\tr W_0$ in $12$ plane} \label{fig:f}
\end{subfigure}
\caption{Instanton core size in dYM (left column) and YM (right column) for three different lattice sizes. {\small \underline{\it Top line:} $L_0 L_3 \ll L_1 L_2$ - abelianization directions due to deformation in dYM and twist in  YM are aligned. \underline{\it Middle line:} $L_0 L_3 \lesssim L_1 L_2$ dYM abelianizes due to deformation, no abelianization in YM. \underline{\it Bottom line:} $L_0 L_3 \gtrsim L_1 L_2$ dYM abelianizes, no abelianization in YM (almost constant solution).}}
\label{fig:sixfigure}
\end{figure}

{\flushleft{L}}et us now discuss the features of the different regimes seen on Figures~\ref{fig25L3} and \ref{fig45L3}:
\begin{enumerate}
\item First, we stress that on both figures, ${L_1 \over L_0} >  1.5$, thus according to eqn.~(\ref{phases}) of Section \ref{sec:competing}, this regime would  correspond  to the flux vacuum of dYM on the $\T^3_{(x_0, x_1, x_2)}$---in the absence of a $n_{03}$ twist or at large enough (strictly infinite) $L_3$.
Thus, for large enough $L_3$, the vacuum 
   surrounding the localized fractional instanton is the dYM flux vacuum; further evidence for this is discussed in the following point 2. Recall also that the dYM flux vacuum (\ref{flux}) approaches the $\R^3 \times \S^1$ dYM vacuum at infinite $L_{1,2}$. We shall see that at large $L_3$, the $|Q|=1/2$ minimum action configurations have an interpretation as the $\R^3 \times \S^1$ monopole-instantons.
\item Next, beginning with large $L_3$, we observe that the  dYM  Wilson action of the instanton is linear in $L_3$, while the deformation action levels off as a function of $L_3$, as seen on the l.h.s. plots on both Figures~\ref{fig25L3} and \ref{fig45L3}.  

The fact that the action of an instanton grows with the volume may be unusual but should not be surprising: this is because it includes the action of the vacuum, which is nonzero for the flux vacuum of dYM, recall  (\ref{energymin2}).\footnote{The instanton interpolates between the two degenerate flux vacua \cite{Unsal:2020yeh}.}
The  fact that the deformation action levels off is also   expected, as $\tr W_0 \ne 0$ only in the core of the instanton (and $\tr W_0=0$ otherwise) and the core size, for large $L_3$, is $L_3$ independent.
Consistent  with the picture above, the linear growth of the total action with $L_3$ seen on the Figures is precisely that of the action   (\ref{energymin2}) of the  flux vacuum (\ref{flux}).\footnote{For example, on Fig.~\ref{fig25L3}, the slope, for $L_3 > 40$, is easily estimated as $S/L_3 \simeq  0.3$, while (\ref{energymin2}) gives $S/L_3 = 2 \pi^2 L_0 L_1^{-2} = 0.31$. A similar estimate holds for Fig.~\ref{fig45L3} with a slope $\sim 0.1$. On both Figures, for large $L_3$, the total action can be fit by a fixed value, roughly of order the BPS action (but we stress that a more precise study of interpolations and scaling is needed, which is beyond our scope here) plus the action of the flux vacuum (\ref{energymin2}) accounting for the linearity in $L_3$.}  \item On the bottom left figure we plot the total action in dYM vs YM, for the same twists and lattice sizes. The YM action equals the BPS action, $4\pi^2 \sim 39.5$, for all values of $L_3$, while the total action in dYM linearly grows with $L_3$, as per the discussion above.
 \item On the bottom right in both Figures \ref{fig25L3} and \ref{fig45L3}, we plot the width of the ``Higgs field'' $\tr W_0$ in the instanton background, measured in the $12$ plane (this is really the core size of the instanton). The width of $\tr W_0$ in the $12$ plane is determined by a simple fit, sufficient for our purposes (details are explained in the caption of Figure~\ref{fig:HiggsWidth}). We see that the core size of the instanton levels off in dYM, consistent with the fact that the instanton size is set by the deformation potential, for large $L_3$. 
    \item Regarding the pure YM core size, we note that on Figure \ref{fig25L3}, the $W_0$ width in YM grows as $L_3$ increases and becomes bigger than $L_1$ around $L_3 \sim 40$$-$$50$ (here $L_1L_2 = 625$ while $L_0 L_3 = 500$, which is close to the $\T^4$ shape leading to the constant-$F$ self-dual solution \cite{tHooft:1981nnx}). The similarity of the core sizes in dYM and YM at small $L_3$, and the  growth of the YM core size with $L_3$ are also seen on Figure \ref{fig45L3}. Notice that the core size growth beyond $L_1$, seen on Fig.~\ref{fig25L3}, does not appear on Fig.~\ref{fig45L3}: here $L_1 L_2 = 2025$  and in order that $L_0L_3$ exceeds $L_1L_2$ one must go to $L_3 \sim 200$, beyond the scope of our numerics. (We also refer to Figure~\ref{fig:sixfigure} and the discussion that follows.) 
  \item At smaller $L_3$, outside of the analytic regime (\ref{hierarchy1}), we notice that the core sizes (as measured by the $W_0$ width defined in Figure~\ref{fig:HiggsWidth}) in YM and dYM appear to coincide.

Now, we recall the features of YM discussed in Section \ref{sec:rvwym}, in particular our eqn.~(\ref{asympt}), arguing that for $L_0 L_3 \ll L_1 L_2$, abelianization due to $W_0$ around the localized instanton also occurs in YM theory. Precisely in this limit, the pure YM abelianization is aligned with the one due to the deformation term. Thus, at the smallest values of $L_3$ the $\tr W_0$ widths seen are similar in YM and dYM (this  is seen even more clearly on our next Figure~\ref{fig:sixfigure}).
The smallness of the deformation action for $L_3=5,10$ is due to the small width of $\tr W_0$.\footnote{In fact, the width shown on the bottom r.h.s. figure can be used to obtain a reasonable estimate of the numerical value of the deformation action of the top r.h.s. figure (assuming e.g. that $|\tr W_0|=2$ inside a region of order the width and $=0$ otherwise).}
We note that the vacuum surrounding the solution in the $12$ plane is now $\tr W_0 = \tr W_3 = 0$.\footnote{This background can, in fact, be thought as the vacuum of dYM on $\T^3_{(L_0,L_1,L_3)\vert_{n_{03}=1}}$, with $L_2$ taken as time (this is relevant since we are considering the large $L_2$ limit); now $\tr W_0=\tr W_3=0$ and $\tr W_1 = \pm 2$ are the two zero energy vacua. These are  also the pure YM classical vacua in the same geometry. } At $L_0 \sim L_3$ the limit where  the two-stage  breaking (\ref{twostage}) becomes the one-stage 
(\ref{onestage}) (with the obvious interchange $03 \leftrightarrow 12$ in the latter). See also Figure~\ref{fig:YMvsdYM4} and its caption.
 \end{enumerate}

  To further elaborate on the similarity and difference between dYM and pure YM, at the same twists and lattice sizes, we now move to Figure ~\ref{fig:sixfigure}. Here, we plot the Higgs field localization in the $12$ plane (or the core size or  the fractional instanton) in YM and dYM for three different lattice sizes. The point is to show that at $L_0 L_3 \ll L_1 L_2$, the abelianization in YM due to twists, discussed in Section~\ref{sec:rvwym}, and in dYM due to the deformation potential are aligned in a manner consistent with eqn.~(\ref{asympt}). The fractional instanton solutions in this regime are qualitatively similar and the role of the deformation potential is only to raise the action above the BPS limit (and to slightly decrease the core size in dYM compared to YM, as  a careful look at the top two plots shows). 
 However, as one increases $L_3$, going through the ``transition'' of $L_0 L_3 = L_1 L_2$ (where the YM fractional instanton is position independent), abelianization in YM disappears, while it persists in dYM. The abelianized regime in dYM on Figures~\ref{fig:c} and \ref{fig:e}, the ones where $L_3 > L_0$ (recall (\ref{hierarchy1}))  is where the considerations of \cite{Hayashi:2024yjc,Guvendik:2024umd} are valid. 

\begin{figure}[p]
 \captionsetup{width=\textwidth}
 \begin{subfigure}[b]{\textwidth}\centering
\includegraphics[width=10cm]{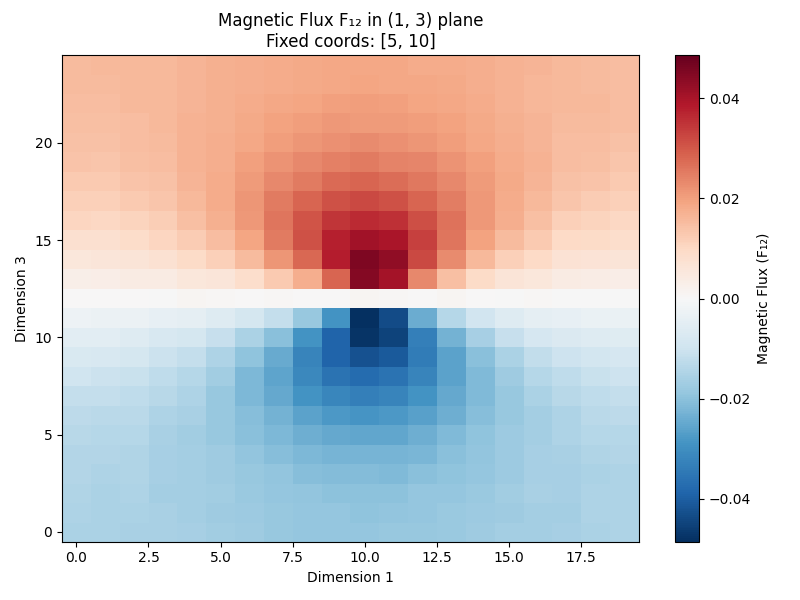}
\caption{\small $F_{12}^{U(1)}$, or the $x_3$-component (on the plot, upward pointing) of the 		``magnetic'' field of a dYM monopole-instanton as a function of $x_1$ and $x_3$ on a $(10,20,20,25)$ lattice at $x_0$$=$$5$, $x_2$$=$$10$. There is cylindrical symmetry in the $12$ plane. The  BPS monopole-instanton flux is absorbed by  its oppositely charged image upon $L_3$ translations. The picture is almost identical to the one in pure YM theory with the same twists, due to the aligned abelianizations in YM and dYM for this lattice size (see fig.~25 of \cite{Wandler:2024hsq} for a YM monopole-instanton on a $(6,18,18,24)$ lattice). }
\label{fig:flux10202025}
\end{subfigure} \\
\begin{subfigure}[b]{\textwidth}\centering
\includegraphics[width=12cm]{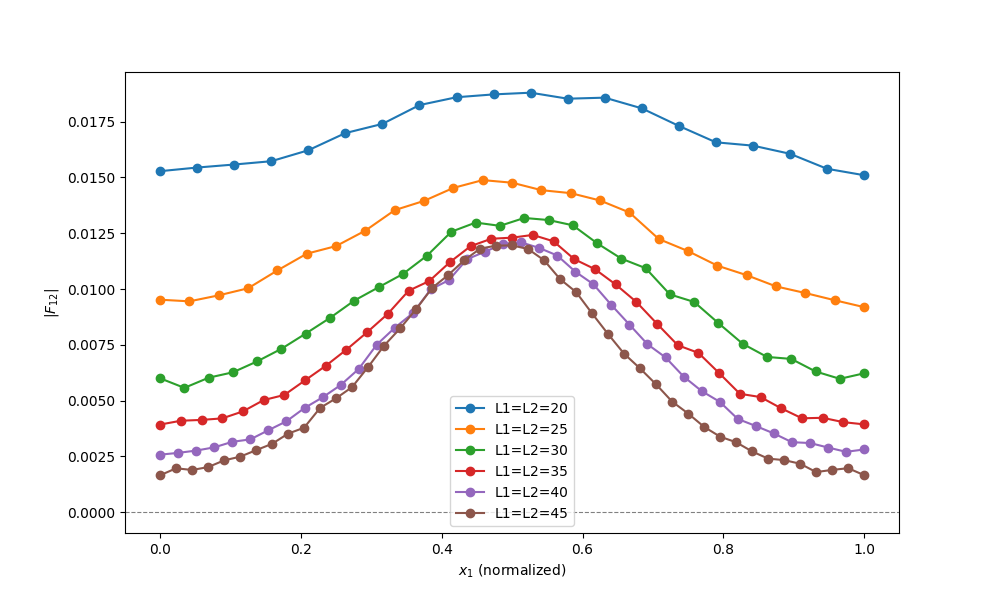}
\caption{\small Collimation of the magnetic flux $F_{12}^{U(1)}$ of a monopole-instanton in dYM into a center vortex on a $(10, L_1, L_1, 25)$ lattice. The monopole-instanton is\ at $x_3$$=$$L_3/2$, $x_1$$=$$x_2$$=$$L_1/2$ and is extended in $x_0$. The value of $F_{12}^{U(1)}$ at $x_3$$=$$0$, $x_2$$=$$L_1/2$, halfway between the BPS and its KK image, is plotted as a function of $x_1/L_1$ for $L_1 = 20,..., 45$. The infinite-$L_1$ limit calculation shows the collimation of the flux into a tube, a center-vortex  of an estimated radius  $\sim L_3/\pi \sim 8$. This   value is consistent with our plots  for  $L_1=40,45$; for smaller $L_1$, there are additional images in the $x_{1,2}$ direction. See the    Gaussian fit on Fig.~\ref{fig:collimationfit} for the $L_1=45$ curve. }
\label{fig:fluxcollimating}
\end{subfigure}
\caption{The chain of monopole-instantons (recall Fig.~\ref{fig:picture1}) and their center-symmetry/translation images collimate the flux into a center vortex; see also Fig.~\ref{fig:collimationfit}.}
\label{fig:collimating1}
\end{figure}

We continue, on Figure \ref{fig:flux10202025}, by studying the monopole-instanton's magnetic field, $F_{12}^{U(1)}$ defined in eqn.~(\ref{field_strength}), in dYM on a $(10,20,20,25)$ lattice. This is, in fact,  very similar to the monopole-instantons in pure YM theory on a lattice of a similar size and with the same twists studied in \cite{Wandler:2024hsq}; as in that reference, the total magnetic flux surrounding the monopole-instanton can be seen to be $4\pi$ (in particular, the integral of the flux of  $F_{12}^{U(1)}$ over each $12$ plane taken half-way (in $x_3$) between the monopole instanton and its $x_3$-translation image
equals $2\pi$, as in that reference and as suggested in \cite{Unsal:2020yeh}). 
As already discussed, the similarity is due to the fact that for this size YM theory also abelianizes in a direction aligned with  the deformation. 
  The image of the BPS monopole-instanton upon $L_3$ translations is a KK monopole-instanton absorbing its flux, etc.. The collimation of the flux in the $12$ plane in a region of size estimated in\cite{Hayashi:2024yjc,Guvendik:2024umd} as $L_3/\pi \sim 8$ is difficult to see for this lattice size (as  the  result for the collimation size of is obtained in the infinite $L_1, L_2$ limit, while here we also have images upon translations in $L_1$, $L_2$).

\begin{figure}[p]
 \centering
  \captionsetup{width=\textwidth}
 \begin{subfigure}[b]{0.45\textwidth}
 \centering
\includegraphics[width=1.1\textwidth]{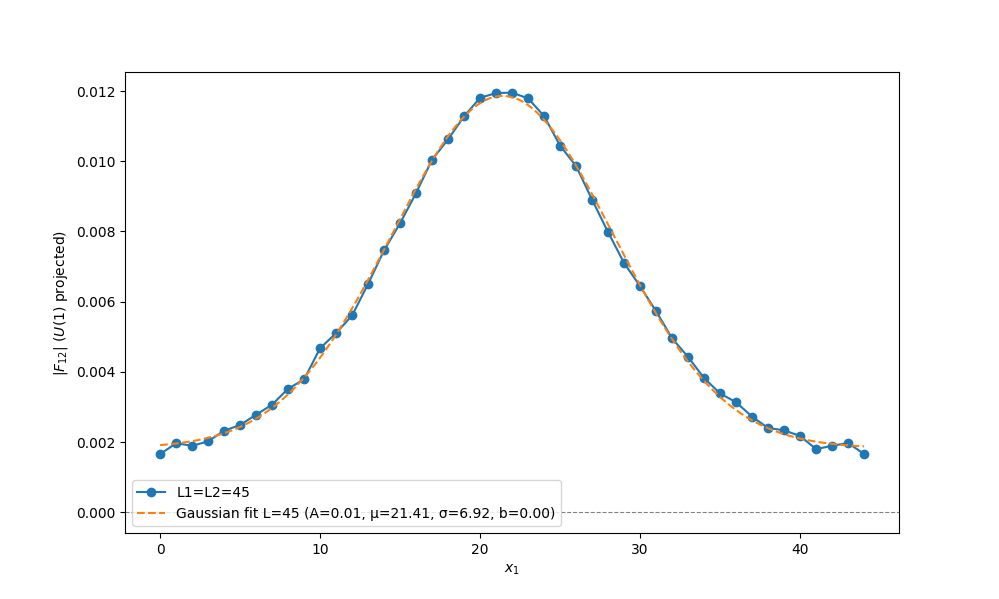}
\caption{\small $F_{12}^{U(1)}$ in dYM on $(10,45,45,25)$ lattice fitted to a Gaussian: $A=0.01$, $\mu=21.41$, $\sigma=6.92$, $b=0$.} \label{fig:a8}
\end{subfigure}
\begin{subfigure}[b]{0.45\textwidth}
\centering
\includegraphics[width=1.1\textwidth]{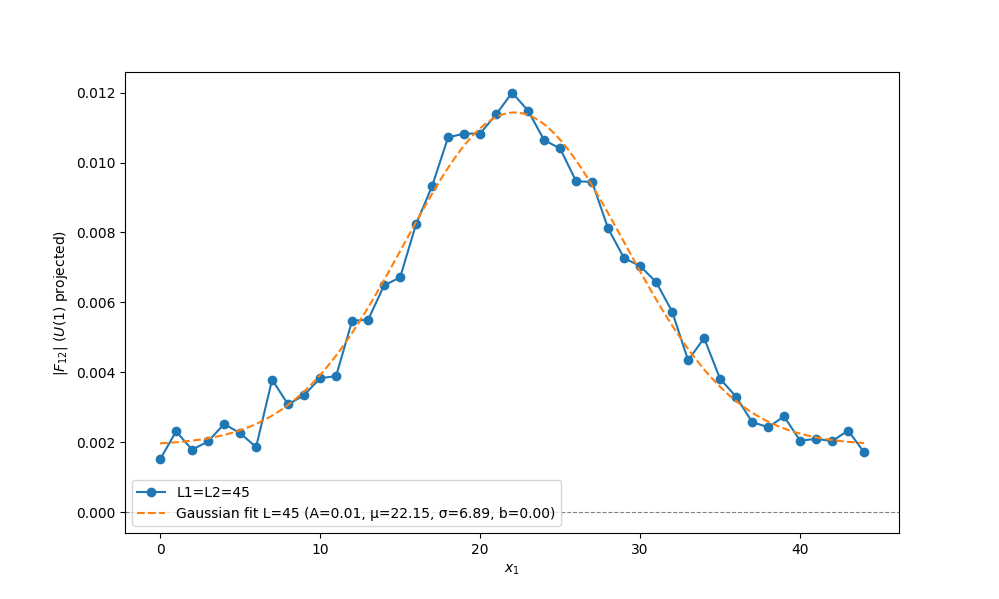}
\caption{\small   $F_{12}^{U(1)}$ YM on $(10,45,45,25)$ lattice fitted to a Gaussian: $A=0.01$, $\mu=22.15$, $\sigma=6.89$, $b=0$.} \label{fig:b8}
\end{subfigure} \\
 \begin{subfigure}[b]{0.45\textwidth}
 \centering
\includegraphics[width=1.1\textwidth]{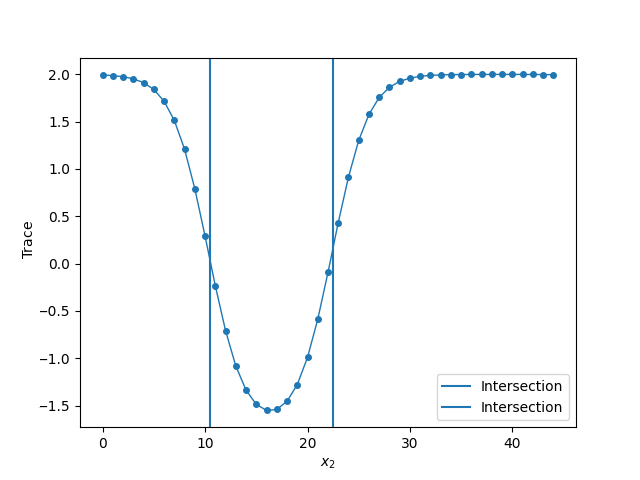}
\caption{\small The center vortex disordering of the Wilson loop: values of $x_2$$<$$10$, $x_2$$>$$25$ correspond to loops $A$ and $C$ (see Fig.~\ref{fig:b4})   not surrounding the center vortex, while $x_2$$\sim$$17$ corresponds to the loop $B$. } \label{fig:a4}
\end{subfigure} 
\begin{subfigure}[b]{0.45\textwidth}
\centering
 \captionsetup{width=\textwidth}
\includegraphics[width=.86\textwidth]{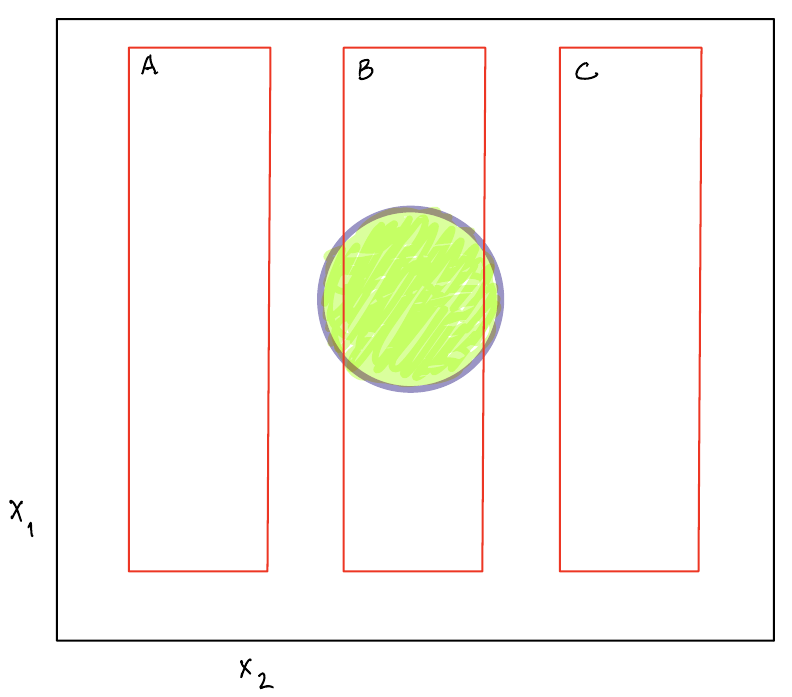}
\caption{\small  The center vortex disordering of the Wilson loop: three positions of the $12$ plane Wilson loop, $A$, $B$, $C$, with respect to the center vortex (schematically indicated by the circle).} \label{fig:b4}
\end{subfigure}
\caption{ The profile of the field of the center vortex and the disordering of  the Wilson loop. {\small{\underline{\it Top row, figs.~\ref{fig:a8}, \ref{fig:b8}:}} A Gaussian fit of the magnetic field, midway between the monopole-instanton and its image upon $L_3$ translation, for the $L_1=45$ curve of Figure \ref{fig:fluxcollimating}. $F_{12}^{U(1)}$, at $x_3=0, x_2=L_2/2$, is fitted to  $A e^{- {1 \over 2}({ x_1-\mu\over \sigma})^2} + b$. The best fit parameters shown are essentially the same in  YM and dYM, due to the alignment of abelianizations, as per (\ref{asympt}). 
\underline{\it Bottom row, figs.~\ref{fig:a4}, \ref{fig:b4}:}  The left Figure \ref{fig:a4} shows the value of trace of a $43\times 12$ Wilson loop in the $12$-plane, evaluated in the ``center vortex'' minimum action configuration in dYM on a $(10, 45, 45, 25)$ lattice. The loop is dragged along the $x_2$ direction  across the center vortex localized in $x_1, x_2$. The Wilson loops at $x_2 < 10$ and $x_2 > 25$ do not enclose the center vortex, but the ones at $10< x_2< 25$ do  (partially), as in the examples $A, B, C$ on Figure~\ref{fig:b4}. There is clear indication of the sign difference between the loops that enclose the center vortex and the ones that do not. The vertical lines on Figure~\ref{fig:a4} represent the points where the edges of the Wilson loop cross the peak of the center vortex, which is located at the midpoint of the lattice.} }
\label{fig:collimationfit}
\end{figure}

\begin{figure}[p]
 \centering
  \captionsetup{width=\textwidth}
 \begin{subfigure}[b]{0.45\textwidth}
 \centering
\includegraphics[width=\textwidth]{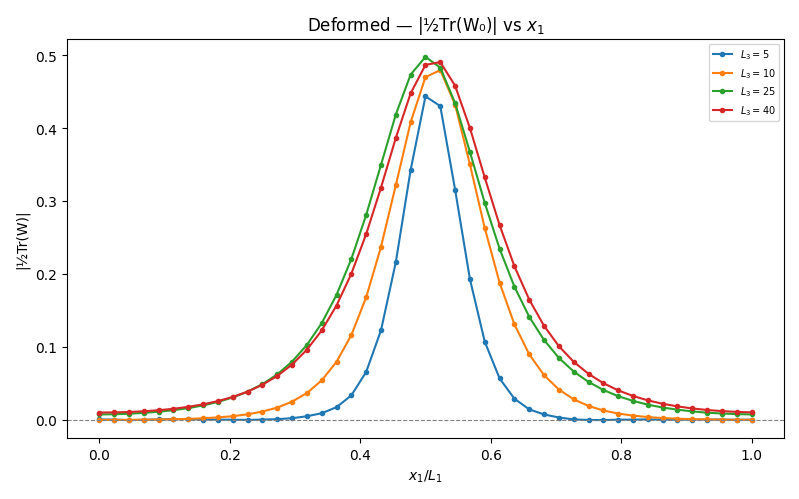}
\caption{\small dYM: $\tr W_0({x_1\over L_1})$ for $L_3=5,15,25,40$} \label{fig:a91}
\end{subfigure}
\begin{subfigure}[b]{0.45\textwidth}
\centering
\includegraphics[width=\textwidth]{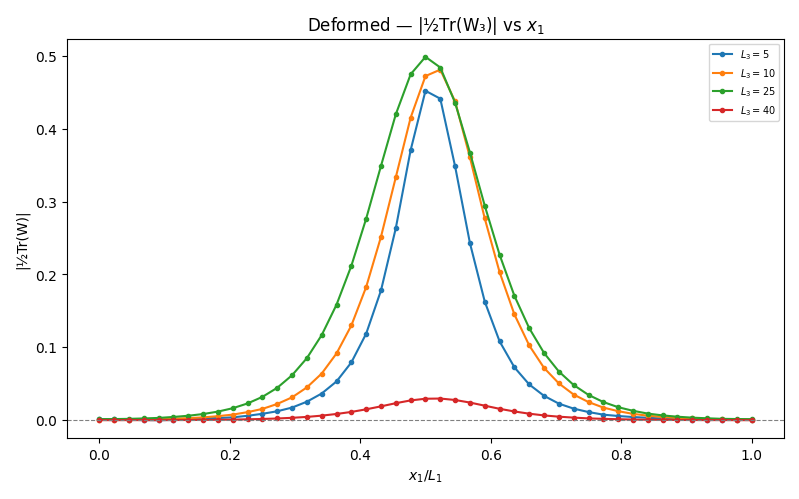}
\caption{\small   dYM: $\tr W_3({x_1\over L_1})$ for $L_3=5,15,25,40$} \label{fig:b91}
\end{subfigure}\\
 \begin{subfigure}[b]{0.45\textwidth}
 \centering
\includegraphics[width=\textwidth]{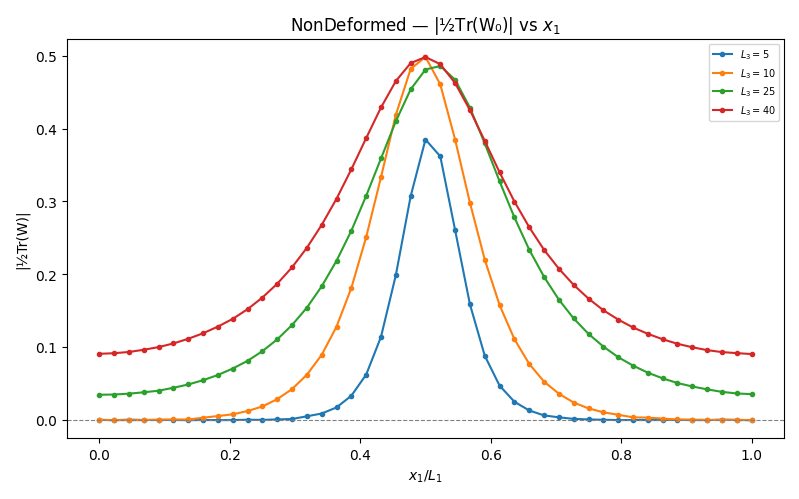}
\caption{\small YM: $\tr W_0({x_1\over L_1})$ for $L_3=5,15,25,40$} \label{fig:c91}
\end{subfigure}
\begin{subfigure}[b]{0.45\textwidth}
\centering
\includegraphics[width=\textwidth]{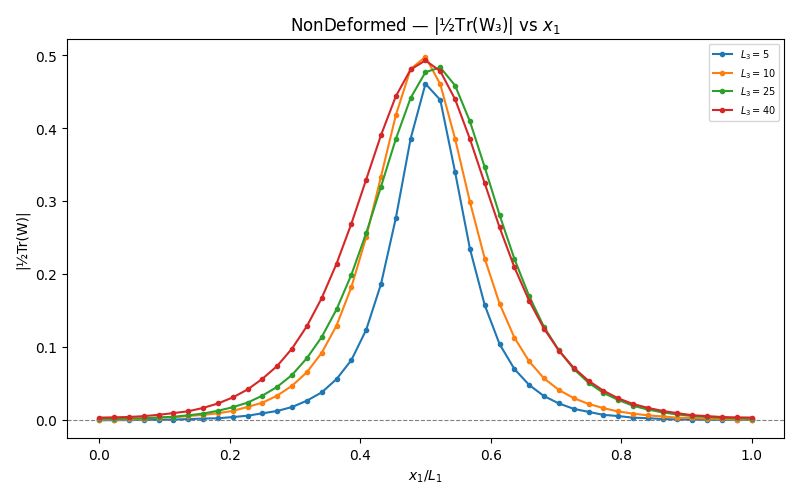}
\caption{\small YM: $\tr W_3({x_1\over L_1})$ for $L_3=5,15,25,40$} \label{fig:d91}
\end{subfigure}\caption{Smoothly interpolating from large to small $L_3$ on a $(10, 45, 45, L_3)$ lattice. Recall that $L_3 \gg L_0$ is the regime where the analytic construction of \cite{Hayashi:2024yjc,Guvendik:2024umd} and the picture of Figure~\ref{fig:picture1} hold. On the other hand, for small $L_3 \sim L_0$, there is no qualitative difference between the YM and dYM center vortex with $L_0 L_3 \ll L_1 L_2$, due to the alignment of abelianizations.  {\small Here we  compare $\tr W_0({x_1\over L_1})$ and $\tr W_3({x_1\over L_1})$ (the other coordinates are taken at the point of maximal action density) for dYM (top row) and YM (bottom row) for four different values of $L_3$. It is clearly seen that for small $L_3 \sim L_0$, blue curve, $\tr W_0$ and $\tr W_3$ in YM and dYM  are qualitatively similar.  On the other hand, for the largest $L_3= 40$, the YM solution becomes more delocalized, as seen on Fig.~\ref{fig:c91}, while  dYM remains localized, as per the behaviour of $\tr W_0$ on Fig.~\ref{fig:a91}, but with $\tr W_3$, on Fig.~\ref{fig:b91},  becoming  small.}   }
\label{fig:YMvsdYM4}
\end{figure}

Next, we study the collimation of the monopole-instanton flux into a tube stretched along $x_3$ and localized in the $12$ plane. We first refer to  Figure \ref{fig:fluxcollimating}, where we plot $F_{12}^{U(1)}$ halfway between the BPS and its KK image, as a function of $x_1/L_1$---equivalently, due to the approximate cylindrical symmetry, this can be though of as the dimensionless radial coordinate in the $12$ plane. We see that for the larger $L_1$ sizes (so that images in $x_1, x_2$ directions can be neglected), an exponential localization occurs, within, roughly, a $L_3/\pi$ radius. 
Further, on Figure~\ref{fig:a8} and \ref{fig:b8} we show that the flux in both in dYM and YM fits well to a Gaussian and determine its parameters, shown in the caption. 
That this tube of flux with a finite thickness is a center vortex is seen clearly by studying the disordering effect on Wilson loops surrounding the tube, shown on Figure \ref{fig:a4}, \ref{fig:b4}.  

We note that refs.~\cite{Hayashi:2024yjc,Guvendik:2024umd}, working in the $L_{1,2} \rightarrow \infty$ limit, used the long-distance abelian field of a monopole-instanton to argue that the effect of the infinite ...-BPS-KK-BPS-... chain of Figure~\ref{fig:picture1} is to collimate the magnetic flux into a center-vortex-like configuration. They showed that the   this collimation  is such that, in the midpoint between a monopole-instanton and its image, as $|x_{1,2}| \rightarrow \infty$, the magnetic field falls off as $e^{- {|x| \pi \over L_3}}$ (this falloff can serve as an approximate measure of the localization of the flux).  The actual Gaussian shape seen on Figure~\ref{fig:a8} and \ref{fig:b8} and its width were not  determined. Thus, their observation and  determination are   new results of this (finite $L_{1,2}$) study. 
  

 \section{From the ``flux'' to the ``no-flux'' vacuum: from monopole-instantons on $\mathbf{\R^3 \times \S^1}$ to  fractional instantons on $\mathbf{\R \times \T^3}$}
\label{sec:morphing2}

We now study the transition of   monopole-instantons in dYM on $\R^3 \times \S^1$ to fractional instantons on $\R \times \T^3$. In this case,  there is no analytical picture of the transition between the corresponding semiclassical configurations  as explicit as   the one of Figure~\ref{fig:picture1} (studied numerically in our previous Section \ref{sec:morphing1}). 

\begin{figure}[p]
 \centering
  \captionsetup{width=\textwidth}
\includegraphics[width=.9\textwidth]{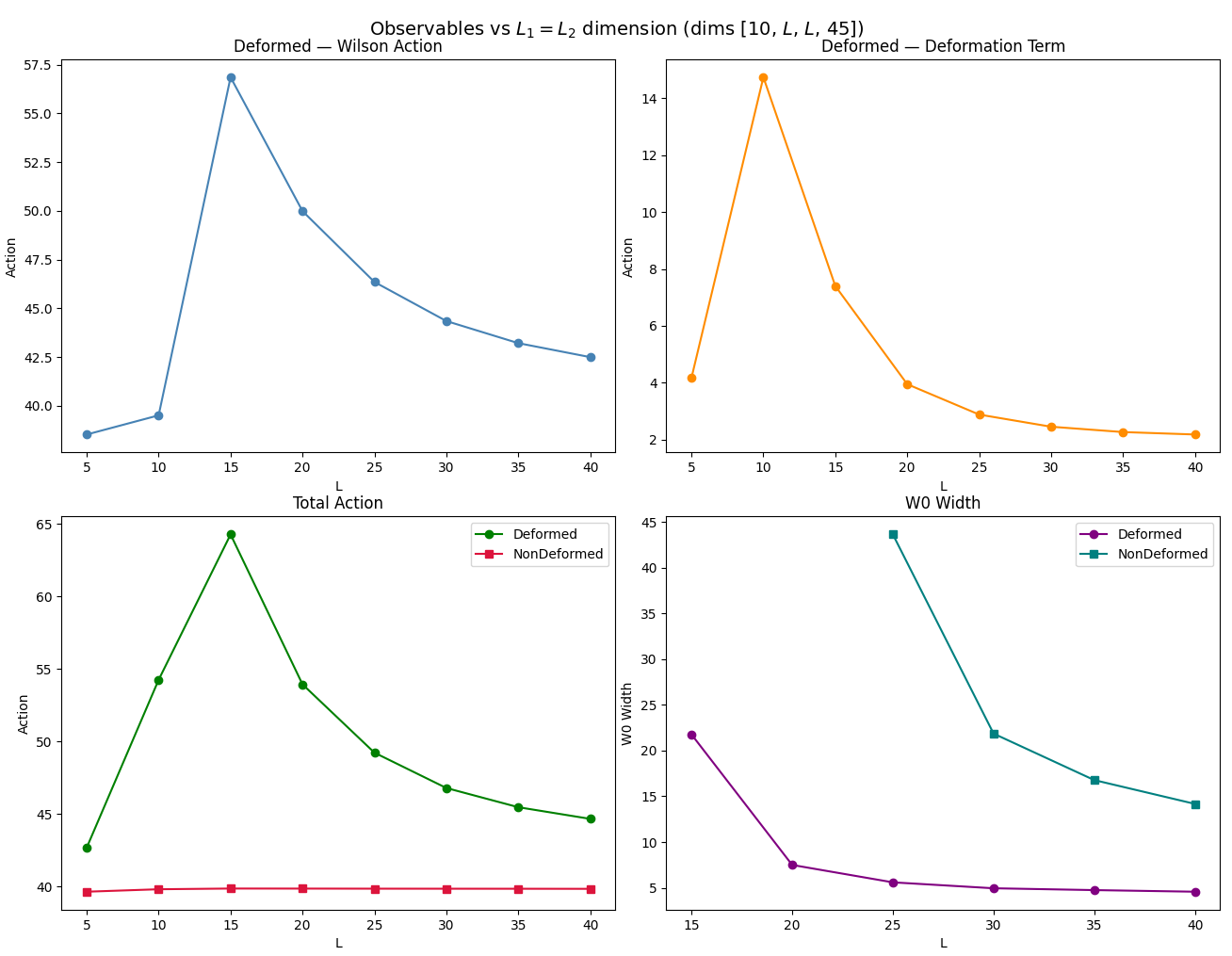}
\caption{Results for a $(10, L,L, 45)$ lattice interpolating between $\T^3 \times \R$ for small $L$ and $\S^1 \times\R^3$ for large $L$. All quantities are shown for $5 \le$$L$$\le 40$. {\small \underline{\it Top row:} The  Wilson and deformation contributions to the action of dYM. \underline{\it Bottom row:} The total actions of YM and dYM, as well as the $\tr W_0$ width in the $x_1,x_2$-plane, determined as in Fig.~\ref{fig:HiggsWidth}. 
  The $W_0$ width in the $12$ plane is not shown for $L<15$. For dYM, this is because the theory transitions to the no-flux vacuum. For YM  minimum action configurations, the $W_0$ width becomes larger than $L$ as $L \rightarrow 25$ from above, showing that the solutions delocalize on the $\T^3$ and abelianization in YM along $\tr W_0=0$, i.e. aligned with dYM flux vacuum, fails. On the other hand, for $L < 15$, the no-flux vacua of dYM and YM are aligned, as per (\ref{asympt}) (see the discussion in the text).}}
\label{fig:L345L1varies}
\end{figure}

We begin our numerical study of the   $\R^3_{(x_1,x_2,x_3)} \times \S^1_{(x_0)}$ to   $ \R_{(x_3)} \times \T^3_{(x_0,x_1,x_2)}$ transition in dYM by  focusing on 
 Figure \ref{fig:L345L1varies}, presenting results for a ($10, L, L, 45$) lattice.\footnote{As always, we use the notation and choice of twists from (\ref{choicesoftwists}).} The plots shown are similar to the ones on  Figures \ref{fig25L3} and \ref{fig45L3}. However,  while there we varied $L_3$ instead, keeping $L_0$ and $L_1$$(=L_2)$ fixed, here  we vary $L=L_1$, from $5$ to $40$, keeping fixed the small $L_0$ and the large $L_3$.  For large  values of $L_1\sim L_3$, our lattice can  be thought of  as  $\R^3_{(x_1, x_2, x_3)} \times \S^1_{(x_0)}$, while for $L_1 \sim L_0$ we approach $\T^3_{(x_0, x_1, x_2)} \times \R_{(x_3)}$; this will be substantiated by the results that we present below.

  The transition from the asymmetric (large $L \sim L_3$) to the symmetric (small $L \sim L_0$) $\T^3$ studied on Figure~\ref{fig:L345L1varies} is related to the transition from the flux to no-flux vacua in dYM as determined by (\ref{phases}).  Recall that the transition from the flux vacuum (large $L_1/L_0$) of dYM on $\T^3_{(x_0, x_1, x_2)}$ to the no-flux vacuum (small $L_1/L_0$) occurs at $L_1 \simeq 1.5 L_0$, or  $L_1= 15$ for the value of $L_0$ chosen on the figures. A quick glance on   Figure \ref{fig:L345L1varies} shows that both the Wilson and the total action of dYM   show a  maximum at  values of  $L_1 = L_c \sim 15$. 
The discontinuous nature of  the flux to the no-flux transition from Section \ref{sec:competing} (recall the level crossing   from the flux to the no-flux vacuum seen on Figure \ref{fig:transition}) suggests that the associated change of the nature of the minimal action fractional instantons upon changing the $\T^4$ shape is also discontinuous. We begin our discussion by first discussing the small- and large-$L$ limits and then focusing on the transition between the two.

   \begin{figure}[p]
 \centering
  \captionsetup{width=\textwidth}
 \begin{subfigure}[b]{\textwidth}
 \centering
\includegraphics[width=.6\textwidth]{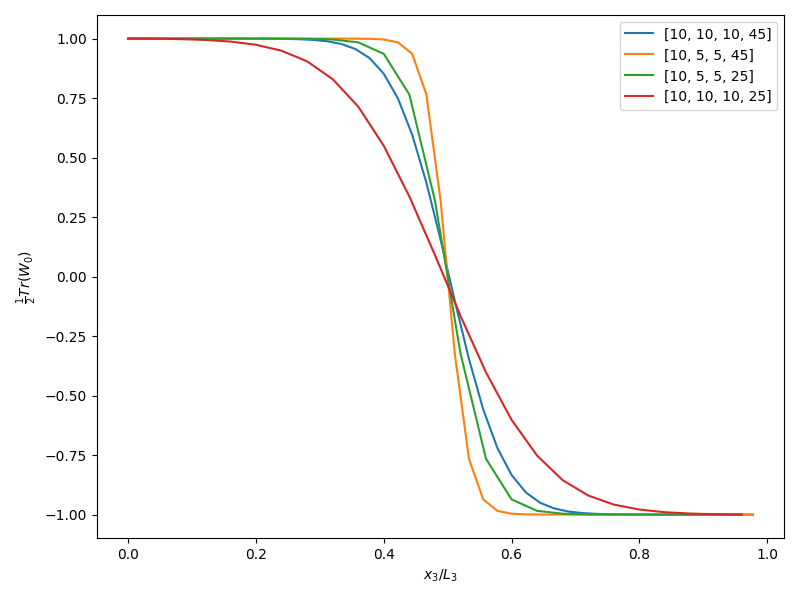}
\caption{\small The dYM fractional instanton on $\T^3 \times \R$ disordering the Wilson loop $\tr W_0$.}\label{fig:c9}
\end{subfigure}\\
\begin{subfigure}[b]{\textwidth}
\centering
\includegraphics[width=.7\textwidth]{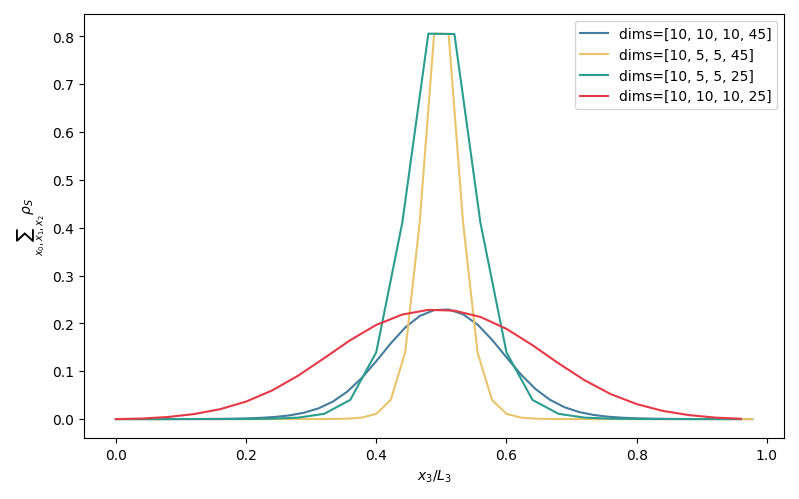}
\caption{\small   The action density of the fractional instanton in dYM integrated over $\T^3$.} \label{fig:d9}
\end{subfigure}
\caption{\small \underline{\it Top figure:} the disordering effect of the monopole instanton on $\R_{(x_3)} \times \T^3_{(x_0, x_1, x_2)\vert_{n_{12}=1}}$ on the Wilson loop winding  in the $x_0$ direction, for  $\T^3$ of sizes $(10,10,10)$ and $(10,5,5)$ for two different values of $L_3=25, 45$. This is the counterpart of the center-vortex disorder of Figure \ref{fig:a4}. \underline{\it Bottom figure:} The instanton is localized at $x_3/L_3=0.5$. The action density plots indicate that the instanton more strongly localized at smaller $L_1 L_2=L_1^2 $, with its scale set by $L_1$. The small-$\T^3$ regime studied here is where the YM and dYM abelianizations align and the effect of the deformation is small. (The continuous curves on the plot are obtained by  interpolating between discrete values of $x_3$, which are not shown.)}
\label{fig:disorderT3}
\end{figure}

{\flushleft\bf Small-$L_1$,  $\T^3 \times \R$ regime:}
Let us first   focus on the small-$L_1$ regime on Figure \ref{fig:L345L1varies}. The $L_1 = 5, 10$ data can be thought as approximating $\T^3_{(x_0,x_1,x_2)\vert_{n_{12}=1}} \times \R_{(x_3)}$.  The first feature we observe is that the deformation action is very small at   $L_1 = 5, 10$. 

We can  estimate the value of the deformation action as follows. Let us assume that the configuration around the localized solution is the no-flux vacuum of dYM with 
$|\tr W_0|=2$ and let us take the deformation action in (\ref{action2}) to be $4 (L_3 - L_1) {L_1^2 \over L_0^3}$. In writing this expression, we replaced $|\tr W_0|^2 \rightarrow 4$  in (\ref{action2}). We also assumed that $\tr W_0$ vanishes for some segment along $x_3$, whose length is of order $L_1$ (for now, this is just our guess for the width in $x_3$ of the finite action solution). Otherwise, $\tr W_0$ equals $\pm 2$, for an extent in the $x_3$ direction of length $L_3 - L_1$. Remarkably, 
for $L_1 = 5$    we find that this expression for the deformation action equals $4$,  precisely matching the values shown on Fig.~\ref{fig:L345L1varies}. Likewise, for $L_1 = 10$    we find for the deformation action  $14$, also perfectly matching the numerics.\footnote{We also studied the smaller lattice with $L_3=25$, where similar estimates at the smallest $L_1$ values work as well.}  Thus, the dYM no-flux vacuum surrounding the localized (in $x_3$) solution explains the value of the deformation action for both $L_1=5$ and $L_1=10$,    the two points below the transition at $L_1 = 15$.

Next, we can check the assumptions used above to explain the value of the deformation action for $L_1 = 5, 10$. We do this by plotting,  on Figure~\ref{fig:disorderT3}, the action density and $\tr W_0$ for the solution as a function of $x_3$. These plots show that the solution localizes in a segment of length of order $L_1$ on the $x_3$ axis, as we now discuss.
On Figure~\ref{fig:d9}, we plot the results for the dYM action density (integrated over $x_0,x_1,x_2$) as a function of $x_3/L_3$, for two values of $L_3=25, 45$ and for $L_1=5, 10$. It is easy to infer from the plot that the action density is localized over a region of size $L_1$.

Further, we note that despite the fact that we are plotting the properties of the dYM minimum action configuration, the resulting configuration is very close to the one in pure YM theory, due to the fact that for lattices such as plotted here, the abelianizations due to the twist $n_{12}$ in YM and in dYM align, as per (\ref{asympt}). 
 Concretely, as in our discussion in Section \ref{sec:morphing1}, we notice that for our choice of parameters,  $L_1=5$, $L_1 L_2 = 25 \ll L_0L_3$ ($=450$), values for which  the pure YM argument of eqn.~(\ref{asympt}) implies that far from the core of the solution, $\tr W_0 = \pm 2$ (a similar argument albeit leading to a somewhat less strong inequality applies for $L_1=10$). 
Thus,  in the small-$L_1$ regime of Figures \ref{fig:L345L1varies} and \ref{fig:disorderT3}, the deformation is not qualitatively important, as the same structure surrounding the solution  localized in $x_3$, the no-flux vacuum  (\ref{noflux}),  is implied by both the dYM criterion (\ref{phases}), and the YM one   (\ref{asympt}), due to the $12$-plane twist.

Finally, we study the disordering of the $\tr W_0$ Wilson loop by the $\T^3 \times \R$ fractional instanton. The top plot,   Figure~\ref{fig:c9}, shows  the variation of $\tr W_0$,  taken at the point of maximal action in $x_1, x_2$, as a function of $x_3/L_3$, on $\T^3 \times \S^1_{(x_3)}$ (the circle size is $L_3$, approximating $\R$) for  $\T^3$ of sizes $(10,10,10)$ and $(10,10,5)$. We see, in particular, that $\tr W_0$ jumps from $+2$ to $-2$ as one crosses the solution. This is the one-dimensional analogue (i.e. on $\T^3 \times \R$) of the disordering of the Wilson loop by a center vortex on $\R^2 \times \T^2$,  observed long ago in \cite{RTN:1993ilw, Gonzalez-Arroyo:1995ynx} and argued to lead to semiclassical center symmetry restoration/confinement.

{\bf\flushleft{The large-$L_1$, $\R^3 \times\S^1$  regime:}} We now move to the large-$L_1 \ge 20$ regime on Fig.~\ref{fig:L345L1varies}. 
In this large-$L_1$ regime, the solutions are the monopole-instantons in the flux vacuum of dYM already discussed in Section \ref{sec:morphing1}. That this is so is already clear from the plot of the Higgs field ($\tr W_0$) width in the $12$ plane shown on the bottom r.h.s.~plot on Figure~\ref{fig:L345L1varies}, which, in dYM, remains finite and smaller than $L$ even for $L=20$. 

It is also easy to see that for the values of $L_1$ and $L_3$ that match the ones of Figs.~\ref{fig25L3} and \ref{fig45L3} all quantities agree with Figure~\ref{fig:L345L1varies}.
Finally, we note that here,   $L_0 L_3 = 450$ and $L_1L_2 = L_1^2 = (400,...,1600)$ as $L_1 = (20,..., 40)$. Thus, the lower range of the $20 < L_1 < 40$ regime plotted,  is similar to the one on Figs.~\ref{fig:c}, \ref{fig:e}, where dYM abelianizes but YM does not. This is not entirely surprising since the level crossing transition at $L_1 = 1.5. L_0$ from flux to no-flux vacuum is one appearing in dYM. In YM, on the other hand, the transition is through the completely delocalized constant field strength minimal action solutions at $L_0L_3 = L_1L_2$.
 \begin{figure}[h]
 \centering
  \captionsetup{width=\textwidth}
 \begin{subfigure}[b]{0.45\textwidth}
 \centering
\includegraphics[width=\textwidth]{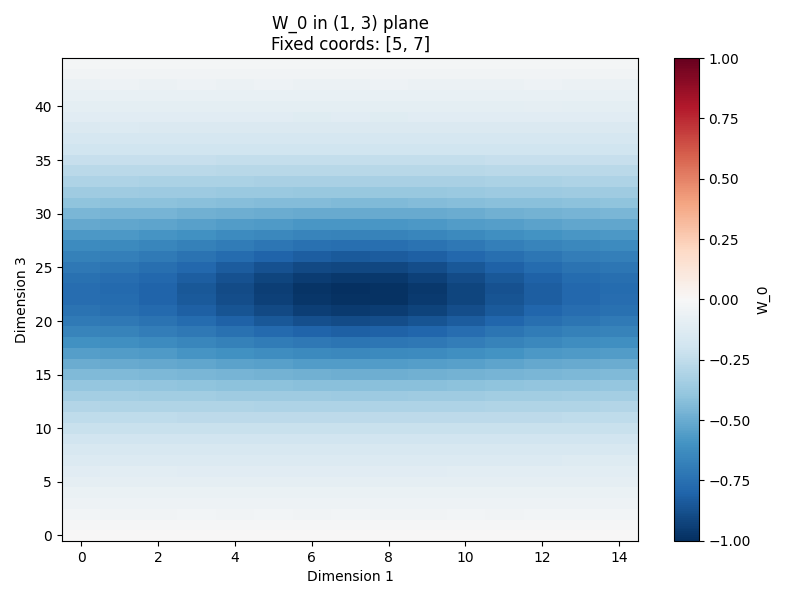}
\caption{\small (10, 15, 15, 45): $\tr W_0$ in $13$ plane}\label{fig:a11}
\end{subfigure}
\begin{subfigure}[b]{0.45\textwidth}
\centering
\includegraphics[width=\textwidth]{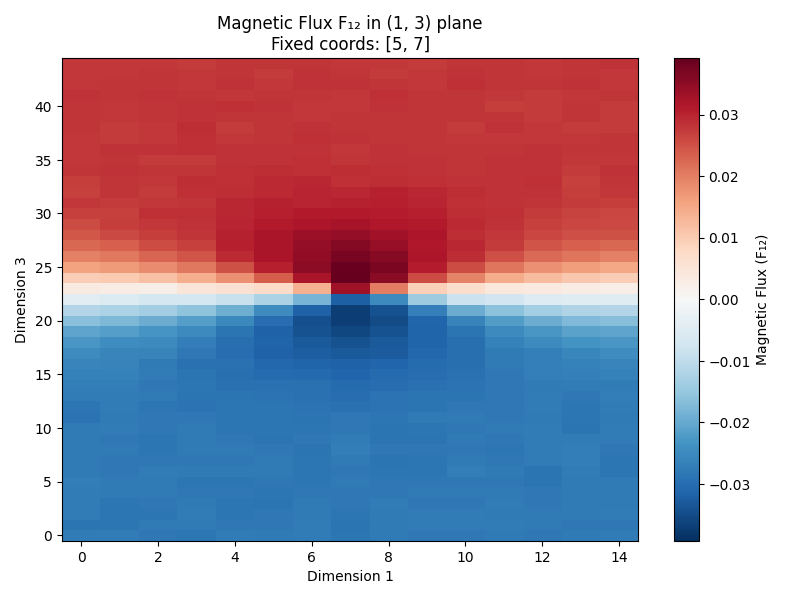}
\caption{\small   (10, 15, 15, 45): $F_{12}^{U(1)}$ in $13$ plane} \label{fig:b11}
\end{subfigure}
\caption{A small number ($2$ out of $9$) of  minimum action dYM configurations with $|Q|=1/2$ at ${L_1 \over L_0} = 1.5$, on a $(10,15, 15, 45)$ lattice, at the transition point between the flux and no-flux vacua of dYM, are found to look like monopole instantons in the flux vacuum of dYM. $\tr W_0$ shows localization in $x_3$, with width of order $L_1$, and approaches $0$ far away from the core of the monopole instanton showing abelianization in dYM. }
\label{fig:dYMtransition12}
\end{figure}

{\bf \flushleft{The transition region, near $L_1 = L_c \sim 15$.}}
We now move on to discuss the transition seen at $L_1 \sim 15$, associated, as per discussion above, with the dYM transition from the flux to the no-flux vacuum.

First we note that on Fig.~\ref{fig:L345L1varies},   the action peak for the $(10,L,L,45)$ lattice is at $L_c =15$ is at $L_1 L_2 = 225$ while $L_0 L_3 = 450$. In pure YM these values would be not far from the ones where the minimal action configuration is the constant-$F$ one (which would occur at $L_1=L_2 \simeq 21$). Thus in YM,  one would expect delocalized configurations, as was found in \cite{Wandler:2024hsq} (this is clear already from the $\tr W_0$ $12$-plane localization plot for YM on the bottom r.h.s.).

\begin{figure}[h]
 \centering
  \captionsetup{width=\textwidth}
 \begin{subfigure}[b]{0.55\textwidth}
 \centering
\includegraphics[width=\textwidth]{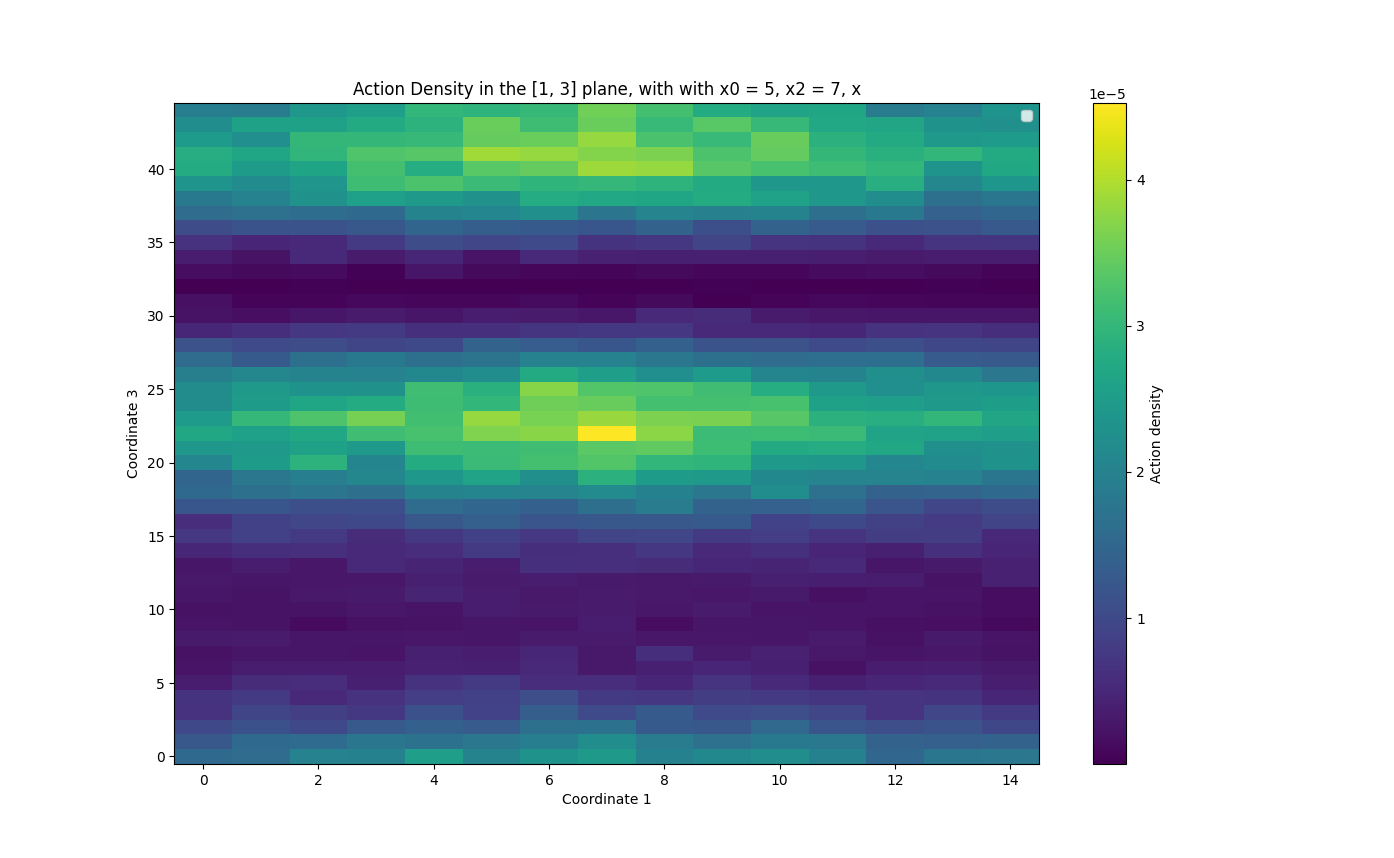}
\caption{\small The Wilson term action density for a $(10,15,15,45)$ lattice in the $13$ plane. Its two-hump profile as a function of $x_3$, taken at the point of maximal action in $12$ plane, is also shown on the right, albeit for a $(10,15,15,64)$ lattice with larger $L_3$.}\label{fig:a12}
\end{subfigure}
\begin{subfigure}[b]{.44\textwidth}
\centering
\includegraphics[width=\textwidth]{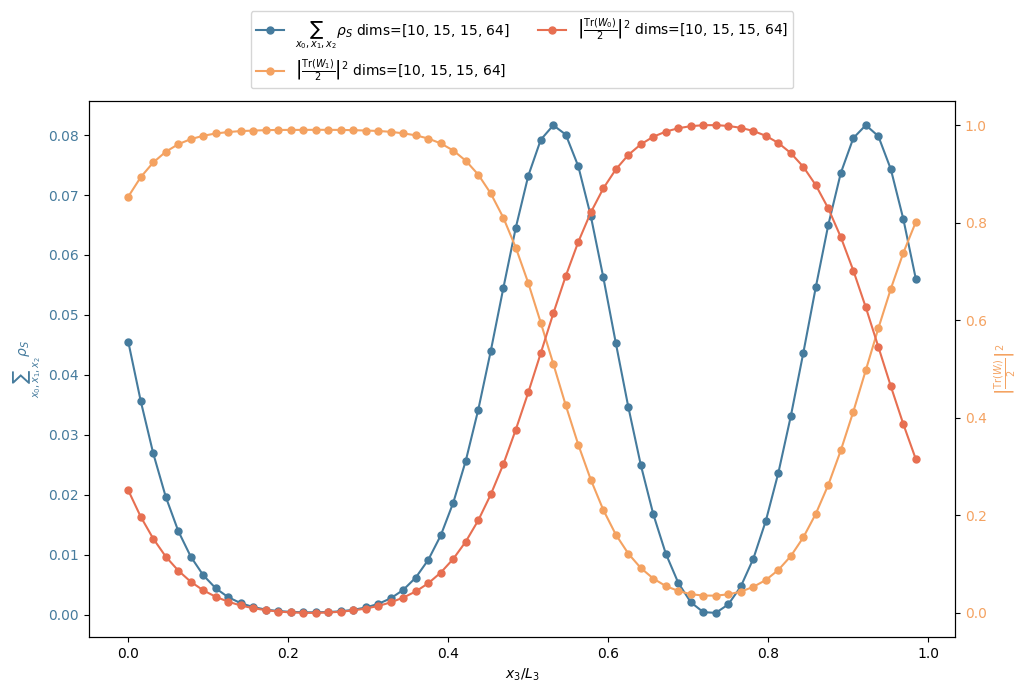}
\caption{\small  The deformation action, $|\tr W_0/2|^2$ (the curve with a single maximum, with scale given on the r.h.s.), along with $|\tr W_1/2|^2$  and the Wilson term action density, displaying a two maxima structure, as a function of $x_3$} \label{fig:b12}
\end{subfigure}
\caption{The majority ($7$ out of $9$) of the  minimum action dYM configurations with $|Q|=1/2$, on a $(10,15, 15, 45)$ lattice, at the transition point  ${L_1 \over L_0} = 1.5$ between the flux and no-flux vacua of dYM, are found to have the two maxima structure of their Wilson action density, shown on the left figure, but a single deformation action peak, seen on the r.h.s. On the r.h.s., we compare  $|\tr W_0/2|^2$, $|\tr W_1/2|^2$, and the Wilson action density as a function of  $x_3$, for a lattice with a  larger $L_3$ at the same critical value of $L_1$ (we do not show $|\tr W_2|^2$ as its behaviour is identical to that of $|\tr W_1|^2$ shown). These configurations show no localization in $x_3$, with the region of nonzero Wilson and deformation action density taking roughly half the $L_3$ length. (This is true for $L_3=45$ shown as well as for $L_3=65$, which we do not show).  These configurations have, within our precision, roughly the same (or a bit smaller) total action as the flux vacua ones on Fig.~\ref{fig:dYMtransition12}, but a smaller Wilson and larger deformation action, due to the $\tr W_0$ behaviour shown.}
\label{fig:dYMtransition13}
\end{figure}

In dYM, the nature of the configurations appearing at $L=15$ is as follows. In a small number ($2$ out of $9$) of the minimum action configurations for a $(10,15,15,45)$ lattice found by our minimization algorithm, we observe configurations with properties like those of the flux vacuum monopole instantons, as shown  on Figure \ref{fig:dYMtransition12}.
However,  the majority ($7$ out of $9$) of minimal action configurations identified by our algorithm, are configurations of similar action, which are not localized in $x_3$. These configurations  show a two-peak Wilson action density structure in $x_3$, see Figure~\ref{fig:dYMtransition13}, but a single-peak deformation action density---with $|\tr W_0|^2/2$ shown on the r.h.s. of the same Figure.\footnote{We note that on the r.h.s. we show the data for the $(10,15,15,64)$ lattice of larger $L_3$ (where we found that all minimum action configurations at $L=15$ were of the ``two-hump'' variety shown). We stress, however, that the identical  identical structure of all quantities shown appears for the $7$ out of $9$ configurations on the $(10,15,15,45)$ lattice.}

Our final comment on the transition region is that its more detailed study requires significantly more resources: in particular, a more fine-grained study of the transition region, as well as studies of larger lattices and possibly different values of $c$ may be warranted, a task that goes beyond the scope of this work. However, the point made clear by our results is that the small-$L$ and large-$L$ values correspond to the no-flux and flux vacua surrounding the localized finite action solution, with the detailed study of the transition region left for future work. 

{\flushleft{\bf Acknowledgments:}} This work is supported by an NSERC Discovery Grant. We also acknowledge the Digital Research Alliance of Canada for giving us access to computer resources. We thank Rajamani Narayanan for helpful discussions of numerical methods.

\appendix
\section{Implementing the gradient flow}
\label{appx:flow}

Typically minimization of the Wilson action in $SU(2)$ pure Yang-Mills is done using cooling techniques, see \cite{GarciaPerez:1993lic,deForcrand:1995qq,Montero:2000mv} or  the recent \cite{Wandler:2024hsq, Anber:2025yub}. However, cooling techniques rely on plaquettes in the Wilson action being first order in the link variables, and therefore will not work for more general actions. Hamiltonian Monte Carlo \cite{Duane:1987216} techniques are more flexible, but reproduce the statistics of the path integral rather than generating minimum action configurations. To circumvent this limitation, we used a modified version of the HMC algorithm, where rather than giving a configuration random momentum in between evolution periods, the configuration's momentum was set to zero. If the Hamiltonian evolution periods are short enough, this ensures that each evolution period will reduce the action of the configuration, unless numerical errors have caused the algorithm to not conserve energy, in which case the configuration is rejected and the evolution is tried again with a smaller $\delta \tau$. The procedure will be summarized below. 

The overall action used was the same as in (\ref{action1}), repeated here for convenience:
\begin{equation}
S_{total} = A(S_{Wilson} + S_{def.}) = A \left( \sum_{x} \sum_{\mu \nu} {\rm tr}\left( \mathbf{1} - B_{\mu \nu}(x)  {\Box}_{\mu \nu}(x) \right) +  {c \over L_0^3} \sum_{\vec{x}} \left| {\rm tr} W_0(\vec{x})\right|^2\right)  ~.
\end{equation}
Following the procedure in \cite{Duane:1987216} and \cite{Lippert:1997qx} we derive equations of motion, where $x\in \mathbb{R}^4$ and $\tau$ being the fictitious time evolution parameter:
\begin{align}
\dot{U}_{\mu}(x,\tau) &= \pi_\mu(x,\tau)U_{\mu}(x,\tau), \notag \\
\dot{\pi}_{\mu}(x,\tau) &= A\left(\underbrace{\frac{c}{L_0^3}(Tr(W_0^\dagger)W_0 - Tr(W_0)W_0^\dagger)}_{\text{Contribution from $S_{def.}$}} + \underbrace{(V^\dagger_\mu \; U_\mu^\dagger -U_\mu\; V_\mu )}_{\text{Contribution from $S_{Wilson}$}}\right)(x,\tau) \label{HMC-EOM} ~.
\end{align}
where $W_0 := W_0(x)$, i.e. the Wilson loop winding in the $0$ direction starting at site $x$, and the staple $V_\mu(x)$ is defined by
\begin{equation}
V_{\mu}(x) =\sum_{\nu \neq \mu} \left( B_{\mu \nu}(x) U_{\nu}(x+\hat{e}_\mu) U^\dagger_{\mu}(x + \hat{e}_\nu)U^\dagger_{\nu}(x) + B_{\mu \nu}(x - \hat{e}_\nu) U_{\nu}^\dagger(x - \hat{e}_\nu + \hat{e}_\mu)U_{\nu}^\dagger(x - \hat{e}_\nu)U_\nu(x - \hat{e}_\nu)\right) ~. 
\end{equation}
where $B_{\mu \nu}$ is the 2-form topological background (\ref{twist}),  as in (\ref{singletwist}) or (\ref{choicesoftwists}). 
The $\dot{\pi}_\mu(x)$ equation is derived by defining the Hamiltonian 
\begin{equation}\mathcal{H} = \sum_{x, \mu} \tr \left( \frac{\pi_\mu \pi_\mu^\dagger}{2} \right) +S_{total}[U_\mu] ~. \label{HMC-Hamiltonian}\end{equation}
and then enforcing $\dot{\mathcal{H}} = 0$ and $\dot{U}_\mu(x, \tau) = \pi_\mu(x,\tau)U_{\mu}(x,\tau)$. 
Then, the procedure for generating a configuration is as follows:
\begin{enumerate}
\item{Generate a lattice with random link variables $U_\mu(x)$.}
\item{Assign each link variable an $\mathfrak{su}(2)$-valued momentum $\pi_\mu(x) = \mathbf{0}$. The initial total energy is thus $\mathcal{H} = S_{total}$.}
\item{Numerically evolve the links and momenta for some length of fictitious time $\tau_0$ according to equations (\ref{HMC-EOM}), implemented using a leapfrog time-stepping method to conserve $\mathcal{H}$.}
\item{Verify that $S_{total}$ has decreased. If it has, accept the configuration. Otherwise, increase numerical precision (by reducing $A$ and shrinking the time step $\delta \tau$) and repeat.}
\item{If accepted, set all $\pi_\mu(x) = \mathbf{0}$ and repeat steps 3 and 4 for the new configuration of links.}
\item{Repeat this process until action has stopped changing or a predetermined number of evolution periods have elapsed.}
\end{enumerate}
This procedure is the same as in standard HMC, with the exception that momenta are set to zero rather than randomized between evolution periods and the acceptance criterion is  a reduction in the action, not of $\mathcal{H}$. In contrast with HMC, this acceptance step plays no statistical role and is simply a check that the algorithm is numerically stable. 

To speed up the rate of convergence, the parameter $A$ can be adjusted, which increases the ``forces'' present in equation (\ref{HMC-EOM}). However, the ``true'' action of a configuration, the value shown in figures above, was always calculated by dividing out the parameter $A$. Each time the change in true action after a period of evolution dropped below 1\%, the parameter $A$ was multiplied by 10 for the next period of evolution. This continued until the program started to encounter numerical instability, at which point the multiplication stopped. Any subsequent numerical instability resulted in $A$ being halved.

This procedure was able to easily reproduce the  known numerical results on pure Yang-Mills twisted lattices found in \cite{Wandler:2024hsq}. For most lattices, if a true minimum of the action was found, the program had reduced the action to that minimum within around 300 periods of evolution for a fictitious time $\tau = 10$. Frequently a local minimum of the action would be encountered far above the true minimum. These local minima appeared frequently when reducing the action of dYM configurations, and became more common as lattice sizes increased. Pure Yang-Mills configurations very rarely encountered local minima, if at all. 

Given the lack of a BPS bound for dYM, the numerical value of actions designated ``true'' vacua were determined by performing this minimization process on a large number of configurations, and taking the smallest action found to be the true minimum. Usually, a significant fraction of configurations clustered around this ``true'' minimum value, and others clustered around various local minima above the true vacuum.

\label{appx:gradientflow}
 
\section{On the classification of finite action Euclidean solutions on $\R^2 \times \T^2_{n_{12}}$}
\label{appx:classify}

Let us work on $\R^2_{(x_0,x_3)} \times \T^2_{(x_1,x_2)}$ with a twist $n_{12}=1$, in an $SU(2)$ pure YM theory.\footnote{For brevity we take $SU(2)$ gauge group, noting that the argument is easily generalized to any $SU(N)$ with a general twist $n_{12}$, see the comment at the end.  Also, it generalizes to dYM, with the deformation either in the $x_3$ direction (which then would have to be taken compact, replacing $\T^2$ with $\T^3$) or in one of the directions of $\T^2$, e.g.the $x_1$.}
We want to classify the finite Euclidean action configurations in this setup. We begin by taking the $A_0=0$ gauge, which, we recall, is also very convenient in   the usual analysis of finite action instantons on $\R \times \R^{3}$, with $\R$ denoting Euclidean time.
A finite action field configuration should have the property that at $\R^2$-infinity, it should satisfy $F_{ij}=0$ and $F_{0i}=0$, where $i=1,2,3$, i.e. should go to a pure gauge. The second condition, in the $A_0=0$ gauge, implies that $\partial_0 A_i=0$ at infinity in $\R^2_{(x_0,x_3)}$. It is convenient to think of infinity in $\R^2$ as a square, whose four sides correspond to $x_0 \rightarrow \pm \infty$ (for any $x_3$) or $x_3  \rightarrow \pm \infty$ (for any $x_0$).

We then ask what are the minimum (zero) energy classical configurations, $F_{ij}=0$, on $\R_{(x_3)} \times \T^2_{(x_1,x_2)\vert_{n_{12}=1}}$, which should be the configurations the finite action Euclidean solutions should approach as $x_0 \rightarrow \pm \infty$. To proceed, we take constant transition functions in the $12$ plane, $\Gamma_1, \Gamma_2 \in SU(2)$,  with the gauge field $A \equiv A_i dx^i$ (a sum over $i=1,2,3$ is  implied) obeying
twisted boundary conditions on $\T^2$: \begin{eqnarray}
A(\vec{x} + \vec{e}_1 L_1) &=& \Gamma_1A(\vec{x}) \Gamma_1^\dagger,\\
A(\vec{x} + \vec{e}_2 L_2) &=& \Gamma_2 A(\vec{x}) \Gamma_2^\dagger, \\
 \Gamma_1 \Gamma_2&=& e^{i \pi} \Gamma_2 \Gamma_1
\end{eqnarray}
Clearly, $F_{ij}=0$ implies that $A \equiv A_i dx^i$ is pure gauge at $x_0 \rightarrow  \pm\infty$. The solution to this problem has been known for a long time \cite{Witten:1982df}:
\begin{eqnarray}\label{two}
A^{(k)} =- i T^k d T^{-k}, ~k=0,1, ~ \text{where} \; T = T(x_1, x_2, x_3) \equiv T(\vec{x}) \in SU(2).
\end{eqnarray}
In other words, the pure-gauge configurations which the finite action instanton should approach as $x_0 \rightarrow \pm \infty$ are either $A^{(0)}=0$, or $A^{(1)} = -i T d T^{-1}$.
The properties that $T(\vec{x})$ obeys are:
\begin{eqnarray}\label{T3}
T(\vec{x} + \vec{e}_1 L_1) &=& \Gamma_1 T(\vec{x}) \Gamma_1^\dagger \nonumber \\
T(\vec{x} + \vec{e}_2 L_2) &=& \Gamma_2 T(\vec{x}) \Gamma_2^\dagger \\
T(x_1, x_2, x_3 \rightarrow \infty) &=&e^{i \pi} T(x_1,x_2, x_3 \rightarrow - \infty) \nonumber \end{eqnarray}
An explicit expression of a  representative\footnote{\label{footnote:fractional}This is because $T$ itself is defined up to large gauge transformations with integer winding number, which can be thought of as maps from $\S^3$ to $SU(2)$, see \cite{Cox:2021vsa} for a recent discussion. In other words, as written, the two configurations (\ref{two})  only capture the fractional part of the topological charge. The integer action classification is the same as on $\R^4$.} of  $T$ can be written using the function $g(x_1,x_2)$ and $f(x_3/L_1)$ obeying the following properties \begin{eqnarray}\label{fg}
T(x_1,x_2,x_3) &=& g(x_1, x_2) e^{- i \pi \sigma^3 f({x_3\over L_1})} g^{-1}(x_1,x_2), \\
\text{where} ~ g(x_1 +L_1, x_2) &=& \Gamma_1 g(x_1, x_2), \nonumber \\
g(x_1, x_2 + L_2) &=& \Gamma_2 g(x_1, x_2) e^{- i \pi \sigma^3 {x_1\over L_1}}, \nonumber \\
f(y \rightarrow +\infty) &=& 1, \nonumber \\
f(y \rightarrow - \infty) &=& 0.\nonumber 
\end{eqnarray}
An explicit expression for the function $g(x_1,x_2)$ can be found in \cite{Poppitz:2022rxv},\footnote{The meaning of $g(x_1,x_2)$ is that it is the function (itself a rather complicated map from the covering space of $\T^2$ to $SU(2)$) that maps the constant to the abelian transition functions. This will not be important in our discussion here. We also note that if $L_3$ is taken finite, then in $T$ we replace $f(x_3/L_1)$  by $x_3/L_3$; $T$ then has the  meaning as the center symmetry generator in the $x_3$ direction. The two configurations $A^{(0)}$ and $A^{(1)}$ from (\ref{two}) are distinguished by the value of the winding Wilson loop in $x_3$ $\tr W_3$: it equals  $2$ for $k=0$ and $-2$ for $k=1$. We note that with the definition (\ref{fg}) this distinction remains true in the infinite $L_3$ limit, where $\tr W_3$ now includes an $x_3$ integral over the real axis.} where $\Gamma_1 = i \sigma^1$ and $\Gamma_2 = i \sigma^3$, but all we need to verify that 
$T(\vec{x})$ of (\ref{fg}) obeys (\ref{T3}) is its existence and properties. The function $f(y)$ can be any smooth function with the given limits, for example  $f(y) = {1 \over 2}(1+ \tanh y)$ and we note that the scale $L_1$ appearing in  $f(x_3/L_1)$ on the first line in (\ref{fg}) can be replaced by any fixed scale.

Further, we note that as $x_3 \rightarrow \pm \infty$, the two configurations $A^{(0)}$ and $A^{(1)}$ from (\ref{two}) (the two limits of the finite action solution at $x_0 \rightarrow \pm \infty$) approach zero exponentially fast because $e^{i \pi \sigma^3 f}$ approaches $\pm 1$ times the unit matrix at spatial infinity, thus $T$ becomes trivial and $A$ vanishes as $|x_3| \rightarrow \infty$.  In each case, as $A=0$ at $|x_3| \rightarrow \infty$, $\partial_0 A_i = F_{0i} =0$ is obeyed there, as required. The configurations at $x_0 = \infty$ and $x_0 = - \infty$ could both be the same of differ, i.e. correspond to the same or different value of $k$ from (\ref{two}). Thus, a finite action Euclidean configuration on $\R^2 \times \T^2$ with $n_{12}=1$ can have one of four possible boundary conditions at the $\R^2$ infinity.
In the case where  the gauge field approaches the same limit at $x_0 \rightarrow \pm \infty$, i.e.~either both have $k=0$ (or  $k=1$), the topological charge is integer (owing to the fact that the large gauge transformations $\S^3 \rightarrow SU(2)$, not included in (\ref{two}), can have different integer winding in the two limits, recall Footnote \ref{footnote:fractional}). 

Consider then the gauge field approaching different limits, for definiteness, $k=1$ at $x_0 \rightarrow \infty$ (i.e. $A\vert_{x_0 \rightarrow \infty} =- i T d T^{-1}$)  and $k=0$ at $x_0 \rightarrow - \infty$ (i.e. $A\vert_{x_0 \rightarrow - \infty} =0$). The (fractional part of the) topological charge of such a configuration is then calculated, upon integrations by parts, in terms of its asymptotics, following the steps outlined below:
\begin{eqnarray}\label{qon2}
Q\big\vert^{k=1}_{k=0} &=& {1 \over 8 \pi^2} \int_{\R^2_{(x_0,x_3)} \times \T^2_{(x_1,x_2)}} \tr F \wedge F = {1 \over 24 \pi^2} \int\limits_{\R_{(x_3)} \times \T^2_{(x_1,x_2)}} \tr (T d T^{-1})^3 
\end{eqnarray}
The   equality above follows from  (\ref{two}) upon  integration by parts in $x_0$, similar to \cite{tHooft:1979rtg,tHooft:1981sps} (also given in detail in  \cite{Cox:2021vsa,Poppitz:2022rxv}; we stress that the use of constant transition functions is important). 

To further calculate the winding number (which appears  on the r.h.s. in (\ref{qon2})) of $T$, considered a map  $\R \times \T^2 \rightarrow SU(2)$ obeying the boundary conditions (\ref{T3}), we use the expression for $T$ from (\ref{fg}) and introduce  the shorthand notation $X_\alpha \equiv g^{-1} \partial_\alpha g$, $\alpha=1,2$ (explicit expressions for $X_\alpha$ are given in Appendix A in \cite{Poppitz:2022rxv}): \begin{eqnarray}\label{qon21}
&&{1 \over 24 \pi^2} \int\limits_{\R_{(x_3)} \times \T^2_{(x_1,x_2)}} \tr (T d T^{-1})^3 \nonumber \\
 &=&{ i  \over 4 \pi}\int\limits_{\R_{(x_3)} \times \T^2_{(x_1,x_2)}} dx_1 dx_2 dx_3 \; \partial_3 f \; \tr\left[ \sigma_3  (X_\alpha (1 - \cos^2 \pi f) - \sigma^3 X_\alpha  \sigma^3 \sin^2 \pi f) \epsilon^{\alpha\beta}  X_\beta \right]\\
&=& \int\limits_{-\infty}^\infty dx_3  (1- \cos 2\pi f)  \partial_3 f \int\limits_{0}^1 dx_1 \int\limits_{0}^1 dx_2 \; \partial_2 \tilde{f}^2(x_2)=( f\vert_{x_3 \rightarrow \infty} - f\vert_{x_3 \rightarrow -\infty}) (\tilde{f}^2(1)- \tilde{f}^2(0))= {1 \over 2}.\nonumber 
\end{eqnarray}
 Obtaining the  the last line requires using the properties of $g(x_1,x_2)$, along with behaviour of $f(x_3)$ from (\ref{fg}), as well as the already  mentioned definitions of $X_\alpha$. We  rescaled all coordinates appropriately so that $L_1=L_2=1$. The function $\tilde{f}(x_2)$ on the last line is the ``bump'' function entering the definition of $g(x_1,x_2)$, see Appendix A in \cite{Poppitz:2022rxv}; most importantly, it has the property that $\tilde{f}^2(1)- \tilde{f}^2(0)= {1 \over 2}$. The entire calculation follows the one given in Section 3.1 of \cite{Poppitz:2022rxv} for the $\R \times \T^3$ case,  the only difference being that the function $f(x_3)$ is taken to be the one appropriate for  $\R^2 \times \T^2$. Combining (\ref{qon2}) and (\ref{qon21}), we obtain that the fractional part of the topological charge with different asymptotics of the gauge field at $x_0 \rightarrow \pm \infty$ is $Q\big\vert^{k=1}_{k=0}= 1/2$.
The generalization to $\R \times \T^3$ with a twist $n_{12}$ in $\T^3$ follows essentially the same steps. 

The moral of the discussion here is that one can argue for the fractionality of the topological charge already in the semi-infinite limit $\R^k \times \T^{4-k}_{n_{12}}$, $k=1,2$. The argument is familiar from $\R^4$ and is based on the conditions on the gauge field configurations at infinity ensuring finite Euclidean action (made convenient by choosing $A_0=0$) and on the understanding of the kinds of locally pure-gauge configurations that are allowed in the presence of 't Hooft twists. This  had been understood already by the authors of \cite{RTN:1993ilw, Gonzalez-Arroyo:1995ynx,Gonzalez-Arroyo:1998hjb,Montero:2000pb} who observed that the fractional instantons persist as localized configurations of finite action when the appropriate semi-infinite volume limit is taken. As already noted, we are not aware of a more formal argument, along the lines given here, in the published literature,\footnote{However, A. Gonz\' alez-Arroyo has told us that he is familiar with the  argument given here.} which is why we included it for completeness. 

The argument generalizes to dYM as well, in both the flux and no-flux vacua. We shall not give any details, which can be filled in by the reader, but will only mention that if one considers the no-flux vacuum
 of dYM, the argument is essentially the same as the one already given. On the other hand, in the flux vacuum, the fractional topological charge corresponds to tunnelling events between the two different flux vacua, a fact that has been recognized already in \cite{Unsal:2020yeh}.
 
 Our final comment concerns the generalization to $SU(N)$.\footnote{The only advantage for considering  $SU(2)$  is that all quantities entering Eqns.~(\ref{fg})-(\ref{qon21}) are explicitly worked out in \cite{Poppitz:2022rxv}.}  Here,  one can similarly show that the topological charges are $Q={p \over N'}(\rm{mod}\; 1)$ with $p=0,...,N'-1$ and $N'  = N/{\rm{gcd}}(n_{12},N)$. The explicit detailed expression for the map $T$, which we used in (\ref{qon21}) to calculate the winding number (\ref{qon2}), is not really needed in order to obtain the fractional part of $Q$, see for example \cite{vanBaal:1982ag} or the Appendices of \cite{Cox:2021vsa}.

\bibliography{draft3.bib}

@article{Anber:2025yub,
    author = "Anber, Mohamed M. and Cox, Andrew A. and Poppitz, Erich",
    title = "{On the moduli space of multi-fractional instantons on the twisted ${T}^4 $}",
    eprint = "2504.06344",
    archivePrefix = "arXiv",
    primaryClass = "hep-th",
    doi = "10.1007/JHEP02(2026)169",
    journal = "JHEP",
    volume = "02",
    pages = "169",
    year = "2026"
}

@article{Bonati:2026ljv,
    author = "Bonati, Claudio and Caselle, Michele and Negro, Alessio and Panfalone, Dario and Verzichelli, Lorenzo",
    title = "{Confining Flux Tube in the Trace Deformed (2+1) Dimensional SU(2) Gauge Theory}",
    eprint = "2606.20270",
    archivePrefix = "arXiv",
    primaryClass = "hep-lat",
    month = "6",
    year = "2026"
}

@article{Poppitz:2026gfa,
    author = "Poppitz, Erich",
    title = "{D-branes and fractional instantons on a twisted ${T}^4 $: the moduli space as an ${N}$ = 2 supersymmetric Higgs branch}",
    eprint = "2604.21980",
    archivePrefix = "arXiv",
    primaryClass = "hep-th",
    doi = "10.1007/JHEP07(2026)240",
    journal = "JHEP",
    volume = "07",
    pages = "240",
    year = "2026"
}

@article{Poppitz:2017ivi,
    author = "Poppitz, Erich and Shalchian T., M. Erfan",
    title = "{String tensions in deformed Yang-Mills theory}",
    eprint = "1708.08821",
    archivePrefix = "arXiv",
    primaryClass = "hep-th",
    doi = "10.1007/JHEP01(2018)029",
    journal = "JHEP",
    volume = "01",
    pages = "029",
    year = "2018"
}

@article{GarciaPerez:2013idu,
    author = "Garc{\'\i}a P{\'e}rez, Margarita and Gonz{\'a}lez-Arroyo, Antonio and Okawa, Masanori",
    title = "{Spatial volume dependence for 2+1 dimensional SU(N) Yang-Mills theory}",
    eprint = "1307.5254",
    archivePrefix = "arXiv",
    primaryClass = "hep-lat",
    reportNumber = "IFT-UAM-CSIC-13-066, FTUAM-13-12, HUPD-1306",
    doi = "10.1007/JHEP09(2013)003",
    journal = "JHEP",
    volume = "09",
    pages = "003",
    year = "2013"
}

@article{Hayashi:2024gxv,
    author = "Hayashi, Yui and Tanizaki, Yuya and Watanabe, Hiromasa",
    title = "{Non-supersymmetric duality cascade of QCD(BF) via semiclassics on $R^2 \times T^2$ with the baryon-'t Hooft flux}",
    eprint = "2404.16803",
    archivePrefix = "arXiv",
    primaryClass = "hep-th",
    reportNumber = "YITP-24-41",
    doi = "10.1007/JHEP07(2024)033",
    journal = "JHEP",
    volume = "07",
    pages = "033",
    year = "2024"
}

@article{Hayashi:2024qkm,
    author = "Hayashi, Yui and Tanizaki, Yuya",
    title = "{Semiclassics for the QCD vacuum structure through T$^{2}$-compactification with the baryon-'t Hooft flux}",
    eprint = "2402.04320",
    archivePrefix = "arXiv",
    primaryClass = "hep-th",
    reportNumber = "YITP-24-15",
    doi = "10.1007/JHEP08(2024)001",
    journal = "JHEP",
    volume = "08",
    pages = "001",
    year = "2024"
}

@article{Hayashi:2023wwi,
    author = "Hayashi, Yui and Tanizaki, Yuya and Watanabe, Hiromasa",
    title = "{Semiclassical analysis of the bifundamental QCD on~$R^2 \times T^2$ with 't Hooft flux}",
    eprint = "2307.13954",
    archivePrefix = "arXiv",
    primaryClass = "hep-th",
    reportNumber = "YITP-23-96",
    doi = "10.1007/JHEP10(2023)146",
    journal = "JHEP",
    volume = "10",
    pages = "146",
    year = "2023"
}

@article{Junior:2025gxg,
    author = "Junior, David R. and Krein, Gast{\~a}o and Oxman, Luis E. and Soares, Bruno R.",
    title = "{Study of the Emergence of a Gluon Mass Scale from Center Vortices Using a Wave-Functional Formalism}",
    eprint = "2510.19103",
    archivePrefix = "arXiv",
    primaryClass = "hep-th",
    doi = "10.1103/fs1k-rjbg",
    journal = "Phys. Rev. Lett.",
    volume = "136",
    number = "11",
    pages = "111902",
    year = "2026"
}

@article{ArabiArdehali:2026kvt,
    author = "Arabi Ardehali, Arash and Resnick, Daniel J.",
    title = "{Perturbative Coulomb branches on $R^3 \times S^1$: the global D-term potential}",
    eprint = "2604.27066",
    archivePrefix = "arXiv",
    primaryClass = "hep-th",
    month = "4",
    year = "2026"
}

@article{Nguyen:2023rww,
    author = {Nguyen, Mendel and {\"U}nsal, Mithat},
    title = "{Refined instanton analysis of the 2D $CP^{N-1}$ model: mass gap, theta dependence, and mirror symmetry}",
    eprint = "2309.12178",
    archivePrefix = "arXiv",
    primaryClass = "hep-th",
    doi = "10.1007/JHEP03(2025)162",
    journal = "JHEP",
    volume = "03",
    pages = "162",
    year = "2025"
}

@article{Anber:2017pak,
    author = {Anber, Mohamed M. and Vincent-Genod, Lo{\"\i}c},
    title = "{Classification of compactified $su(N_c)$ gauge theories with fermions in all representations}",
    eprint = "1704.08277",
    archivePrefix = "arXiv",
    primaryClass = "hep-th",
    doi = "10.1007/JHEP12(2017)028",
    journal = "JHEP",
    volume = "12",
    pages = "028",
    year = "2017"
}

@article{Lippert:1997qx,
    author = "Lippert, Thomas",
    editor = "Meyer-Ortmanns, H. and Klumper, A.",
    title = "{The Hybrid Monte Carlo algorithm for quantum chromodynamics}",
    eprint = "hep-lat/9712019",
    archivePrefix = "arXiv",
    reportNumber = "HLRZ-1997-72",
    doi = "10.1007/BFb0106881",
    journal = "Lect. Notes Phys.",
    volume = "508",
    pages = "122",
    year = "1998"
}

@article{Nguyen:2025voy,
    author = {Nguyen, Mendel and {\"U}nsal, Mithat},
    title = "{Self-dual monopole loops, instantons and confinement}",
    eprint = "2509.09625",
    archivePrefix = "arXiv",
    primaryClass = "hep-th",
    month = "9",
    year = "2025"
}

@article{Bonati:2018rfg,
    author = "Bonati, Claudio and Cardinali, Marco and D'Elia, Massimo",
    title = "{$\theta$ dependence in trace deformed $SU(3)$ Yang-Mills theory: a lattice study}",
    eprint = "1807.06558",
    archivePrefix = "arXiv",
    primaryClass = "hep-lat",
    doi = "10.1103/PhysRevD.98.054508",
    journal = "Phys. Rev. D",
    volume = "98",
    number = "5",
    pages = "054508",
    year = "2018"
}

@article{Bonati:2020lal,
    author = "Bonati, Claudio and Cardinali, Marco and D'Elia, Massimo and Giordano, Matteo and Mazziotti, Fabrizio",
    title = "{Reconfinement, localization and thermal monopoles in $SU(3)$ trace-deformed Yang-Mills theory}",
    eprint = "2012.13246",
    archivePrefix = "arXiv",
    primaryClass = "hep-lat",
    doi = "10.1103/PhysRevD.103.034506",
    journal = "Phys. Rev. D",
    volume = "103",
    number = "3",
    pages = "034506",
    year = "2021"
}

@article{Bonati:2025hik,
    author = "Bonati, Claudio and Caselle, Michele and Negro, Alessio and Panfalone, Dario and Verzichelli, Lorenzo",
    title = "{Effective string description of the reconfined phase in the trace deformed SU(2) Yang-Mills theory in (2+1) dimensions}",
    eprint = "2501.13684",
    archivePrefix = "arXiv",
    primaryClass = "hep-lat",
    doi = "10.22323/1.466.0394",
    journal = "PoS",
    volume = "LATTICE2024",
    pages = "394",
    year = "2025"
}

@article{Athenodorou:2020clr,
    author = "Athenodorou, Andreas and Cardinali, Marco and D'Elia, Massimo",
    title = "{Spectrum of trace deformed Yang-Mills theories}",
    eprint = "2010.03618",
    archivePrefix = "arXiv",
    primaryClass = "hep-lat",
    doi = "10.1103/PhysRevD.104.074510",
    journal = "Phys. Rev. D",
    volume = "104",
    number = "7",
    pages = "074510",
    year = "2021"
}

@article{GarciaPerez:1993lic,
    author = "Garcia Perez, Margarita and Gonzalez-Arroyo, Antonio and Snippe, Jeroen R. and van Baal, Pierre",
    title = "{Instantons from over - improved cooling}",
    eprint = "hep-lat/9309009",
    archivePrefix = "arXiv",
    reportNumber = "INLO-PUB-11-93, FTUAM-93-31",
    doi = "10.1016/0550-3213(94)90631-9",
    journal = "Nucl. Phys. B",
    volume = "413",
    pages = "535--552",
    year = "1994"
}

@article{deForcrand:1995qq,
    author = "de Forcrand, Philippe and Garcia Perez, Margarita and Stamatescu, Ion-Olimpiu",
    editor = "Kieu, T. D. and McKellar, B. H. J. and Guttmann, A. J.",
    title = "{Improved cooling algorithm for gauge theories}",
    eprint = "hep-lat/9509064",
    archivePrefix = "arXiv",
    reportNumber = "IPS-95-21, INLO-PUB-11-95, HD-THEP-95-41",
    doi = "10.1016/0920-5632(96)00172-7",
    journal = "Nucl. Phys. B Proc. Suppl.",
    volume = "47",
    pages = "777--780",
    year = "1996"
}

@article{Montero:2000mv,
    author = "Montero, Alvaro",
    title = "{Numerical analysis of fractional charge solutions on the torus}",
    eprint = "hep-lat/0004009",
    archivePrefix = "arXiv",
    reportNumber = "MIT-CTP-2972",
    doi = "10.1088/1126-6708/2000/05/022",
    journal = "JHEP",
    volume = "05",
    pages = "022",
    year = "2000"
}

@article{Kraan:1998sn,
    author = "Kraan, Thomas C. and van Baal, Pierre",
    title = "{Monopole constituents inside SU(n) calorons}",
    eprint = "hep-th/9806034",
    archivePrefix = "arXiv",
    reportNumber = "INLO-PUB-9-98",
    doi = "10.1016/S0370-2693(98)00799-0",
    journal = "Phys. Lett. B",
    volume = "435",
    pages = "389--395",
    year = "1998"
}

@article{Gross:1980br,
    author = "Gross, David J. and Pisarski, Robert D. and Yaffe, Laurence G.",
    title = "{QCD and Instantons at Finite Temperature}",
    reportNumber = "PRINT-80-0538 (PRINCETON)",
    doi = "10.1103/RevModPhys.53.43",
    journal = "Rev. Mod. Phys.",
    volume = "53",
    pages = "43",
    year = "1981"
}

@article{Luscher:1982ma,
    author = "Luscher, M.",
    title = "{Some Analytic Results Concerning the Mass Spectrum of Yang-Mills Gauge Theories on a Torus}",
    reportNumber = "BUTP-23/1982",
    doi = "10.1016/0550-3213(83)90436-4",
    journal = "Nucl. Phys. B",
    volume = "219",
    pages = "233--261",
    year = "1983"
}

@article{Bjorken:1979hv,
    author = "Bjorken, J. D.",
    title = "{Elements of Quantum Chromodynamics}",
    reportNumber = "SLAC-PUB-2372",
    doi = "10.1007/978-1-4899-6691-9_5",
    journal = "Prog. Math. Phys.",
    volume = "4",
    pages = "423--561",
    month = "12",
    year = "1979"
}

@article{Dunne:2016nmc,
    author = {Dunne, Gerald V. and {\"U}nsal, Mithat},
    title = "{New Nonperturbative Methods in Quantum Field Theory: From Large-N Orbifold Equivalence to Bions and Resurgence}",
    eprint = "1601.03414",
    archivePrefix = "arXiv",
    primaryClass = "hep-th",
    doi = "10.1146/annurev-nucl-102115-044755",
    journal = "Ann. Rev. Nucl. Part. Sci.",
    volume = "66",
    pages = "245--272",
    year = "2016"
}

@article{Kraan:1998pm,
    author = "Kraan, Thomas C. and van Baal, Pierre",
    title = "{Periodic instantons with nontrivial holonomy}",
    eprint = "hep-th/9805168",
    archivePrefix = "arXiv",
    reportNumber = "INLO-PUB-5-98",
    doi = "10.1016/S0550-3213(98)00590-2",
    journal = "Nucl. Phys. B",
    volume = "533",
    pages = "627--659",
    year = "1998"
}

@article{vanBaal:1982ag,
    author = "van Baal, Pierre",
    title = "{Some Results for SU(N) Gauge Fields on the Hypertorus}",
    reportNumber = "Print-82-0072 (UTRECHT)",
    doi = "10.1007/BF01403503",
    journal = "Commun. Math. Phys.",
    volume = "85",
    pages = "529",
    year = "1982"
}

@article{Unsal:2020yeh,
    author = {{\"U}nsal, Mithat},
    title = "{Strongly coupled QFT dynamics via TQFT coupling}",
    eprint = "2007.03880",
    archivePrefix = "arXiv",
    primaryClass = "hep-th",
    doi = "10.1007/JHEP11(2021)134",
    journal = "JHEP",
    volume = "11",
    pages = "134",
    year = "2021"
}

@article{GarciaPerez:1999hs,
    author = "Garcia Perez, Margarita and Gonzalez-Arroyo, Antonio and Montero, Alvaro and van Baal, Pierre",
    title = "{Calorons on the lattice: A New perspective}",
    eprint = "hep-lat/9903022",
    archivePrefix = "arXiv",
    reportNumber = "FTUAM-99-6, IFT-UAM-CSIC-99-8, INLO-PUB-6-99",
    doi = "10.1088/1126-6708/1999/06/001",
    journal = "JHEP",
    volume = "06",
    pages = "001",
    year = "1999"
}

@inproceedings{Gonzalez-Arroyo:1997ugn,
    author = "Gonzalez-Arroyo, Antonio",
    title = "{Yang-Mills fields on the four-dimensional torus. Part 1.: Classical theory}",
    booktitle = "{Advanced Summer School on Nonperturbative Quantum Field Physics}",
    eprint = "hep-th/9807108",
    archivePrefix = "arXiv",
    reportNumber = "FTUAM-97-18",
    pages = "57--91",
    month = "6",
    year = "1997"
}

@article{Hayashi:2024psa,
    author = "Hayashi, Yui and Misumi, Tatsuhiro and Tanizaki, Yuya",
    title = "{Monopole-vortex continuity of $N$ = 1 super Yang-Mills theory on $R^2 \times S^1 \times S^1$ with 't Hooft twist}",
    eprint = "2410.21392",
    archivePrefix = "arXiv",
    primaryClass = "hep-th",
    reportNumber = "YITP-24-136",
    doi = "10.1007/JHEP05(2025)194",
    journal = "JHEP",
    volume = "05",
    pages = "194",
    year = "2025"
}

@article{Hayashi:2025mgk,
    author = "Hayashi, Yui and Misumi, Tatsuhiro and Tanizaki, Yuya",
    title = "{On 2d perimeter law in $N$=1 super Yang-Mills theory on $R^2 \times S^1 \times S^1$ with 'tt Hooft twist from Monopole-Vortex continuity}",
    doi = "10.1142/S0217751X25480069",
    journal = "Int. J. Mod. Phys. A",
    volume = "41",
    number = "07",
    pages = "2548006",
    year = "2026"
}

@article{Anber:2015wha,
    author = "Anber, Mohamed M. and Poppitz, Erich",
    title = "{On the global structure of deformed Yang-Mills theory and QCD(adj) on $R^3 \times S^1$}",
    eprint = "1508.00910",
    archivePrefix = "arXiv",
    primaryClass = "hep-th",
    doi = "10.1007/JHEP10(2015)051",
    journal = "JHEP",
    volume = "10",
    pages = "051",
    year = "2015"
}

@article{Nguyen:2024ikq,
    author = {Nguyen, Mendel and Sulejmanpasic, Tin and {\"U}nsal, Mithat},
    title = "{Phases of Theories with ZN 1-Form Symmetry, and the Roles of Center Vortices and Magnetic Monopoles}",
    eprint = "2401.04800",
    archivePrefix = "arXiv",
    primaryClass = "hep-th",
    doi = "10.1103/PhysRevLett.134.141902",
    journal = "Phys. Rev. Lett.",
    volume = "134",
    number = "14",
    pages = "141902",
    year = "2025"
}

@article{Witten:1982df,
    author = "Witten, Edward",
    title = "{Constraints on Supersymmetry Breaking}",
    reportNumber = "PRINT-82-0163 (PRINCETON)",
    doi = "10.1016/0550-3213(82)90071-2",
    journal = "Nucl. Phys. B",
    volume = "202",
    pages = "253",
    year = "1982"
}

@article{GonzalezArroyo:1987ycm,
    author = "Gonzalez Arroyo, Antonio and Korthals Altes, C. P.",
    title = "{The Spectrum of Yang-Mills Theory in a Small Twisted Box}",
    reportNumber = "FTUAM/87-20",
    doi = "10.1016/0550-3213(88)90068-5",
    journal = "Nucl. Phys. B",
    volume = "311",
    pages = "433--449",
    year = "1988"
}

@article{Unsal:2008ch,
    author = "Unsal, Mithat and Yaffe, Laurence G.",
    title = "{Center-stabilized Yang-Mills theory: Confinement and large N volume independence}",
    eprint = "0803.0344",
    archivePrefix = "arXiv",
    primaryClass = "hep-th",
    reportNumber = "SLAC-PUB-13144, NSF-KITP-08-16",
    doi = "10.1103/PhysRevD.78.065035",
    journal = "Phys. Rev. D",
    volume = "78",
    pages = "065035",
    year = "2008"
}

@article{Poppitz:2022rxv,
    author = "Poppitz, Erich and Wandler, F. David",
    title = "{Gauge theory geography: charting a path between semiclassical islands}",
    eprint = "2211.10347",
    archivePrefix = "arXiv",
    primaryClass = "hep-th",
    doi = "10.1007/JHEP02(2023)014",
    journal = "JHEP",
    volume = "02",
    pages = "014",
    year = "2023"
}

@article{Cox:2021vsa,
    author = "Cox, Andrew A. and Poppitz, Erich and Wandler, F. David",
    title = "{The mixed 0-form/1-form anomaly in Hilbert space: pouring the new wine into old bottles}",
    eprint = "2106.11442",
    archivePrefix = "arXiv",
    primaryClass = "hep-th",
    doi = "10.1007/JHEP10(2021)069",
    journal = "JHEP",
    volume = "10",
    pages = "069",
    year = "2021"
}

@article{Lee:1997vp,
    author = "Lee, Ki-Myeong and Yi, Piljin",
    title = "{Monopoles and instantons on partially compactified D-branes}",
    eprint = "hep-th/9702107",
    archivePrefix = "arXiv",
    reportNumber = "CU-TP-817",
    doi = "10.1103/PhysRevD.56.3711",
    journal = "Phys. Rev. D",
    volume = "56",
    pages = "3711--3717",
    year = "1997"
}

@book{Greensite:2011zz,
    author = "Greensite, Jeff",
    title = "{An introduction to the confinement problem}",
    doi = "10.1007/978-3-642-14382-3",
    publisher="Springer",
    volume = "972",
    year = "2020"
}

@article{Gaiotto:2017yup,
    author = "Gaiotto, Davide and Kapustin, Anton and Komargodski, Zohar and Seiberg, Nathan",
    title = "{Theta, Time Reversal, and Temperature}",
    eprint = "1703.00501",
    archivePrefix = "arXiv",
    primaryClass = "hep-th",
    doi = "10.1007/JHEP05(2017)091",
    journal = "JHEP",
    volume = "05",
    pages = "091",
    year = "2017"
}

@article{Hayashi:2024yjc,
    author = "Hayashi, Yui and Tanizaki, Yuya",
    title = "{Unifying Monopole and Center Vortex as the Semiclassical Confinement Mechanism}",
    eprint = "2405.12402",
    archivePrefix = "arXiv",
    primaryClass = "hep-th",
    reportNumber = "YITP-24-59",
    doi = "10.1103/PhysRevLett.133.171902",
    journal = "Phys. Rev. Lett.",
    volume = "133",
    number = "17",
    pages = "171902",
    year = "2024"
}

@article{Guvendik:2024umd,
    author = {G{\"u}vendik, Canberk and Schaefer, Thomas and {\"U}nsal, Mithat},
    title = "{The metamorphosis of semi-classical mechanisms of confinement: from monopoles on $R^3 \times S^1$ to center-vortices on $R^2 \times T^2$}",
    eprint = "2405.13696",
    archivePrefix = "arXiv",
    primaryClass = "hep-th",
    doi = "10.1007/JHEP11(2024)163",
    journal = "JHEP",
    volume = "11",
    pages = "163",
    year = "2024"
}

@article{Gonzalez-Arroyo:1998hjb,
    author = "Gonzalez-Arroyo, Antonio and Montero, A.",
    title = "{Selfdual vortex - like configurations in SU(2) Yang-Mills theory}",
    eprint = "hep-th/9809037",
    archivePrefix = "arXiv",
    reportNumber = "FTUAM-98-17, IFT-UAM-CSIC-98-12",
    doi = "10.1016/S0370-2693(98)01229-5",
    journal = "Phys. Lett. B",
    volume = "442",
    pages = "273--278",
    year = "1998"
}

@article{Montero:2000pb,
    author = "Montero, Alvaro",
    title = "{Vortex configurations in the large N limit}",
    eprint = "hep-lat/0004002",
    archivePrefix = "arXiv",
    reportNumber = "MIT-CTP-2967",
    doi = "10.1016/S0370-2693(00)00572-4",
    journal = "Phys. Lett. B",
    volume = "483",
    pages = "309--314",
    year = "2000"
}

@article{Tanizaki:2022plm,
    author = {Tanizaki, Yuya and {\"U}nsal, Mithat},
    title = "{Semiclassics with 
't Hooft flux background for QCD with 2-index quarks}",
    eprint = "2205.11339",
    archivePrefix = "arXiv",
    primaryClass = "hep-th",
    reportNumber = "YITP-22-45",
    doi = "10.1007/JHEP08(2022)038",
    journal = "JHEP",
    volume = "08",
    pages = "038",
    year = "2022"
}

@article{Poppitz:2021cxe,
    author = "Poppitz, Erich",
    title = "{Notes on Confinement on $R^3 \times S^1$: From Yang-Mills, Super-Yang-Mills, and QCD (adj) to QCD(F)}",
    eprint = "2111.10423",
    archivePrefix = "arXiv",
    primaryClass = "hep-th",
    doi = "10.3390/sym14010180",
    journal = "Symmetry",
    volume = "14",
    number = "1",
    pages = "180",
    year = "2022"
}

@article{Unsal:2007jx,
    author = "Unsal, Mithat",
    title = "{Magnetic bion condensation: A New mechanism of confinement and mass gap in four dimensions}",
    eprint = "0709.3269",
    archivePrefix = "arXiv",
    primaryClass = "hep-th",
    reportNumber = "SLAC-PUB-12825",
    doi = "10.1103/PhysRevD.80.065001",
    journal = "Phys. Rev. D",
    volume = "80",
    pages = "065001",
    year = "2009"
}

@article{Unsal:2007vu,
    author = "Unsal, Mithat",
    title = "{Abelian duality, confinement, and chiral symmetry breaking in QCD(adj)}",
    eprint = "0708.1772",
    archivePrefix = "arXiv",
    primaryClass = "hep-th",
    reportNumber = "SLAC-PUB-12726",
    doi = "10.1103/PhysRevLett.100.032005",
    journal = "Phys. Rev. Lett.",
    volume = "100",
    pages = "032005",
    year = "2008"
}

@article{Gonzalez-Arroyo:1995ynx,
    author = "Gonzalez-Arroyo, Antonio and Martinez, P.",
    title = "{Investigating Yang-Mills theory and confinement as a function of the spatial volume}",
    eprint = "hep-lat/9507001",
    archivePrefix = "arXiv",
    reportNumber = "FTUAM-95-15A",
    doi = "10.1016/0550-3213(95)00601-X",
    journal = "Nucl. Phys. B",
    volume = "459",
    pages = "337--354",
    year = "1996"
}

@article{tHooft:1979rtg,
    author = "'t Hooft, Gerard",
    title = "{A Property of Electric and Magnetic Flux in Nonabelian Gauge Theories}",
    reportNumber = "PRINT-79-0117 (UTRECHT)",
    doi = "10.1016/0550-3213(79)90595-9",
    journal = "Nucl. Phys. B",
    volume = "153",
    pages = "141--160",
    year = "1979"
}

@article{tHooft:1981sps,
    author = "'t Hooft, Gerard",
    title = "{Aspects of Quark Confinement}",
    doi = "10.1088/0031-8949/24/5/007",
    journal = "Phys. Scripta",
    volume = "24",
    pages = "841--846",
    year = "1981"
}

@article{Wandler:2024hsq,
    author = "Wandler, F. David",
    title = "{Numerical fractional instantons in SU(2): center vortices, monopoles, and a sharp transition between them}",
    eprint = "2406.07636",
    archivePrefix = "arXiv",
    primaryClass = "hep-lat",
    month = "6",
    year = "2024"
}

@article{Anber:2025vjo,
    author = "Anber, Mohamed M. and Poppitz, Erich",
    title = "{Mass-deformed super Yang-Mills theory on $T^4$: sum over twisted sectors, $\theta$-angle, and CP violation}",
    eprint = "2509.00157",
    archivePrefix = "arXiv",
    primaryClass = "hep-th",
    doi = "10.1007/JHEP10(2025)182",
    journal = "JHEP",
    volume = "10",
    pages = "182",
    year = "2025"
}

@article{Anber:2024mco,
    author = "Anber, Mohamed M. and Poppitz, Erich",
    title = "{Higher-order gaugino condensates on a twisted ${T}^4 $}",
    eprint = "2408.16058",
    archivePrefix = "arXiv",
    primaryClass = "hep-th",
    doi = "10.1007/JHEP02(2025)114",
    journal = "JHEP",
    volume = "02",
    pages = "114",
    year = "2025"
}

@article{tHooft:1981nnx,
    author = "'t Hooft, Gerard",
    title = "{Some Twisted Selfdual Solutions for the Yang-Mills Equations on a Hypertorus}",
    reportNumber = "CALT-68-835",
    doi = "10.1007/BF01208900",
    journal = "Commun. Math. Phys.",
    volume = "81",
    pages = "267--275",
    year = "1981"
}

@article{RTN:1993ilw,
    author = "Garcia Perez, M. and others",
    collaboration = "RTN",
    title = "{Instanton like contributions to the dynamics of Yang-Mills fields on the twisted torus}",
    eprint = "hep-lat/9302007",
    archivePrefix = "arXiv",
    reportNumber = "FTUAM-92-39",
    doi = "10.1016/0370-2693(93)91069-Y",
    journal = "Phys. Lett. B",
    volume = "305",
    pages = "366--374",
    year = "1993"
}

@article{Gonzalez-Arroyo:2023kqv,
    author = "Gonz\'alez-Arroyo, Antonio",
    title = "{On the fractional instanton liquid picture of the Yang-Mills vacuum and Confinement}",
    eprint = "2302.12356",
    archivePrefix = "arXiv",
    primaryClass = "hep-th",
    reportNumber = "IFT-UAM/CSIC-23-20",
    month = "2",
    year = "2023"
}

@article{Tanizaki:2022ngt,
    author = {Tanizaki, Yuya and \"Unsal, Mithat},
    title = "{Center vortex and confinement in Yang\textendash{}Mills theory and QCD with anomaly-preserving compactifications}",
    eprint = "2201.06166",
    archivePrefix = "arXiv",
    primaryClass = "hep-th",
    reportNumber = "YITP-22-04",
    doi = "10.1093/ptep/ptac042",
    journal = "PTEP",
    volume = "2022",
    number = "4",
    pages = "04A108",
    year = "2022"
}

@article{Anber:2023sjn,
    author = "Anber, Mohamed M. and Poppitz, Erich",
    title = "{Multi-fractional instantons in SU(N) Yang-Mills theory on the twisted four-torus}",
    eprint = "2307.04795",
    archivePrefix = "arXiv",
    primaryClass = "hep-th",
    doi = "10.1007/JHEP09(2023)095",
    journal = "JHEP",
    volume = "09",
    pages = "095",
    year = "2023"
}

@article{Anber:2022qsz,
    author = "Anber, Mohamed M. and Poppitz, Erich",
    title = "{The gaugino condensate from asymmetric four-torus with twists}",
    eprint = "2210.13568",
    archivePrefix = "arXiv",
    primaryClass = "hep-th",
    doi = "10.1007/JHEP01(2023)118",
    journal = "JHEP",
    volume = "01",
    pages = "118",
    year = "2023"
}

@article{Dorey:2002ik,
    author = "Dorey, Nick and Hollowood, Timothy J. and Khoze, Valentin V. and Mattis, Michael P.",
    title = "{The Calculus of many instantons}",
    eprint = "hep-th/0206063",
    archivePrefix = "arXiv",
    doi = "10.1016/S0370-1573(02)00301-0",
    journal = "Phys. Rept.",
    volume = "371",
    pages = "231--459",
    year = "2002"
}

@article{Duane:1987216,
title = {Hybrid Monte Carlo},
journal = {Physics Letters B},
volume = {195},
number = {2},
pages = {216-222},
year = {1987},
issn = {0370-2693},
doi = {https://doi.org/10.1016/0370-2693(87)91197-X},
url = {https://www.sciencedirect.com/science/article/pii/037026938791197X},
author = {Simon Duane and A.D. Kennedy and Brian J. Pendleton and Duncan Roweth}
}

\bibliographystyle{JHEP}

\end{document}